\documentclass[preprint2]{aastex}
\pdfoutput=1

 \usepackage{graphicx}
 \usepackage{float}

\slugcomment{Manuscript abc, submitted 23:59:59.9999, December 31, 2012}
\shorttitle{STIS observations of Fomalhaut}
\shortauthors{Kalas et al.}

\begin{document}

\title{ 
{\small Submitted to ApJ, Dec. 31, 2012}\\
STIS Coronagraphic Imaging of Fomalhaut: \\
    Main Belt Structure and the Orbit of Fomalhaut $b$}

\author{
Paul Kalas\altaffilmark{1,2}, James R. Graham\altaffilmark{1,3}, Michael P.
Fitzgerald\altaffilmark{4}, Mark Clampin\altaffilmark{5}
}
\affil{}

\altaffiltext{1}{Astronomy Department, University of California, Berkeley, CA 94720}
\altaffiltext{2}{SETI Institute, Mountain View, CA 94043}
\altaffiltext{3}{Dunlap Institute for Astronomy and Astrophysics, \\ University of Toronto, Toronto, Canada}
\altaffiltext{4}{Astronomy Department, UCLA, Los Angeles, CA}
\altaffiltext{5}{NASA Goddard Space Flight Center, Greenbelt, MD}

\begin{abstract}
We present new optical coronagraphic data of the bright star Fomalhaut
obtained with the Hubble Space Telescope in 2010 and 2012 using STIS that extend the astrometric baseline of 
previous 2004/2006 observations with ACS/HRC.  
Fomalhaut $b$ is recovered at both epochs to high significance.  
The observations include the discoveries of tenuous nebulosity beyond the main dust belt 
detected to at least 209 AU projected radius
and a $\sim$50 AU wide
azimuthal gap in the belt northward of Fomalhaut $b$.  The two epochs of
STIS photometry exclude optical variability greater than 35\%.
The morphology of Fomalhaut b appears elliptical in the 2010 and 2012 detections.  
We show that residual noise in the processed data can plausibly result in point
sources appearing extended.  A Markov chain Monte Carlo analysis
demonstrates that irrespective of any assumption regarding inclination to the line of sight, 
the orbit of Fomalhaut b is highly eccentric, with $e=0.8\pm0.1$, $a=177 \pm 68$ AU, and $q=32\pm24$ AU.  
Fomalhaut b is apsidally aligned with the belt and 90\% of allowed orbits have mutual inclination $\leq36\degr$.
Fomalhaut $b$'s orbit is belt-crossing in the sky plane projection, but only 12\% of 
possible orbits have ascending or descending nodes within a 25 AU wide belt annulus (133 -158 AU).  
The high eccentricity invokes a dynamical history where Fomalhaut b may have experienced a significant dynamical interaction with a hypothetical planet Fomalhaut c, and the current orbital configuration may be relatively short-lived.
The Tisserand parameter with respect to a hypothetical Fomalhaut planet at 30 AU or at 120 AU lies in the range $2-3$, similar to highly eccentric dwarf planets in our solar system.  
The new value for the periastron distance diminishes the Hill radius of Fomalhaut b and any weakly bound satellite system surrounding a planet would be sheared and
dynamically heated at periapse.  We argue that Fomalhaut b's minimum mass is that of a dwarf planet in order for a circumplanetary satellite
system to remain bound to a sufficient radius from the planet to be consistent with the dust scattered light hypothesis.
Fomalhaut b may be optically bright because the recent passage through periapse and/or the ascending node
has increased the erosion rates of planetary satellites.
In the coplanar case, Fomalhaut b will collide with the main belt around 2032, and the subsequent emergent phenomena
may help determine its physical nature.  We show that if Fomalhaut b
has a bound dust cloud, then the cloud survives hundreds of belt crossings.  
If Fomalhaut b has the mass of a gas giant planet, then belt crossings will erode the belt edges after $\sim10^2$ orbits.  

\end{abstract}

\section{Introduction}
\label{sec:introduction}

The number of candidate exoplanets detected by direct imaging
techniques has recently surpassed the number of solar system planets.  In every case
there are a significant number of challenges in observation and
interpretation.  Observationally, a faint
companion to a star or brown dwarf must be shown to
be a real astrophysical feature instead of an instrumental
artifact.  Precision astrometry over multiple epochs
of observations is also required to establish common
proper motion and orbital motion.  Complementary
data and analyses are essential for estimating the age of the
system, which is required to constrain the mass of the
object from the theory of planet luminosity evolution
as well as dynamics.  And finally, even though an
object may pass all of these tests, appear to
be a bound companion, and have mass under the brown dwarf mass
limit of $\sim$13 $M_J$, the question of planet identity
may persist, given open questions such as  whether or not a planet mass
object needs to be bound to a star (as opposed to a brown dwarf), or if the mode of
formation should be critical to the definition
of a planet \citep{basri06a, soter06a, schneider11a}.

The Fomalhaut system certainly presents a number of
open questions and puzzling observations.  The discovery
of infrared excess due to circumstellar dust was
firmly established by a series of infrared and sub-mm
observations \citep{backman87a, zuckerman93a}.  Resolved thermal infrared images
at 850 $\mu$m demonstrated that
the debris disk was in fact a torus of material in a region 100$-$140 AU
radius from the star \citep{holland98a, dent00a}.
A dust torus could be sustained over the age of the system if planet-mass objects 
served to dynamically halt or delay the inward drift of grains governed
by Poynting-Robertson drag \citep{ozernoy00a, moromartin02}.  Higher
resolution thermal images at 450 $\mu$m suggested the presence
of warmer dust concentrated within 100 AU radius to the southeast
of the star in an arc-like morphology \citep{wyatt02a, holland03a}.

Motivated by these findings, the first Hubble Space Telescope (HST)
coronagraphic observations of Fomalhaut were intended to search
for a planet mass object within the dust belt using the Advanced Camera for Surveys
High Resolution Channel (ACS/HRC;  GO 9862; PI Kalas).  These relatively
shallow observations with the F814W filter (0.8 $\mu$m) yielded
the first detection of the dust belt in optical scattered light.  Deeper, follow-up
observations (GO 10390; PI Kalas) were conducted in both
the F606W (0.6 $\mu$m) and F814W filter in 2004.

The first-epoch, 2004 data revealed that the geometric center of the dust belt is offset
to the northwest of the stellar position by $\sim$15 AU, and the inner edge
of the belt is consistent with a knife-edge.  \citet{kalas05a} proposed
that these two facts were robust indirect evidence for a planet orbiting
interior to the 133 AU inner border of the belt.  At roughly the same time,
new thermal infrared images using the Spitzer Space Telescope
and the Caltech Submillimeter Observatory revealed asymmetry
in the disk emission due to the fact that the southeast side
of the belt is significantly closer to the star than the northwest side
\citep{stapelfeldt04a, marsh05a}.

To search for the putative planet and determine the scattered light
colors of the belt, deeper coronagraphic images with HST in
three filters (F435W, F606W and F814W) were obtained.  
The second-epoch, 2006 observations resulted in the discovery of a point source 18 AU interior to the
inner border of the belt \citep{kalas08a}.  The point source was verified in multiple
data sets, in two filters, and was detected in the F606W data
obtained in 2004.  Due to the high proper motion of Fomalhaut (0.4$''$/yr),
the point source was  easily separated from background stars, but a 
small offset between epochs suggested orbital motion in the
counterclockwise direction as projected on the sky.  Thus, the point
source was physically associated with the central star and designated 
Fomalhaut b.

A key surprise with the Fomalhaut b observations was the
unexpected blue color (i.e., an unexpectedly high luminosity at optical
wavelengths).   \citet{kalas08a} presented ground-based observations at 1.6 $\mu$m and 3.8 $\mu$m that did not detect Fomalhaut b,
establishing that non-thermal sources probably contribute a fraction of its 0.6 $\mu$m brightness.
They proposed that light reflected from a circumplanetary dust disk
could account for the visible-light flux, though the data
also indicated a dimming of Fomalhaut b between 2004
and 2006 in the 0.6 $\mu$m detections.  If the 0.8 $\mu$m 
flux was entirely attributed to thermal emission, then 
the mass of Fomalhaut would be $<$3 M$_J$.  This
upper limit to the mass was also consistent
with dynamical theory that showed the inner edge of the dust belt
could not reside as close as 18 AU from a planet
unless the planet mass is below a few Jupiter
masses  \citep{quillen06a, chiang09a}.  

An alternative model studied quantitatively in Kalas et al. (2008) is that
Fomalhaut b is a rare and short lived dust cloud that has recently appeared
in the system due to the collision of two planetesimals (see also Currie et al. 2012
and Galicher et al. 2013).  The hypothetical conversion 
of a 10 km radius planetesimal into $0.1-0.2~\mu$m water ice grains represented
the minimum mass (4.1$\times10^{18}$ g) that could explain the optical photometry
in terms of reflected light.  An alternate water ice cloud model that assumes a grain size distribution
between 0.08 $\mu$m and 1 mm corresponds to a 67 km radius planetesimal  (1.2$\times10^{21}$ g). 
However, the scenario of a disrupted planetesimal was deemed less likely
than the planetary rings hypothesis because:  
(1) observing a rare and short lived event is 
unlikely,  (2) planetesimal collisions are unlikely
far from the star where dynamical timescales are
relatively long ($P\sim$ 800 yrs), (3) modeling the dimming of the
dust cloud requires a fine-tuning of the model 
such that small-grains are quickly depleted from
the cloud just as the observations are conducted,
(4) the simulated dust cloud predicts optical
colors that do not agree with the observed color.

A new model was proposed by \citet{kennedy11a} where they
examined the origin and collisional evolution of irregular satellites
orbiting solar system planets.  The collisional erosion of
irregular satellites can produce an hourglass-shaped dust
cloud around a planet rather than a flattened circumplanetary
dust disk.  When applied to Fomalhaut b, a few lunar-mass planetesimal
dust cloud orbiting a 2$-$100 M$_\earth$ planet is 
consistent with the theoretical assumptions and observational constraints.

Additional mass limits for Fomalhaut b have been established
by modeling the non-detection of Fomalhaut b at mid-infrared wavelengths
with the Spitzer Space Telescope \citep{marengo09a, janson12a}.
Adopting the new age determination of $\sim$400 Myr for Fomalhaut \citep{mamajek12a}, 
Janson et al. estimate that the mass of Fomalhaut b is $\lesssim$1 M$_J$, consistent with previous findings.
However, their primary conclusion is that the \citet{kalas08a} dust-cloud model 
is the best fit to observations because they 
claim Fomalhaut b resides out of the sky plane, and this geometry
rules out reflection from a circumplanetary disk.  However,
this conclusion has at least three significant problems.  First, they presume
the circumplanetary dust is optically thick, when in fact there is no such constraint.
An optically thin, circumplanetary dust cloud would still permit
forward scattering if the current geometry puts the planet between
the host star and the observer.  Second, the orientation of
planetary ring systems relative to planet orbital planes in our solar system 
varies in the range 0$-$177$\degr$.  Determining the orbital geometry
of Fomalhaut b does not directly translate into knowledge of how the planetary
rings are oriented relative to the star and observer.  
Third, the assumption that Fomalhaut b is situated between the
star and observer is a tentative result from  \citet{lebouquin09a}.
These authors observed Fomalahaut A with VLT optical interferometry and the AMBER instrument,
finding that the spin axis of  the star extends out of the sky plane in the NE.  Given
a counterclockwise spin, the western portion of the belt and Fomalhaut b 
would reside out of the sky plane.  However, \citet{lebouquin09a} clearly state 
``...no check star is available in the dataset to secure the
sign of the AMBER phase...we cannot draw definite conclusions
before a real spectroastrometric reference has been observed.''

The tentative geometry suggested by  \citet{lebouquin09a} means that the brightest
hemisphere of the belt resides behind the sky plane and the
grain scattering phase function is strongly backscattering.  \citet{min10a} suggest that backscattering
can dominate in Fomalhaut's dust belt.   However, the backscatter model is consistent
with the observations $only$ if all grains smaller than 100 $\mu$m are absent from
the system.  
The radiation pressure blowout grain size for Fomalhaut is $8 -13 ~\mu$m, depending
on composition and porosity \citep{artymowicz97a, acke12a}.  Therefore,
if one accepts that debris disks are replenished by the collisional erosion of 
larger bodies, there should be a significant population of grains in the 13-100 $\mu$m
size range that ensures the belt is dominated by forward scattering.

Specifying the belt geometry by finding the orientation of the stellar spin axis is moot in any case because
it has been established that the spin-orbit alignment of exoplanets may be highly oblique or retrograde
\citep{hebrard08a, anderson10a, johnson11a}.  Moreover, debris disks are also known to
be misaligned with the stellar angular momentum orientation \citep{kennedy12a}.  Therefore,
the spin vectors of the host star(s), exoplanet(s) and debris belt(s) within any given
exoplanetary system are not necessarily aligned.

In our view the most significant observable is
that a dust belt with an asymmetric scattering phase function will exhibit preferential forward scattering.
Therefore the bright side of Fomalhaut's belt is out of the sky plane.  Fomalhaut b resides
near the faint part of the belt, which is likely behind the sky plane due to weaker backscattering.
The assumptions underlying the \citet{janson12a} argument that hypothetical planetary
rings surrounding Fomalhaut $b$ would not be seen in reflected light are therefore unsupported.

\citet{janson12a} note that the HST detections may not be trustworthy,
yet two independent groups have analyzed the same HST data and detected 
Fomalhaut b \citep{currie12a, galicher13a}.  Moreover, these groups
claim that  Fomalhaut $b$ is detected in a third optical filter, F435W (0.4 $\mu$m).
Fomalhaut b's optical variability is not confirmed, but the photometry
presented in  \citet{currie12a} and \citet{galicher13a} each
span the range of photometry given in \citet{kalas08a}. 
The photometric uncertainties are evidently greater than previously assumed
and the case for variability requires further work (c.f. Sec.~\ref{sec:acsphotometry}, Sec.~\ref{sec:photometry}).

More recent thermal infrared observations of Fomalhaut's dust belt have been
made with ALMA \citep{boley12a}, Herschel \citep{acke12a} and
the Australian Telescope Compact Array (ATCA) \citep{ricci12a}.
The physical properties of the dust belt derived
from these new observations generally support the results of previous studies.   
One significant new result from a study focused on the parent star is a revised, older age
for Fomalhaut \citep{mamajek12a}.  
The previous value of 200 Myr \citep{byn97a} is superceded by 440$\pm$40 Myr.
The older age means that any Fomalhaut planets have a lower
temperature for a given mass, which has a bearing on predicted
infrared detection limits, and other derivations
involving dynamical lifetime arguments must be revised.
\citet{mamajek12a} also calls attention to a likely
stellar companion 5.7$\times$10$^4$ AU from Fomalhaut called TW PsA.
Fomalhaut could be newly designated as Fomalhaut A, Fomalhaut b is
Fomalhaut Ab, and TW PsA is Fomalhaut B.  For the sake of
continuity with the prior literature, we continue
using the term Fomalhaut $b$.

In the present paper we describe the results of additional imaging
observations of Fomalhaut b using HST/STIS in 2010 and 2012 (Table 1).  Our main
goal is to derive the orbital elements of Fomalhaut b using astrometric
measurements from all four epochs of observation.
The most significant challenge is that 
follow-up observations using the original discovery instrument are precluded
due to a failure in the ACS/HRC electronics, which could not be restored during the
Servicing Mission 4.  The HRC was ideally suited for
high contrast observations, given fine sampling (25 mas pixels), three
coronagraphic occulters in the focal plane, and a Lyot stop which
suppressed light diffracted around telescope elements.  
By changing instruments to STIS, we had to accept a broader
optical bandpass, with different detector and coronagraphic
characteristics.  When Fomalhaut b 
was recovered in 2010 with STIS, the third epoch of astrometry 
indicated that Fomalhaut b's orbit is not nested
within the dust belt \citep{kalas10a, kalas11a}.  The preliminary
orbit was found to have $e\sim$ 0.7.  However, given that these
were the first data obtained with a different instrument, it was
difficult to determine if the 2010 position measurement was
compromised by uncorrected geometric distortion in STIS, or
a systematic uncertainty in the roll angle of the
telescope due to single guide star guiding.
We therefore chose to wait for a second epoch
of STIS observations in 2012 to confirm the new findings
of the 2010 epoch.  

Here we present  Fomalhaut b astrometry from a total of four epochs of HST observations spread over eight years, as well as new discoveries concerning the morphology of the dust belt in optical scattered light.  In Sec.~\ref{sec:acsrevisited} we briefly revisit and reanalyse the ACS/HRC photometry and astrometry.  The STIS data and results are given in Sections~\ref{sec:stisobservations} -- \ref{sec:mainbelt}.  Fomalhaut b's astrometry and newly determined orbital parameters are provided in Sec.~\ref{sec:orbit}.  We discuss the implications of the new orbit in Sec.~\ref{sec:discussion}, which includes an observational inventory of the objects and structures comprising the Fomalhaut system (Sec.~\ref{sec:inventory}).  Discussion of Fomalhaut b's physical nature, dynamical history, relationship to the main belt, and comparison to the solar system are also found in Sec.~\ref{sec:discussion}.

\section{ACS/HRC Observations Revisited}
\label{sec:acsrevisited}

Here we present new work to understand the sources of astrometric and photometric uncertainty in the data previously presented by \citet{kalas05a} and \citet{kalas08a}.
In both the 2004 and 2006 observations, Fomalhaut was 
occulted by the HRC 1.8$''$ diameter occulting spot,
which is located near the center of the field.  To expand
the field of view, in 2006 we also occulted Fomalhaut 
behind the 3.0$''$ diameter occulting spot, which is 
closer to the top edge of the detector.  All of the observations
included multiple telescope roll angles so that the point spread function (PSF)
of Fomalhaut A could be self-subtracted while
recovering astrophysical objects in the field.  The final
PSF-subtracted images for any given sequence of observations
include a significant number of residuals that appear as
point sources or extended features.  The key to separating
Fomalhaut b from residual noise features is that the noise
features vary significantly in position, morphology and intensity 
among different observing sequences, as the telescope roll
angle changes.

\subsection{ACS/HRC Astrometry}
\label{sec:acsastrometry}

\begin{figure}[!ht]
\epsscale{0.9}
\plotone{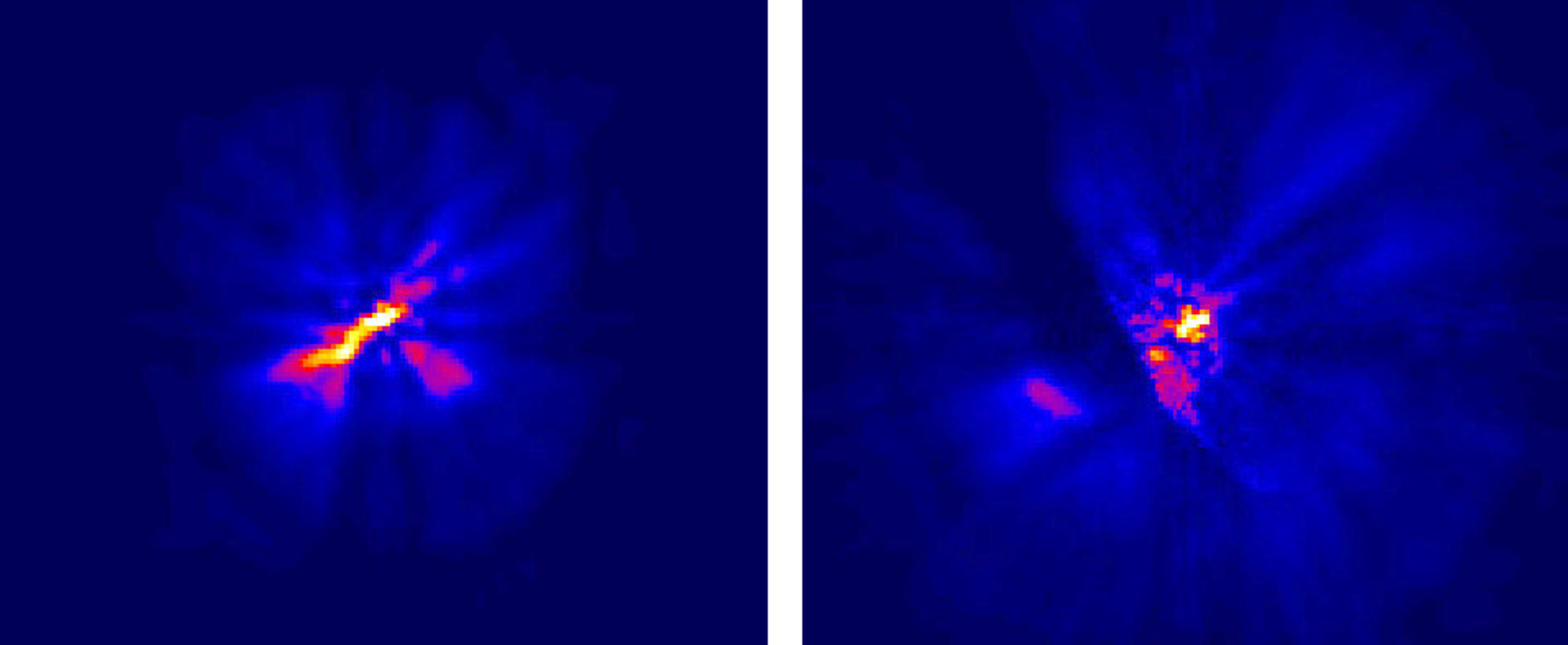}
\caption{\footnotesize Fomalhaut A viewed through the 1.8$''$ occulting spot (left)
and 3.0$''$ occulting spot (right) of the ACS/HRC in the shortest exposure data (0.20 s) from 2006.  The color scale is linear with white pixels representing 128,000 electrons for the 1.8$''$ spot data
and 32,000 electrons for the 3.0$''$ spot data.  Neither
the CCD nor the analog-to-digital conversion is saturated
(gain = 2.2 electrons / DN).  The morphology of the  3.0$''$ spot data includes a 
dark band extending downward from the top left
of the frame, representing the ACS/HRC occulting
bar that resides in the corrected beam between the
spot and the focal plane array. 
\label{fig1}}
\end{figure}

A significant source of astrometric uncertainty in the HRC data, as in various other 
coronagraphic data sets \citep{digby06a}, is determining the position of the central star
behind the occulting spots.  The two techniques that we use are to: 
(1) centroid the core of the stellar image as it appears through the occulting spot, and (2) determine the
centroid of the stellar PSF halo seen beyond the outer edge of the occulting spot.
The former is possible because the ACS/HRC occulting spot transmits 4.5\% of incident light
(ACS Instrument Handbook 2005).  The 2006 observing strategy included three 0.20 second 
integrations (the minimum allowable) at the conclusion of every orbit targeting Fomalhaut.  In
these short exposures, the morphology of the PSF core (after the image distortion correction is applied)
is highly asymmetric, with a tail of light extending to the lower left of a distinct few pixel
PSF core (Fig.~\ref{fig1}).  This asymmetric morphology is less pronounced for the 3.0$''$ spot data.

For the longer exposures (e.g., 340 seconds), the core is significantly saturated.
Our technique for centroiding the PSF involves rotating the image 180$\degr$
and subtracting the rotated image from the unrotated image.  The center of
rotation is adjusted iteratively to symmetrically subtract the PSF.   This technique
provides two additional centroid positions because we choose two different
regions to assess the success of the rotated self-subtraction.  The first is the region
interior to the occulting spot where the PSF core is saturated, but nevertheless the rotation center can
be adjusted so that the core is azimuthally uniformly subtracted.  The second is the region
exterior to the occulting spot which is not saturated.

To summarize, the three methods
for estimating the stellar position in ACS/HRC data use:
(1) the PSF core in short exposure data, (2) the PSF core in 180-degree,
self-subtracted, long-exposure data, and (3) the PSF halo in 180-degree, self-subtracted data,
long exposure data.  The standard deviation of these position measurements
is given in Table 2 as the 1$-\sigma$ uncertainty in the stellar location.  The astrometry cited in \citet{kalas08a}
utilized only the third method, with an estimated 1$-\sigma$ uncertainty of 12.5 mas.  The
larger uncertainties in Table 2 demonstrate that the differences between techniques
account for additional uncertainty in estimating the stellar position.  An important note
is that changing the assumed location of the star propagates throughout the data reduction because
observations made at different telescope orientations must be rotated to a common 
orientation based on this stellar location.  In effect, measurements made in this manuscript are based
on different final versions of the 2004 and 2006 observations and the results are not expected to be identical
with \citet{kalas08a}.
Table 2 also gives the uncertainty in obtaining the location of Fomalhaut b using a variety of
centroiding algorithms ($IRAF/pradprof~and~IDL/IDP3$) and applied to different versions of the final, reduced images at each of the epochs.
Again the standard deviation is larger than quoted in \citet{kalas08a} because the latter
work adopted only one type of centroiding algorithm ($IRAF/pradprof$).  

Another possible source of astrometric error is the position angle uncertainty of 
the observations.  The lack of guide stars means that Fomalhaut observations
are made with single guide star guiding (true for all HST observations discussed in this paper), which
means that the telescope roll angle is maintained using the gyros.  
The drift rate has a typical value of 28 mas per orbit, with a maximum value of 
53 mas per orbit (ACS Long Data Handbook Version 5, 2006).  Given the focal
plane geometry of the ACS/HRC relative to the guide fields, the drift means that
the entire HRC field may suffer a translation as high as 53 mas during an orbit with a small
change in position angle within the field.  In discussion below, we 
determine an $0.06\degr$ uncertainty in determining the position angle.  
Telescope jitter is in the range 3-5 mas.

The positions of Fomalhaut b in 2004 and 2006 and the associated uncertainties are
given in Table 3.  Compared to \citet{kalas08a} these measurements are within their mutual 1-$\sigma$ error bars.
The greatest difference lies in the revised 2004 position, which in this manuscript
is 36 mas Eastward (i.e. closer to the star) and 26 mas South of the position published in \citet{kalas08a}.
As noted above, the 2004 data were re-reduced based on new estimates of the stellar location,
thus accounting for differences in the astrometric analysis.  The independent data reductions
and analyses of these ACS observations performed by \citet{currie12a} and \citet{galicher13a} give measured
positions within the mutual 2-$\sigma$ error bars.

\subsection{ACS/HRC Photometry}
\label{sec:acsphotometry}

\citet{kalas08a} reported that Fomalhaut b 
dimmed by up to 0.8 magnitude between the 2004 and 2006 epochs in the F606W bandpass.  
At the time, this result was validated by (a) checking that the 2004 Fomalhaut PSF halo beyond
the edge of the occulting spot subtracts  the 2006 Fomalhaut PSF halo
without any additional image scaling, and (b) performing aperture photometry on a background
star common to both epochs of data.  The former test showed no more than 2\% change
in calibration (with 2004 as the dimmer of the two epochs), and the latter test showed
that the 2004 image of the background star was no more than 0.1 mag dimmer than the 2006 data.  
The key difference between the control star and Fomalhaut b is that the background star has
higher SNR because it is brighter than Fomalhaut b by $\sim$1.0 mag, and it is located
farther away from the residual speckle halo.  The 2006 location of the background control star was 5.6$''$ West and
13.1$''$ South of Fomalhaut, outside the boundary of the dust belt.   

These results indicated that systematic calibration uncertainties would
make Fomalhaut b appear $dimmer$ in 2004, when in fact \citet{kalas08a} reported that it was
brighter in 2004.  \citet{kalas08a} quoted the standard error as 0.05-0.10 mag for the Fomalhaut b photometry, 
which translates to a standard deviation $\sigma \approx 0.2$ mag.  Therefore the photometric
variability measured for Fomalhaut b was interpreted as significant.  

Independent analyses of the same observations by \citet{currie12a} and \citet{galicher13a} do not confirm astrophysical variability, 
but the differences between these two follow-up studies suggest that the photometric uncertainties are larger and of order the variability given by \citet{kalas08a}.
For the F606W apparent magnitude, \citet{currie12a} give $24.97\pm0.09$ mag and $24.92\pm0.10$ mag for 2006 and 2004, respectively.
These values are consistent with the 2006 measurement of $25.1\pm0.2$ in \citet{kalas08a}.  However, the
photometry presented by \citet{galicher13a} is consistent with the the 2004 photometry given by \citet{kalas08a}.  
Since the observations are the same, these results suggest that
there are systematic photometric uncertainties due to the choices of data reduction and analysis methods for
high contrast imaging.  In an experiment discussed below, we show that the uncertainty in STIS photometry
may be $>35$\% due to the residual speckle noise.

\section{HST/STIS Observations and Data Reduction}
\label{sec:stisobservations}

We observed Fomalhaut in 2010 and 2012 using the coronagraphic
mode of Space Telescope Imaging Spectrograph \citep[STIS;][]{woodgate98a} aboard HST.  STIS 
includes a 1024$\times$1024 pixel CCD with 
two orthogonal occulting wedges and a 3.0$''$ wide occulting
bar located in the focal plane.  STIS does not have a Lyot pupil plane
stop and therefore the diffraction spikes
due to the secondary support spider are evident.  
Also, STIS imaging is conducted without filters.  
The effective sensitivity of the CCD covers the full 
range between 0.20  $\mu$m and 1.03 $\mu$m.  The 0.05077$''$ pixel
scale results in a $52''\times52''$ field of view.  We set the
gain to 4.015 electrons / DN.

Table 1 summarizes the STIS observations of Fomalhaut used
in the present paper.  
Calibration of the data, such as bias subtraction and flatfielding, are
executed by the OPUS pipeline.  We manually processed the 
image frames (with extension flt.fits)
further by identifying cosmic rays or chip defects  and
replacing these pixels with values derived from an interpolation over
neighboring pixels.  For sky subtraction in each exposure, we
take the median value of a 400 pix$^2$ region in the upper left corner of the 
CCD (furthest away from the star).  The images are then divided
by their exposure times.  The geometric distortion correction
is performed manually using the currently recommended calibration
file in the archive (o8g1508do.idc.fits).

The next steps involve subtracting the central PSF using roll deconvolution.
Each orbit has a fixed telescope position angle orientation that differs by
a few degrees from neighboring orbits (Table~ref{log}).  The first step is to
coregister all of the orbits in x,y translation by adopting a single fiducial
orbit and subtracting the PSF's of all the other orbits, iteratively adjusting
the x and y offsets until the PSF subtraction residuals are minimized. 
The registered frames from all of the orbits at different roll angles are then median combined.  
Astrophysical features rotate in the CCD frame whereas the stellar PSF structure is fixed.  
The median value of each pixel effectively
removes astrophysical features and produces a clean image of the
stellar PSF.  This approximation of the Fomalhaut PSF is then
subtracted from each of the frames taken at different position angles.
These PSF-subtracted frames are then rotated so that astrophysical
features are registered, and these frames are then median-combined,
recovering a PSF-subtracted astrophysical scene.  

In the September 2010 data set we discovered time variable distortions that
contribute to our astrometric errors for Fomalhaut b.  In the first two orbits, two field stars are visible at the top and at the bottom of
the full field ($36.5''$ and $27.5''$ radius from Fomalhaut, respectively).  
Splitting each orbit into quarters, we find the top field
star drifts upward (+Y) by 0.2 pixel (10 mas), whereas the bottom field star
remains stationary (1-$\sigma$ = 0.03 pixel).  In the horizontal
direction (X), both field stars remained stationary (1-$\sigma$ = 0.03 pixel) throughout the orbit.  
This effect is observed for both orbits.  The third and fourth orbits
have these field stars outside the field of view due to the changes in
telescope roll orientation.  

One possible source of position drift within an orbit is the fact that 
all of our observations utilized a single guide star.  However, we rule out
this effect because the bottom field star does not appear to drift, nor is
there a systematic offset in the PSF of Fomalhaut that corresponds
in direction and magnitude to the offset of the top field star.  

To estimate astrometric error due to uncorrected
geometric distortion, we report recent private communication with STIS instrument
scientists (J. Duval and A. Aloisi) concerning ongoing astrometric 
calibration measurements.  The calibration program consists of 
observing Omega Cen once a year with STIS and WFC3.  The STIS
observing sequence consists of four 10 second exposures in a four-point
dither pattern obtained within a single HST orbit.  After pipeline processing,
the STIS field is registered to the WFC3 field and the RA and DEC
offsets between the STIS and WFC3 centroids are computed.  Only exposures
with $>$10 stars in common between the STIS and WFC3 fields are used.
In 2011 the mean and standard deviations of
these offsets in STIS pixel units are (1.84, 0.82, 0.96, 1.46, 0.92) $\pm$
(2.13, 1.43, 1.83, 1.74, 1.94), respectively.  In 2012 there are
five different exposures giving  (0.59, 0.52, 0.57, 1.00, 0.97) $\pm$
(0.33, 0.31, 0.30, 1.45, 1.49).  

These data suggest that the uncorrected geometric distortion is variable
not only from year to year, but also within a single orbit.  The 0.2 pixel
drift that is detected in the Fomalhaut data is therefore a general characteristic
of STIS imaging observations and  not necessarily
a consequence of our specific guide star uncertainty.
The time dependence suggests effects such as thermal breathing
are important.
An astrometric calibration program for Fomalhaut would therefore
require observations of fields near Fomalhaut and interspersed
in time with the Fomalhaut observations.  

Since an astrometric calibration program designed specifically
for Fomalhaut is not practical, one option is to adopt the RMS value
of the astrometric uncertainties given by the $entire$ STIS
astrometric calibration from 2001 to the present epoch.
This value is 1.303 pixels, or 66 mas (Table~\ref{astrometryerrors}).  However, as discussed
below and Section~\ref{sec:astrometry}, this overestimates
the uncorrected geometric distortion uncertainty.

As a second method to estimate the error in geometric distortion correction,
we measure the positions of two field stars in the 2012 data before and
after the geometric distortion is applied to the data.  The largest difference
measured between the before and after positions is 0.33 pixels, or 17 mas.
Thus we adopt 17 mas as an estimate of geometric distortion ``measured in data''
(Table~\ref{astrometryerrors}).

The single guide star guiding also introduces the possibility that
the position angle orientation of the telescope contains significant systematic error.  
To quantify the position angle uncertainty, we found two field stars that are contained
in the STIS field, as well as previous observations of Fomalhaut with
WFC3/IR and WFPC2.  All three observations were made with a single
guide star.  The two field stars are separated from each other by 47$''$ and are located East
and North of Fomalhaut.  The measurement of position angle in these
three data sets gives an empirical uncertainty of 0.06$\degr$ (1-$\sigma$)
in the telescope orientation.

Finally, as with the ACS/HRC data, we used several methods
for determining the uncertainty in the position of the star behind
the occulting wedge.  Note that since all of the orbits were
registered to a single fiducial orbit (see above), the stellar
center is determined for the single, fiducial orbit.
Unlike the HRC, the STIS images have
prominent diffraction spikes that can be used to estimate
the stellar position.  However, since the STIS occulting
wedges are not partially transparent, the core of the
PSF is inaccessible using short exposures.  Perhaps the greatest
advantage with STIS is the larger field of view that will 
contain more field stars (however, the occulting wedges often
block the field stars at various telescope roll angles).  
Since the data are obtained at different
roll angles about Fomalhaut, it is possible to verify if the
rotation center is accurate by studying how well the field
stars are coregistered after derotation to a common reference frame.

Our first estimate for the stellar position is derived from the
diffraction spikes.  We fit two straight lines to the spikes,
yielding an RMS residual of 0.15 pixels (8 mas), with no systematic
curvature.  This result can be validated by rotating the image
180$\degr$ to demonstrate that the diffraction spikes self-subtract.  
Using this center position, the frames are rotated so that north is
up and east is left.  The centroids of the field stars can be determined
for each north-rotated orbit, and the procedure is repeated again using 
0.5 pixel and 1.0 pixel deviations from the initial center position determined from
the diffraction spikes.  If the field star positions in the north-rotated frames differ from
each other in separate roll angle observations by more than 0.2 pixels, we 
consider the assumed rotation center position as invalid.  In this way the center position is tested
and refined so that the field star centroids are statistically
identical from orbit to orbit in the registered, de-rotated frames.
A weakness of this technique is that there are only two field
stars for reference, and some frames contain one star and not the
other because the STIS occulting wedges block different portions of the field
at different roll angles.  The center position determined using
the field-star-rotation technique differs from the diffraction spike center position
by 0.5 pixel in x and 0.5 pixel in y.  However, the diffraction
spike center does not violate the 0.2 pixel cut-off defined above.  
Therefore, adopting a conservative, worst-case scenario, we
establish 0.5 pixel (25 mas) as the uncertainty in determining 
the stellar center in the STIS coronagraphic data (``STIS centroiding star'' in Table 2).

\section{2010 Recovery of Fomalhaut b}
\label{sec:2010recovery}

\begin{figure}[!ht]
\epsscale{0.93}
\plotone{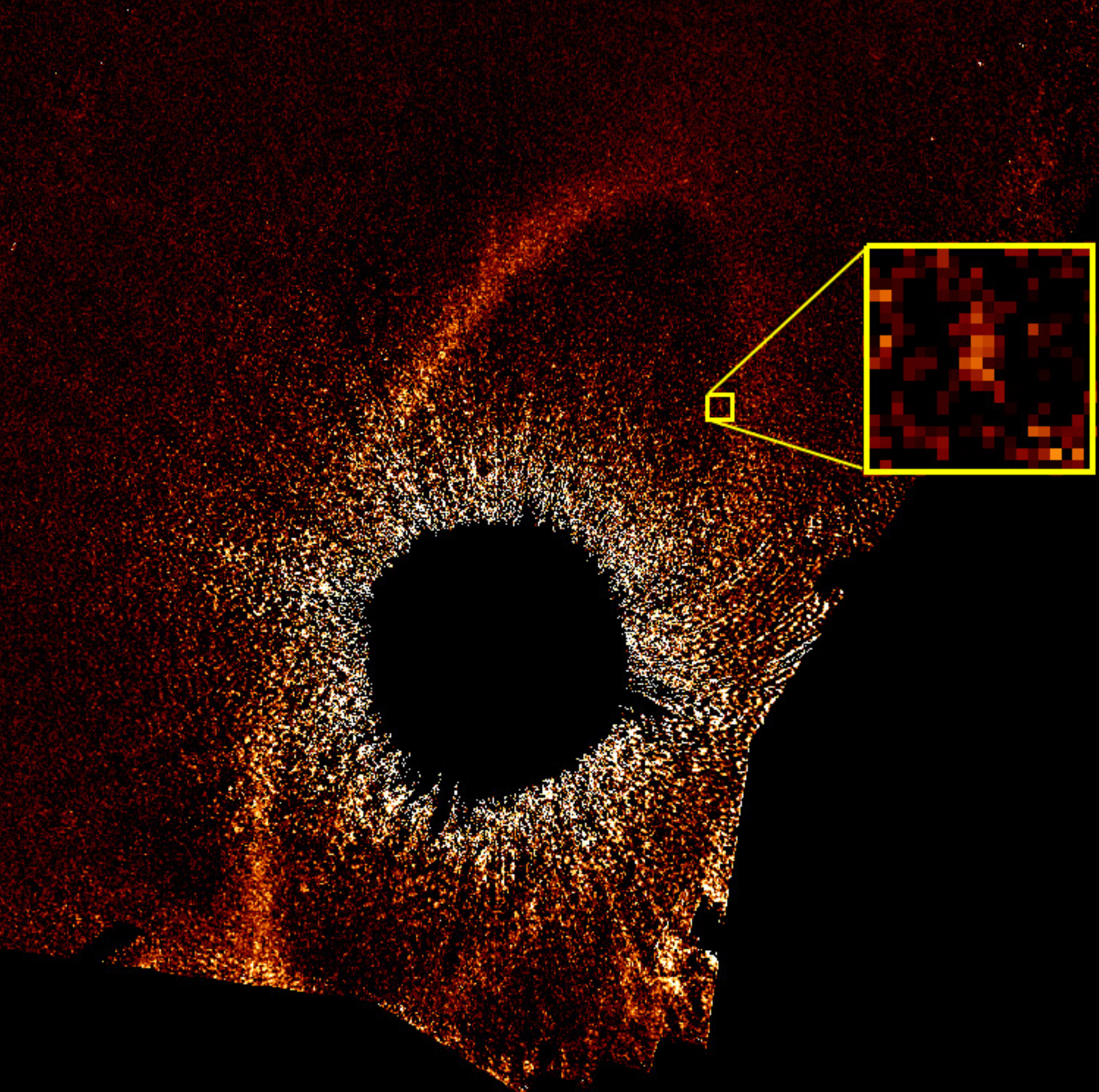}
\caption{\footnotesize
False-color, linear-scale image of a portion of the Fomalhaut field containing the 2010 detection of the belt and Fomalhaut b to the northwest of the star.  North is up, east is left.  Also shown is the discovery of a halo of nebulosity northwest of the main belt.   The box inset is 1$''$ on a side and magnifies the image of Fomalhaut b.  A smoothed version appears in Fig~\ref{hardstretch}.
\label{2010falsecolor}}
\end{figure}

Figure~\ref{2010falsecolor} presents the final, unsmoothed image resulting from the combination of the four orbits executed in September 2010.  
The southeast side of the belt lies outside the STIS field of view. The 
northwest portion of the belt is detected for the first time in scattered light 
since this region was outside the field of view in the ACS/HRC data.  
As reported in \citet{kalas10a} and \citet{kalas11a}, the northwest side of the belt reveals an extended halo of nebulosity, indicating 
that the belt is broader than previously reported from the ACS/HRC observations.

The source we identify as Fomalhaut b is the brightest object in a 1$''$$\times$$1''$
search box at the expected location of Fomalhaut b at
the epoch of observation (Fig.~\ref{2010falsecolor} inset).
The centroid of Fomalhaut b is offset from the central star by $\Delta$RA = $-8.828\pm0.042''$,
$\Delta$Dec=+$9.822\pm0.044''$ (J2000.0), with projected stellocentric
separation $\rho=13.206''$ (Table~\ref{astrometry}).   This 2010 position is consistent
with the independent detection of Fomalhaut b in these data
by \citet{galicher13a}, who measure
$\Delta$RA=$-8.81\pm0.07''$,
$\Delta$Dec=+$9.79\pm0.07''$.

The apparently extended morphology of Fomalhaut b along the vertical axis will be discussed further when we introduce the 2012 detection.
The signal-to-noise of the STIS detection is degraded relative to the ACS/HRC
observations by significant residual speckle noise and subtraction artifacts.
Compared to the ACS/HRC data, STIS has a coarser pixel
scale (51 mas versus 25 mas for the HRC) that results in
poorer sampling of the PSF halo and speckles.   The significantly
broader bandpass of the STIS data also increases the radial
extent of speckles relative to the ACS/HRC observations.

\begin{figure}[!ht]
\epsscale{1.0}
\plotone{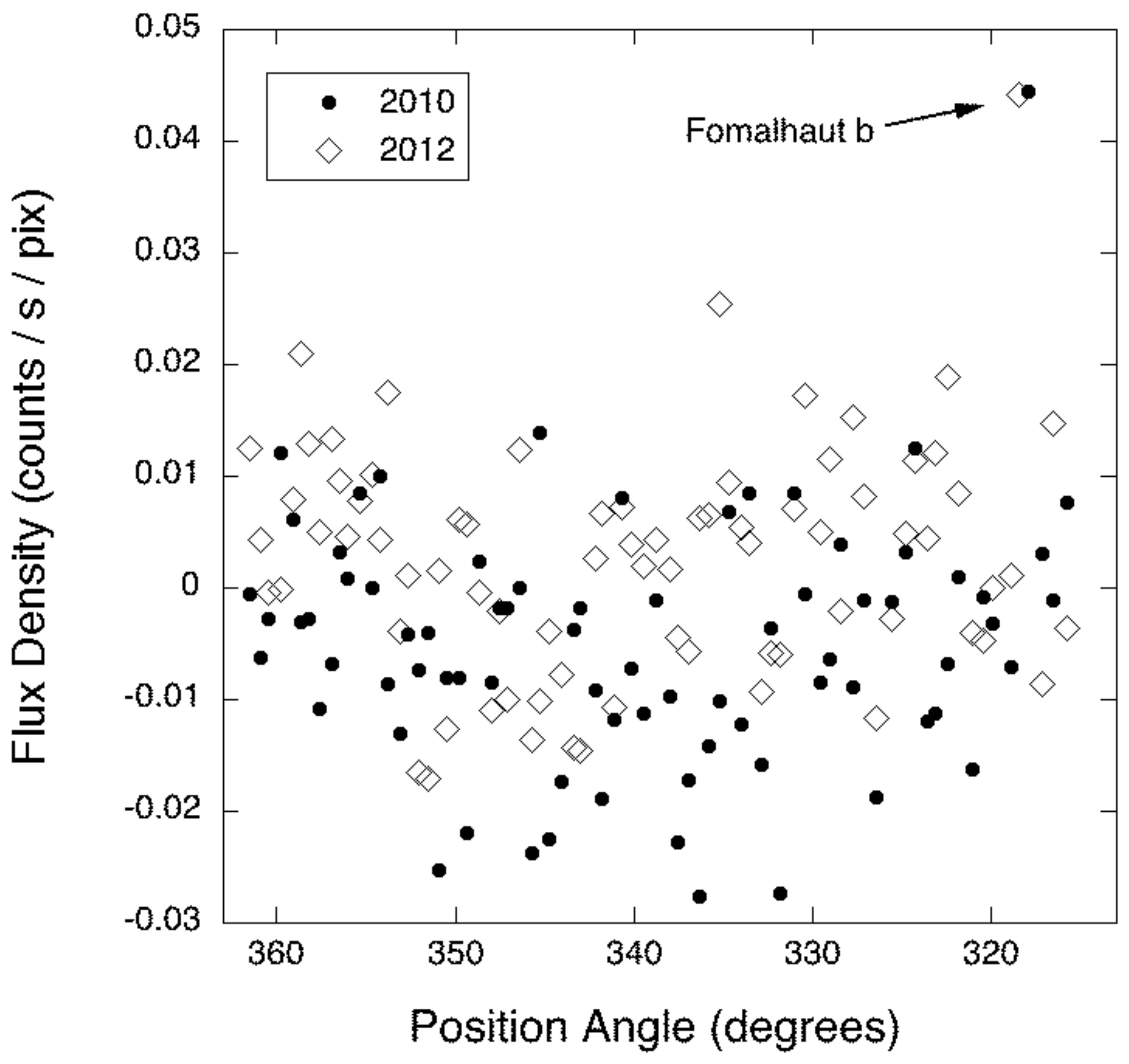}
\caption{\footnotesize
Median pixel values computed in $3\times3$ pixel apertures spread across an arc that has the same radius as Fomalhaut b.
These values have not been background or aperture corrected.
\label{medianarc3x3}}
\end{figure}

To quantify the signal-to-noise of the proposed detection, we compare the median pixel value
in a single $3\times3$ pixel box ($0.15''\times0.15''$) centered on Fomalhaut b to 76 boxes
spread in an arc at the same radial distance from the star as Fomalhaut b (Fig.~\ref{medianarc3x3}).  
We use an arc region because speckle noise decreases with
increasing radial distance from the star and it is the dominant source of noise
at the location of Fomalhaut b.
To avoid the regions dominated by diffraction spikes and Fomalhaut's
dust belt, the arc resides entirely within the inner boundary of the dust belt.
We adopt the relatively small $3\times3$ pixel aperture because the residual
speckles that may produce false positive detections of point sources have this characteristic
size.  Therefore our estimate of signal-to-noise using this method is an
attempt to quantify the speckle noise in the reduced images.
The box containing Fomalhaut b has a median value of 0.044 cts/s/pix.  
whereas the median counts in the 76 comparison boxes have standard deviation 0.010 cts/s/pix.
This metric suggests that Fomalhaut b is a $4.4~\sigma$ detection in the
2010 data. 

The surprising result from the 2010 detection is that Fomalhaut b is detected three pixels (150 mas) westward from the
position that would be expected for a low-eccentricity orbit ($e$$\sim$0.1) nested within the inner boundary of the debris belt.  
\citet{kalas10a} and \citet{kalas11a} reported that the uncertainties in the roll angle of the telescope and the uncorrected geometric distortion
may plausibly account for the three pixel deviation.  Subsequent work reported in Section~\ref{sec:stisobservations} quantifies both the position angle and uncorrected geometric distortion uncertainties (Table~\ref{astrometryerrors}).  Adopting a geometric distortion uncertainty of 66 mas based on the unpublished STIS calibration program leads to an astrometric uncertainty (1-$\sigma$) of 76 mas in Right Ascension.  Therefore the 150 mas Westward deviation observed in 2010 could be considered a 2$\sigma$ result.  We concluded that a fourth epoch of observation was required in order to test the significance of the Westward deviation.  As discussed in subsequent sections, after the 2012 epoch confirmed a highly eccentric orbit, the error analysis given all four epochs of astrometry justifies the adoption of a smaller value for the uncorrected geometric distortion (17 mas instead of 66 mas; Table~\ref{astrometryerrors}).

\section{2012 Confirmation of Fomalhaut b}
\label{sec:2012}

\begin{figure}[!ht]
\epsscale{1.0}
\plotone{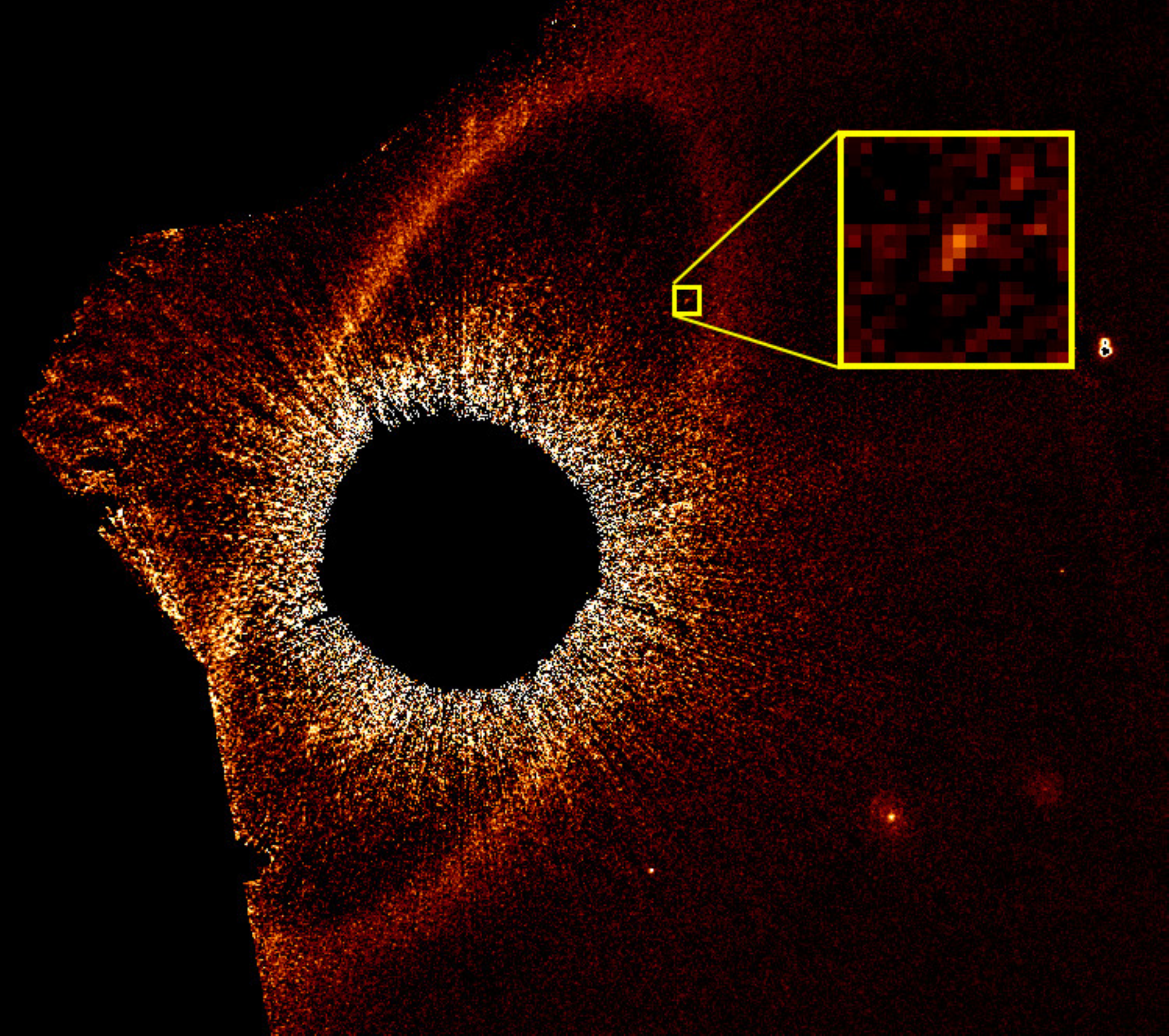}
\caption{\footnotesize
False-color, unsmoothed image of the 2012 STIS data.  A smoothed version appears in Fig~\ref{hardstretch}.
The box inset is 1$''$ on a side and magnifies the image of Fomalhaut b.    North is up, east is left.
\label{2012falsecolor}}
\end{figure}

\begin{figure}[!ht]
\epsscale{1.0}
\plotone{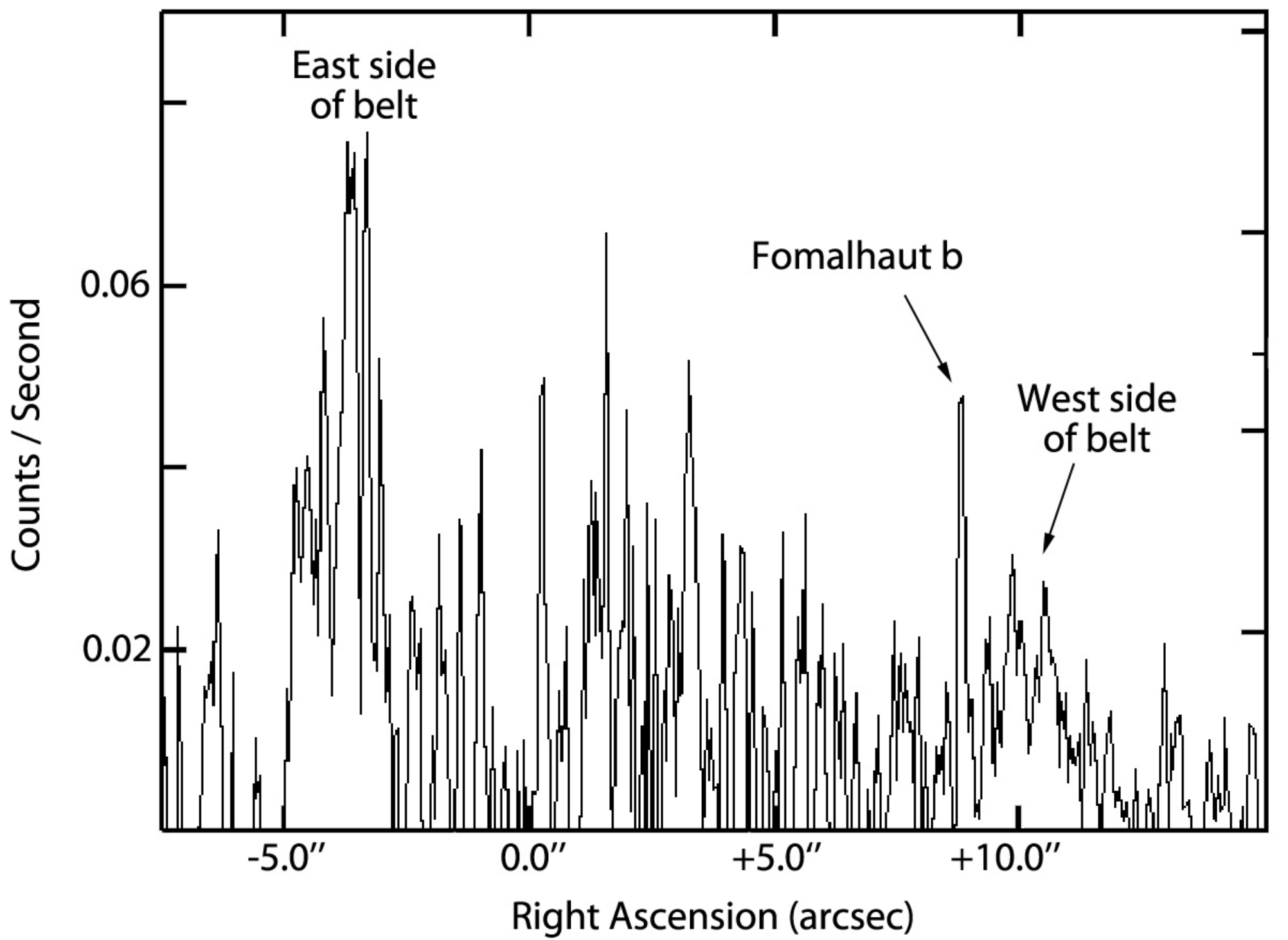}
\caption{\footnotesize
Horizontal cut across the image in Fig.~\ref{2012falsecolor}.  The cut represents the average of three lines
centered on Fomalhaut b, where the center line is 10.00$''$ above the stellar position.  The x-axis
plots the position relative to the star.  Unlike Fig.~\ref{medianarc3x3}, this horizontal cut passes through a range
of radii from the star.  Therefore the residual speckle noise is prominent in the range $-5.0''-5.0''$.
However, Fomalhaut b is a prominent feature relative to the local noise in the region $\sim13''$ radius from the star.
\label{2012horicut}}
\end{figure}

The main difference between the 2012 STIS data and the 2010 STIS data is a factor of three increase in integration time and telescope roll angles.  Figure~\ref{2012falsecolor} shows the 2012 confirmation of Fomalhaut b.   We employ the 3$\times$3 aperture measurements of the previous section and find Fomalhaut b has the identical flux density (0.044 cts/s/pix) as in 2010 (Fig.~\ref{medianarc3x3}).  The standard deviation of the 76 comparison boxes is  also 0.010 cts/s/pix, resulting in the same 4.4-$\sigma$ detection as in 2010.  
Noise due to quasi-static speckles should be reduced by increasing the number of realizations of speckles.
Our result suggests that the $\sim2\degr$ rotation between frames was not sufficiently large to decorrelate
the quasi-static residual speckle structure at the radius of Fomalhaut b.  
Figure~\ref{2012horicut} shows a horizontal line cut through the image that intersects Fomalhaut b.  Fomalhaut b is a prominent feature compared to the local noise and the western portion of the dust belt.  

Fomalhaut b appears to have an elliptical morphology with major axis at PA $\approx137\degr$.
The major axis has full-width half-maximum (FWHM) of
5.2 pixels (264 mas) and full-width quarter-maximum (FWQM) of 7.4 pixels (376 mas).  
The minor axis has FWHM and FWQM 2.5 and 3.5 pixels, respectively.  
The background field star south of Fomalhaut b in the lower quarter of the 
frame has FWHM of 2.4 and 2.1 pixels in the x (RA) and y (DEC) directions (FWQM 3.6 pix and 2.9 pix, respectively).  
Therefore Fomalhaut b is consistent with a point source along its minor axis direction, 
but appears significantly extended along its major axis. If the extended morphology 
is astrophysical, then the 376 mas FWQM of the major-axis corresponds to 2.9 AU.  

However, the radial direction relative to the star at the position of Fomalhaut b is PA=$138\degr$, 
invoking the possibility that the extended morphology is due to the residual speckle noise.  
To assess this possibility, we conduct an experiment using a STIS artificial stellar PSF generated by TinyTim \citep{krist11a}.  
First, we determine if the data reduction steps distort the morphology of the artificial PSF.
The PSF is inserted into blank fields at positions corresponding to the various roll angles of the data.  
When these experimental data are processed in a manner identical to the real data, and measurements conducted in an identical manner, the resulting TinyTim PSF has FWHM 2.0 and 1.8 pixels (FWQM 3.4 and 3.2 pixels) along the x and y directions, respectively.  
The field star is $\sim15\%$ broader than this, most likely because of
noise (demonstrated below), the uncertainty in determining the rotation center in the real data, and the fact that a real PSF
core may be land over several pixels instead of a single pixel. 

\begin{figure}[!ht]
\epsscale{1.0}
\plotone{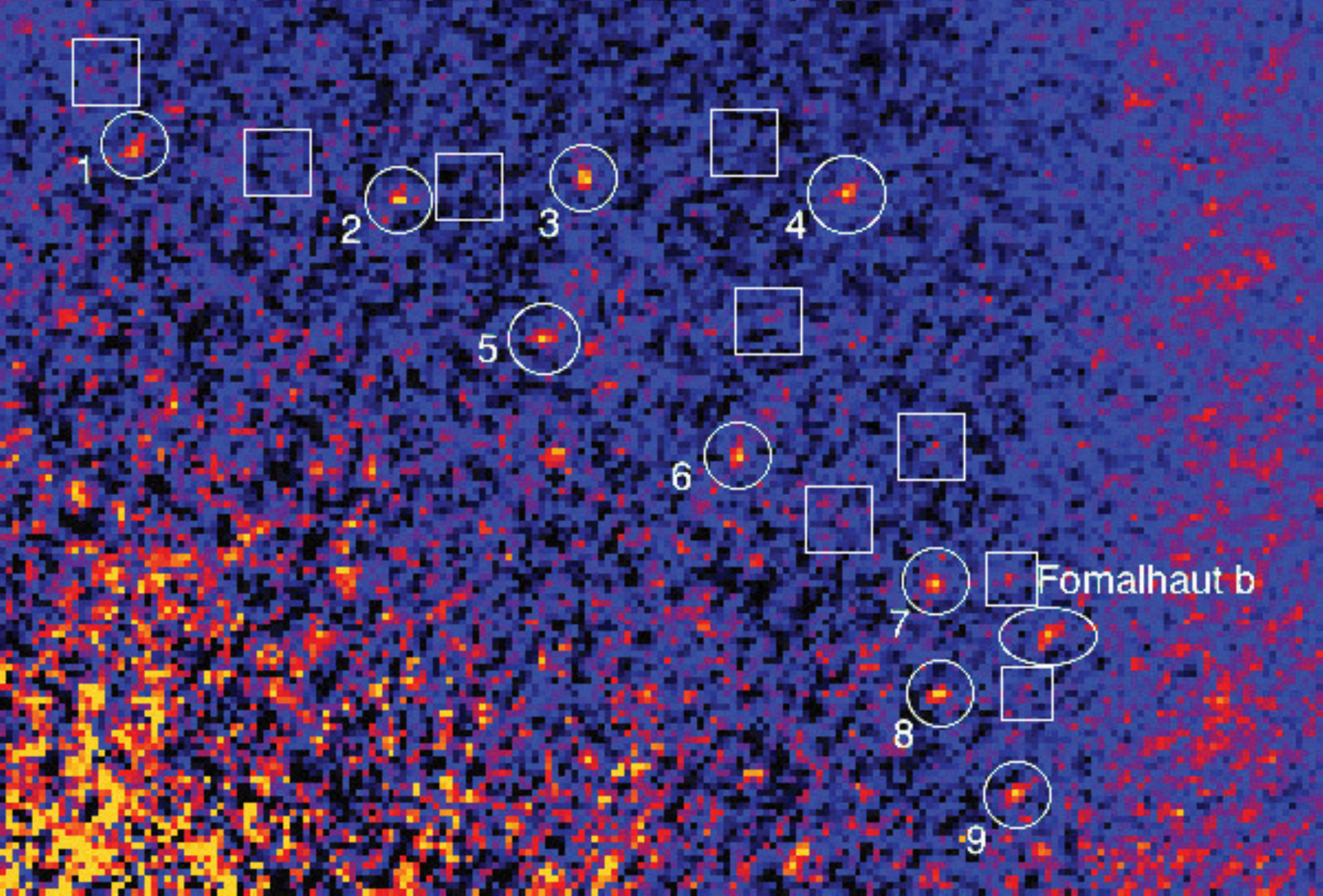}
\caption{\footnotesize
Experiment with implanted point sources in the 2012 data.  Circles mark locations where an artificial point source is inserted at the position of positive noise features in the image.  Squares denote locations where the artificial source is inserted at negative noise features.  The oval marks Fomalhaut b. 
\label{2012implants}}
\end{figure}

\begin{figure}[!ht]
\epsscale{1.0}
\plotone{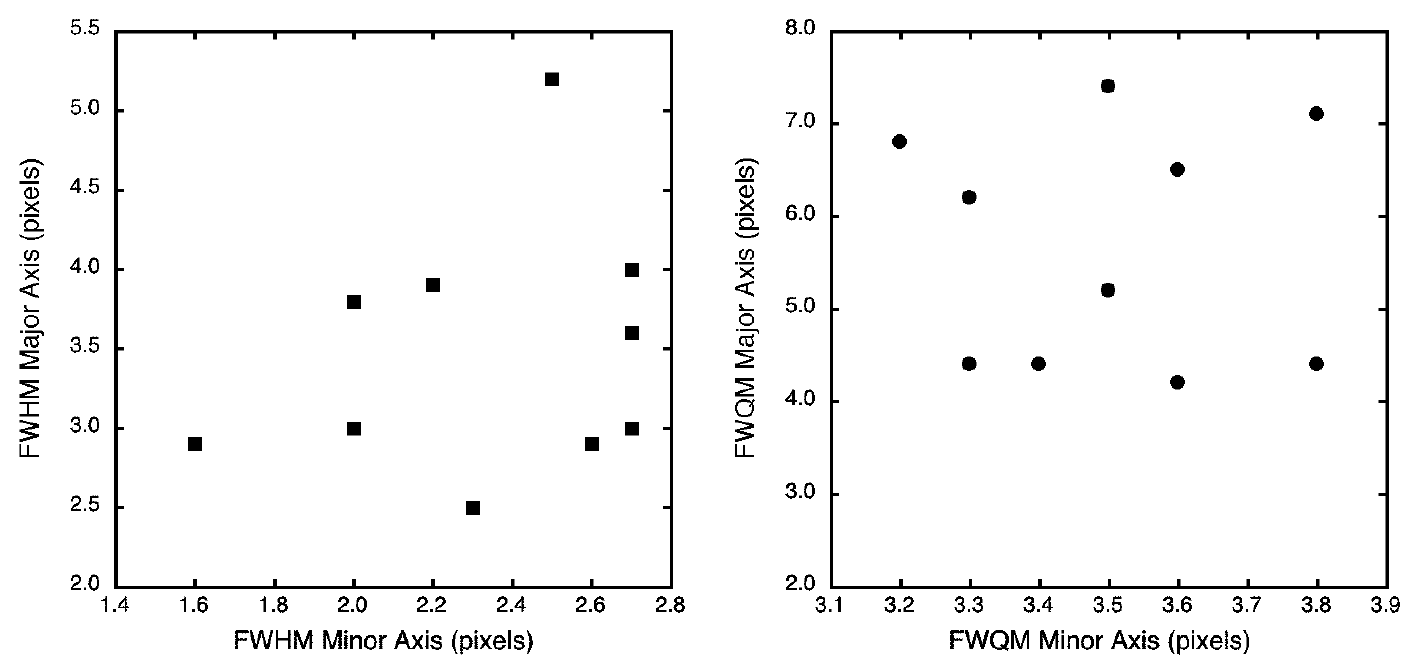}
\caption{\footnotesize
Measurements of the implant sources and Fomalhaut b FWHM (left) and FWQM (right) along the minor and major axes.  Fomalhaut b appears to be an outlier in the measurement of the FWHM of the major axis (it is the topmost data point).  However, Fomalhaut b is indistinguishable in the graph of FWQM.
\label{2012implant-plot}}
\end{figure}

The second step in the experiment is to insert the artificial PSF described above into the 2012 data to quantify the magnitude of PSF distortion due to residual speckle noise.  
We insert nine copies of the artificial PSF at locations in the image where positive noise features of at least four contiguous pixels are apparent, and nine more locations where the noise is negative.  Figure~\ref{2012implants} demonstrates that point sources may appear extended due to positive noise.  For example, source number nine has FWHM and FWQM of 3.9 and 6.8 pixels, respectively.  We plot the respective FWHM and FWQM measurements for the nine positive noise sources and Fomalhaut b in Fig.~\ref{2012implant-plot}.  Fomalhaut b is the most extended source in both cases, but it does not appear to be an outlier when measuring the FWQM.  
The minor axis measurements shown in Fig.~\ref{2012implant-plot} for the artificial implants demonstrate that in most cases the minor axis of a point source is also broadened due to noise.

We conclude that since residual noise is a plausible explanation for Fomalhaut b's extended morphology, other data sets are required to establish whether or not Fomalhaut b is extended.  In the 2010 STIS data, Fomalhaut b appears somewhat extended.  The FWHM's of the minor and major axis are 2.1$\times$7.2 pixels (107$\times$366 mas).  The corresponding FWQM's are 3.5$\times$8.9 pixels (178$\times$452 mas).  However, the orientation of the major axis is north-south in 2010.  Thus, if both the 2010 and 2012 STIS observations detect extended structure, rotation by $\sim45\degr$ in 18 months must be explained.  The deepest, best-sampled images of Fomalhaut b are the 2006 observations obtained with the ACS/HRC and the F606W filter.  In these data, Fomalhaut b appears to be a point source with FWHM = $69 \pm 6$ mas \citep{kalas08a}.  Therefore if the 2012 extended morphology is real, it would require that Fomalhaut b is spreading over time, or has a triaxial shape that occasionally appears point like as the major axis rotates into our line of sight, minimizing the projected size.  Future observations with HST/STIS or other instruments can refute or confirm the extended, time-dependent morphology.

The implant experiment also quantifies the astrometric error for centroiding on Fomalhaut b.  The median difference between the known implant locations and the recovered centroid positions of the implants is 0.1 pix in X and 0.2 pix in Y.  These values are used in quantifying the astrometric error in Table~\ref{astrometryerrors}.

\section{STIS Photometry on Fomalhaut b}
\label{sec:photometry}

As noted above, $3\times3$ pixel apertures centered on Fomalhaut b have a median pixel value of 0.044 cts/sec/aperture/pix for
both epochs of STIS observation.  However, this photometry requires aperture and sky background corrections.  
The arc statistics give a slightly positive sky value (0.004 cts/s/pix) for the 2012 epoch, and a slightly negative value (-0.004 cts/s/pix) for 2010.
Therefore the sky-corrected photometry is 0.040 cts/sec/pix and 0.048 cts/sec/pix for 2012 and 2010 respectively.  
Multiplying by nine square pixels gives 0.360 cts/sec/aperture and 0.432 cts/sec/aperture for 2012 and 2010, respectively.
For the aperture correction, we use the STIS Exposure Time Calculator to build a curve of growth for point source photometry,
yielding a correction factor of 1/0.642.  For an infinitely large aperture we then obtain 0.673 cts/sec and 0.561 cts/sec for
2010 and 2012, respectively.  

We then use the $IRAF/STSDAS$ package $countrate$ to
obtain a STIS zeropoint (i.e. 1 ct/sec) of $m_v$=24.48 mag in the VEGAMAG system for an input spectrum
comprising a T$_{eff}$ = 8500 K, log ($g$) = 4.0 Kurucz model spectrum.  In other words, this
particular experiment with the photometry
assumes that Fomalhaut b has the same spectrum as the host star.
This gives a factor of $4.628\times10^{-4}$ for converting counts to mJy.  
The sky and aperture corrected photometry
on Fomalhaut b therefore corresponds to  0.311 $\mu$Jy and 0.260 $\mu$Jy for 2010 and 2012, respectively.

One measure of the photometric uncertainty is to adopt the speckle noise of 0.010 cts/s/pix, or  0.005 $\mu$Jy  as the 1-$\sigma$ photometric uncertainty.
A second measure derives from the experiment with artificial point sources inserted into the data.  The implants on negative noise regions result in very weak detections or non-detections.  This is because the artificial point source flux was scaled so that the peak pixels, when implanted on positive noise features, result in pixel values close to that of Fomalhaut b in the image (and thereby giving a proper comparison of FWHM and FWQM).  For the nine implants on positive noise features, the standard deviation in photometry gives 25\% uncertainty in the photometry.  If we add in seven more implant locations on negative region (excluding two implants that give none-detections), the standard deviation in photometry is  35\% .  Given these significant uncertainties, the 2010 and 2012 photometry on Fomalhaut b are consistent with each other.

Kalas et al. (2008) report Fomalhaut b photometry of 0.75 and 0.36 $\mu$Jy in the ACS/HRC F606W filter in 2004 and 2006 observations,
respectively.  Using $countrate$ with the same input spectrum as above, we 
convert the ACS F606W fluxes to STIS values, giving 0.63 and 0.30 $\mu$Jy for the 2004 and 2006 photometric points, respectively.  
We therefore find very good agreement in the photometry between the 2006 ACS/HRC data and the two epochs of STIS data.  These results indicate that the 2004 ACS/HRC photometry may be anomalous. We have not found a source of error in our 2004 photometric analysis other than residual speckle noise.

We note that in the scenario where the optical flux of Fomalhaut b originates from light scattering by dust grains in a cloud
orbiting a planet, Fomalhaut b must dim over time, everything else being equal,  because the stellocentric distance is currently increasing.
In 2006, at a stellocentric distance of 119 AU (assuming a co-planar geometry with the belt), the incident 
flux on Fomalhaut b was 1.58 W m$^2$.  At the 2012 epoch the distance has increased to 125.4 AU, and the incident flux
decreased to 1.42  W m$^2$.  In this scenario we expect an 0.12 mag dimming from 2006 to 2012.  This effect is not apparent
in the nominal photometry derived above, nor would the effect be detectable to sufficiently high confidence because
the photometric noise described above is of order a few tenths of a magnitude.  Moreover,
the conversion of instrumentation from ACS to STIS involves observations in different bandpasses, requiring
additional assumptions that need to be made regarding the optical spectrum of Fomalhaut b.  If we instead
consider only the STIS observations, the 2010 position of Fomalhaut b is 124.1 AU from the star and the incident
flux is 1.46 W m$^2$.  Fomalhaut b would be dimmer in 2012 by
only  $\sim$0.02 mag, which is undetectable given the current photometric uncertainties.  The nominal
STIS photometry above shows an 0.2 mag dimming between 2010 and 2012, which if it were astrophysical
cannot be attributed to the motion of Fomalhaut b away from the star in the coplanar case.  Continued photometric
measurements of Fomalhaut b with STIS in future epochs will certainly build a time series of photometry that
would explore whether or not Fomalhaut b's apparent magnitude decreases over time.  

\section{Main Belt Properties from the STIS Observations}
\label{sec:mainbelt}

\begin{figure}[!ht]
\epsscale{1.0}
\plotone{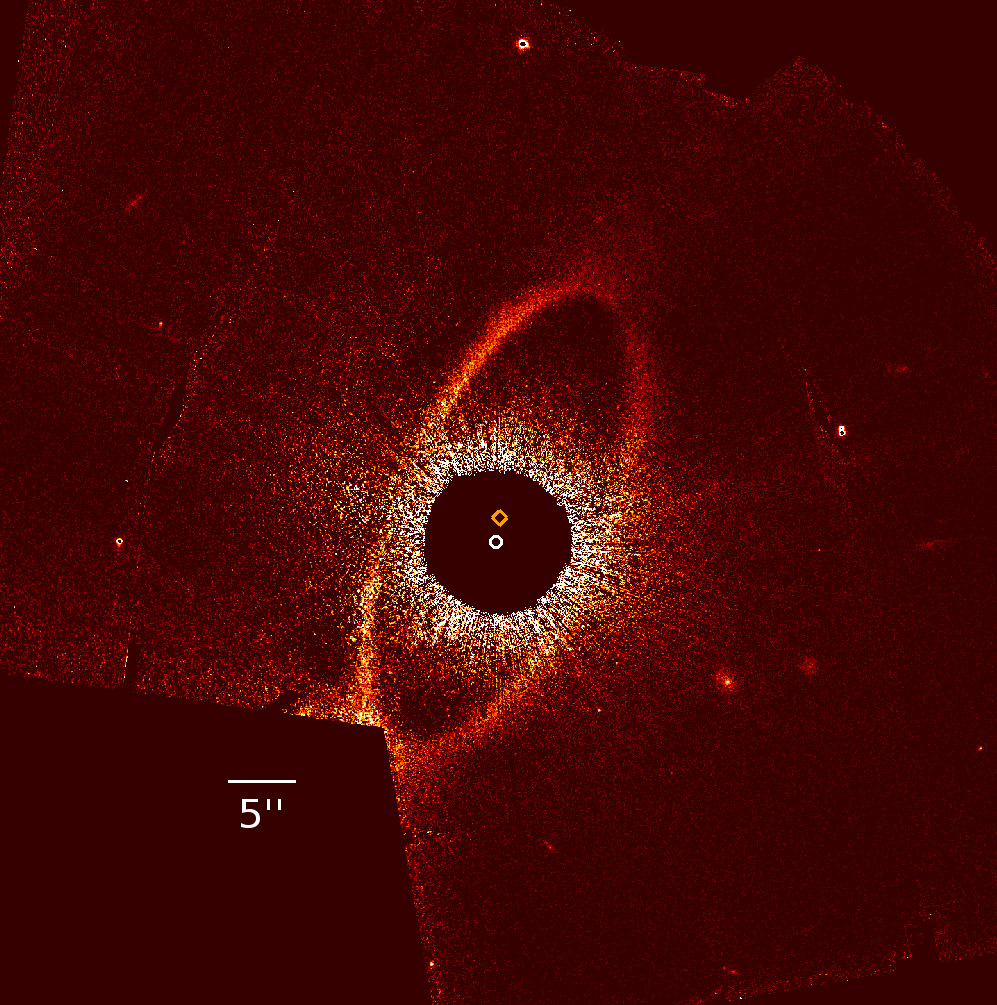}
\caption{\footnotesize
Mosaic of the 2010 and 2012 STIS data registered to the location of Fomalhaut A.  In the central regions of overlap (see Fig.~\ref{2010falsecolor} and Fig.~\ref{2012falsecolor}) this combined figure represents the average value from the two epochs of observation, which means that background objects and Fomalhaut b are blurred due to their motion between epochs.  North is up, east is left.  The circle and diamond mark the stellar center and the geometric center of the belt, respectively.  The data are not smoothed.
\label{combo}}
\end{figure}

Figure~\ref{combo} is a mosaic of the 2010 and 2012 STIS images in the reference frame of the star.  The main difference between this combined STIS image and the 2004/2006 ACS/HRC observations is that the northwestern side of the belt is now fully contained within the field of view, revealing an extended halo of nebulosity to the northwest.  Neither the STIS nor the ACS/HRC observations include the field significantly beyond the southwest side of the belt so as to ascertain if the extended halo is a symmetric feature to either side of the belt, which we refer to as the ``main belt'' to distinguish it from other belts in the system (Section 9.1).  

\subsection{Main Belt Geometry}
\label{sec:mainbeltgeometry}

\begin{figure}[!ht]
\epsscale{1.0}
\plotone{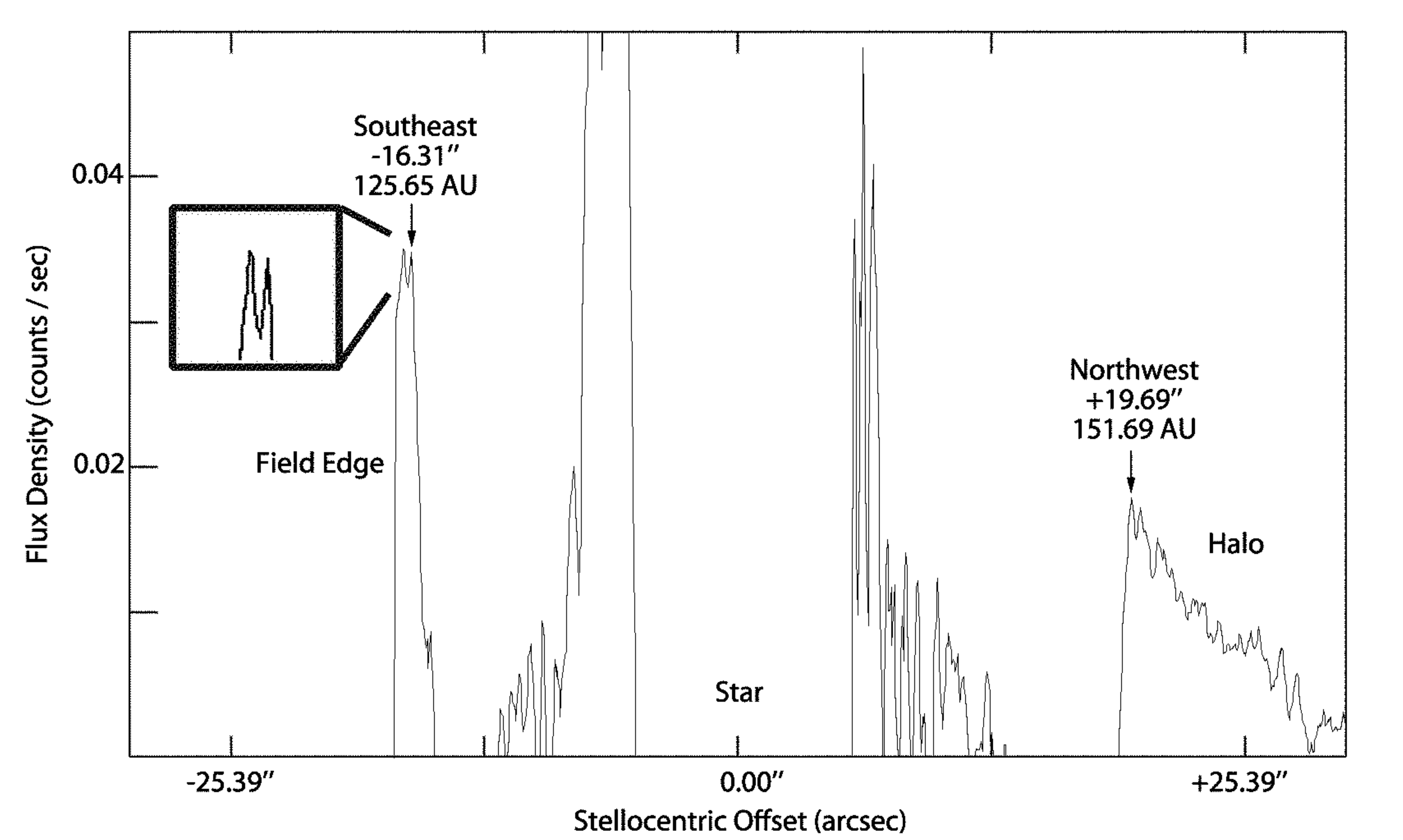}
\caption{\footnotesize
A cut along the major axis of the belt shown in Fig.~\ref{combo} that is the average pixel value in a one arcsecond wide segment along the minor axis direction.  The inset to the left magnifies the characteristic double peak structure of the belt, where the peaks are bisected and separated by $\sim$3.4 AU.
\label{combo-majorcut}}
\end{figure}

Figure~\ref{combo-majorcut} shows a cut along the apparent major axis of the belt.  In the majority of cuts, there is a double peak of maximum brightness (see also Fig. 3 in Kalas et al. 2005).  To avoid confusion with the azimuthal ``gap'' discussed below, we characterize this specific radial morphology as a ``bisected'' plateau.  The apparent major axis measured between the inner peaks of the belt has length 277.34 AU.  The center of the belt bisector is 1.85 AU and 1.62 AU (SE and NW sides, respectively) further outward from the inner peak.  The outer peak is another 1.54 and 1.85 AU (SE and NW sides, respectively) outward from the gap.  Therefore the distance between the two peaks is approximately 3.4$-$3.5 AU.  The major axis is 280.81 AU if we measure the distance between the belt bisectors, which is consistent with the semi-major axis value (140.7$\pm$1.8 AU) derived by \citet{kalas05a}.

To measure the projected (sky plane) shape of the belt, we determine the stellocentric positions of two distinct features in radial cuts through the belt:  (1) the bisector, and (2) the inner edge of the belt, defined as the half-maximum of the line that rises to the inner peak.   These position measurements from radial cuts are not possible in the regions closest to the star that are dominated by speckle noise, or crossing the azimuthal belt gap that is discussed below.  

\begin{figure}[!ht]
\epsscale{1.0}
\plotone{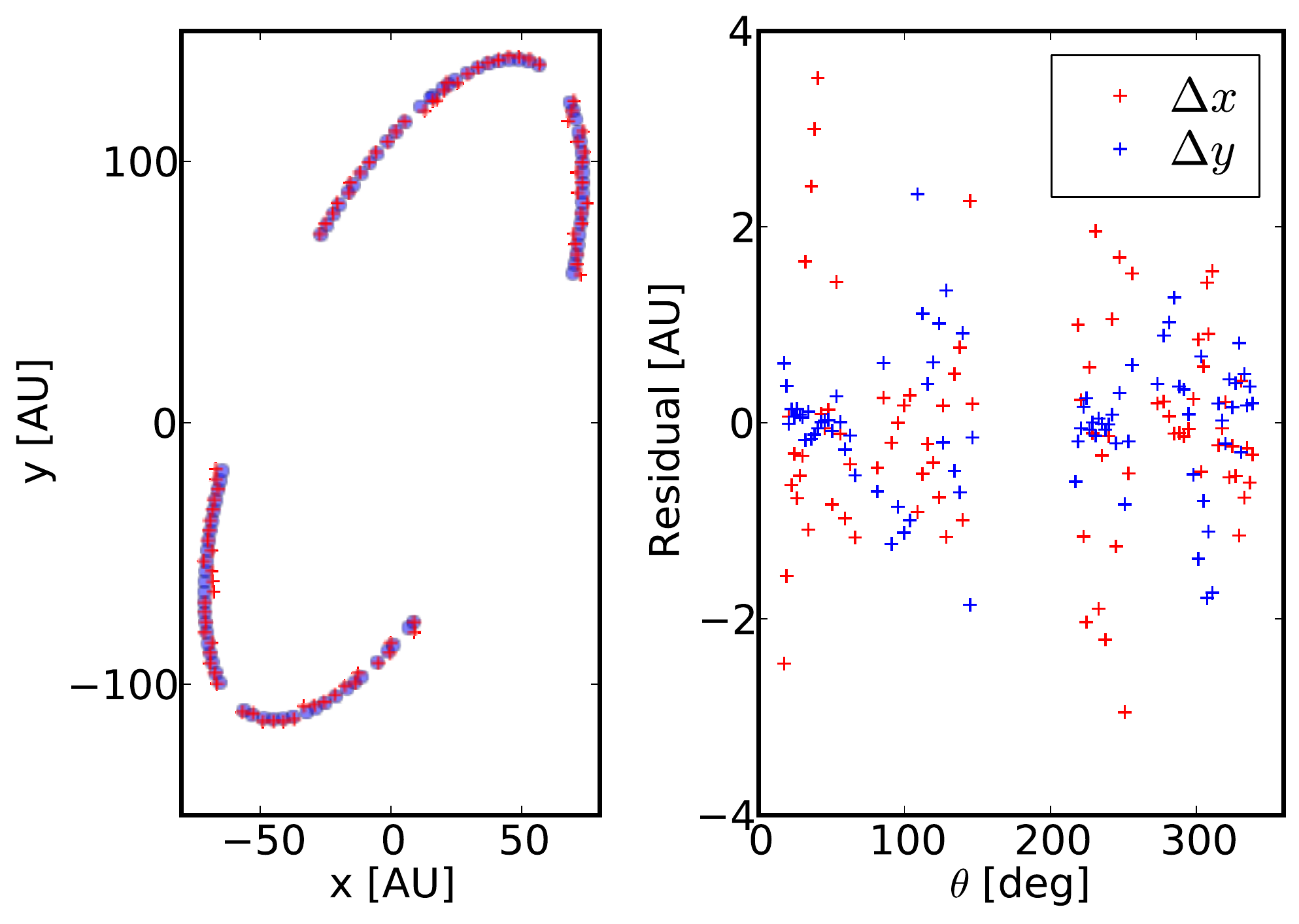}
\caption{\footnotesize
{\bf Left Panel}:  Measurements of the peak scattered light from the main belt (red crosses) and a fit (blue crosses) based on a Keplerian orbit.   {\bf Right Panel:} Residuals between the Keplerian fit and the measured points.  $\theta$ is the best-fit value of $\nu + \omega$ (true anomaly + argument of perapse).
\label{peak-belt-fit}}
\end{figure}

Figure~\ref{peak-belt-fit} plots the bisector positions and a corresponding least-squares Keplerian fit.  The fit assumes that the apparent belt structure traces a simple Keplerian orbit.  We consider the orbital phase at which a hypothetical belt particle would pass each measured point, and solve jointly for the orbital elements and these orbital phases. These phases are ``nuisance parameters'' in the problem, and the posterior distributions are marginalized over these parameters.

\onecolumn
\twocolumn

Table~\ref{apparent} summarizes the belt properties from \citet{kalas05a} and these two new measurements and Keplerian fits for the STIS data.  Even though we use a Keplerian orbit for the apparent structure, the belt is presumed to be a circle inclined to the line of sight.  If the belt represents a non-circular Keplerian orbit, the eccentricity is the ratio of the stellocentric offset of the belt center to the semi-major axis.   Therefore the projected ellipticity of the belt is due to both the inclination and the inherent non-circular morphology.  Table~\ref{derived} gives revised values for the belt's properties assuming a non-circular structure inclined to the line of sight.  We note that \citep{kalas05a} derived these values from the Kowalsky construction using apparent ellipses to find the true orbital elements (from Smart p. 352-353).  Here we have revised the \citet{kalas05a} values using the Keplerian orbit approach. The new values derived from the STIS data are in good agreement with those published by \citet{kalas05a}.

\subsection{Main Belt Outer Halo}
\label{sec:halo}

\begin{figure}[!ht]
\epsscale{1.0}
\plotone{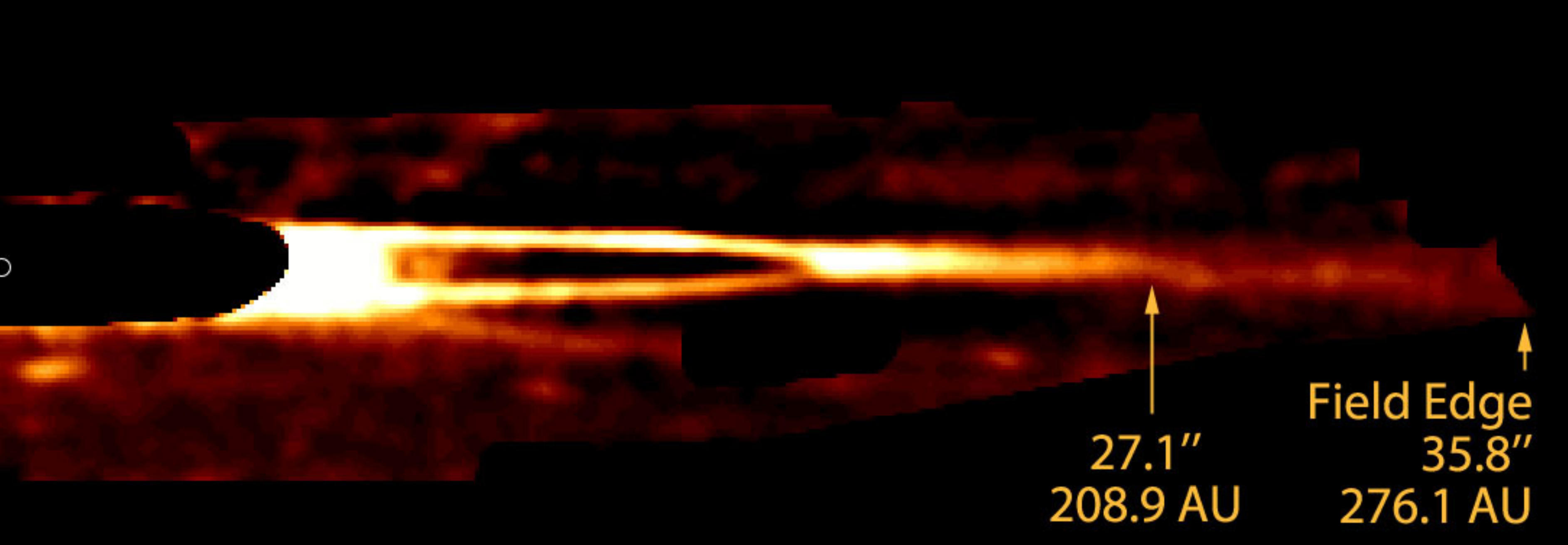}
\caption{\footnotesize
Binned image showing the northwest side of Fomalhaut's belt.  The combined image is rotated such that the belt major axis lies along a horizontal, with the northwest side pointing right.  The data are then binned 2 pixels along the x axis and 20 pixels along the y axis, and then convolved with a gaussian with $\sigma$=2 pixels.  The left edge of the frame represents the stellar position, and the distances mark the stellocentric positions.
\label{combo-collapsed}}
\end{figure}

The slope of the belt halo brightness along the apparent semi-major axis between $20''$ and 27$''$ projected radius (Fig.~\ref{combo-majorcut}) is fit by a power law with index -3.3.  The right portion of Fig.~\ref{combo-majorcut} shows that halo is detected more than $6''$ beyond the inner edge of the belt.  To improve the signal-to-noise of low surface brightness nebulosity along the apparent major axis, we bin the data along the apparent minor axis direction.  Figure~\ref{combo-collapsed} indicates that the extended halo extends at least as far as 209 AU from the star, or 57 AU beyond the inner belt edge (Fig.~\ref{combo-majorcut}).  At 209 AU the isophotes bend westward by $\sim$35$\degr$ (relative to the major axis, when the image is restored from its collapsed state).  Other debris disk midplanes show evidence for bending, such as  HD 32297, which has a $\sim31\degr$ bend \citep{kalas05b}, and HD 61005, which has a $\sim$10$\degr$ bend \citep{maness09a}.   Another possible explanation for the apparent bending of the Fomalhaut belt halo in the STIS data is  that the lowest surface brightness contours are influenced by a small mismatch ($<0.005$ cts/sec) in the background sky levels between frames in the 2012 epoch, and in between epochs.   These mismatches are emphasized when the images are binned and smoothed.  Future observations are required to confirm whether or not the Fomalhaut dust belt is detected beyond 209 AU and with a bend in the position angle.  In any case, the main finding is that the belt is significantly more extended in scattered light than previously known, with a detection out to at least 209 AU radius from the star.

\begin{figure}[!ht]
\epsscale{1.0}
\plotone{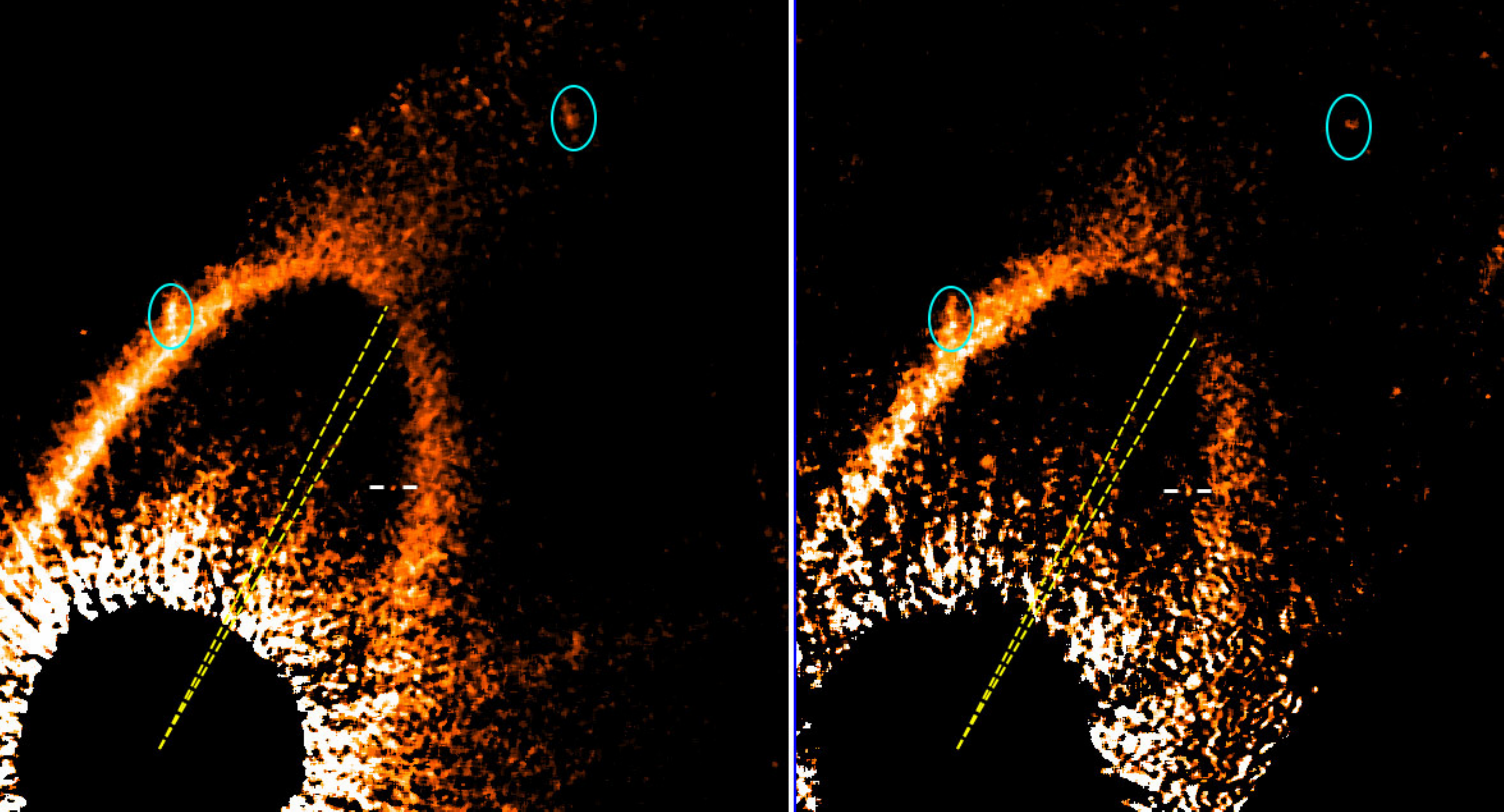}
\caption{\footnotesize
The northwest portions of the 2012 data (left) and 2010 data (right) with 5$\times$5 median binning and a hard grayscale stretch to emphasize the northwest gap.  The dotted lines are at PA = $329.8\degr$ and $332.8\degr$ in both data sets.  Ovals indicate background galaxies and Fomalhaut b is marked between white line segments.
\label{hardstretch}}
\end{figure}

\subsection{Main Belt 331$\degr$ Azimuthal Gap}
\label{sec:gap}

An additional newly discovered belt feature is an azimuthal belt gap approximately 6$''$ north of Fomalhaut b.  Figure~\ref{hardstretch} emphasizes this gap by smoothing the data and displaying the images with a hard stretch.  We classify it as a real astrophysical feature (as opposed to an instrumental artifact) because it is apparent in both epochs of STIS data. In the projected (sky plane) view, the gap appears to be $\sim$3$\degr$ wide, beginning at PA=$329.8\degr$.  Since this is the faintest portion of the belt, we examined the possibility that the gap is an artifact introduced by instrumentation or data reduction.  For the 2012 observations we exclude from data reduction the five orbits where this gap region lands near a diffraction spike, occulting mask, or field edge.  The final image produced with the remaining seven orbits continues to show the belt gap.  We therefore conclude that since artifacts are excluded and the gap appears in both STIS data sets (this region is not in the field of view for the ACS/HRC observations), it is likely a real astrophysical gap.

\begin{figure}[!ht]
\epsscale{1.0}
\plotone{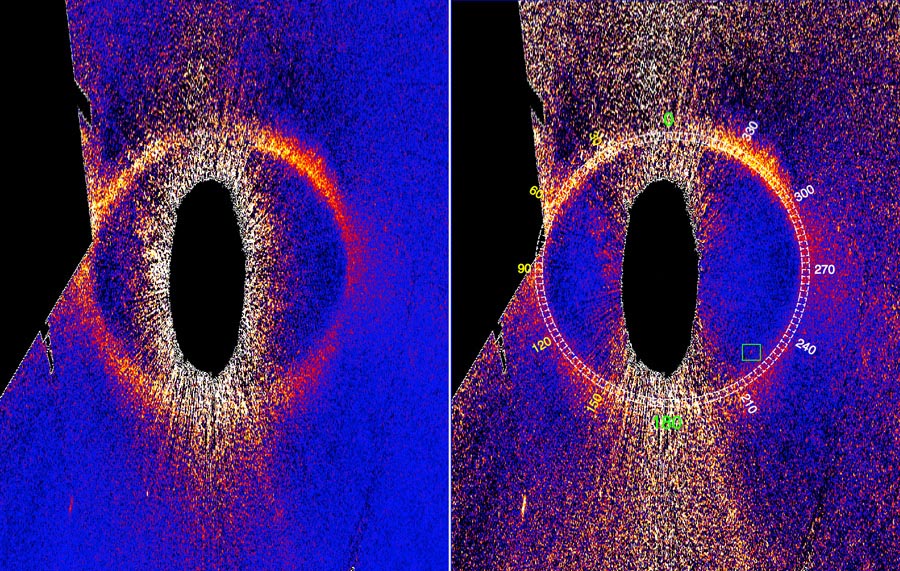}
\caption{\footnotesize
Deprojection of the belt (Fig.~\ref{combo}) by 66.5$\degr$ after the apparent semi-major axis is rotated to horizontal (clockwise by 66.0$\degr$).  The right panel has been normalized by multiplying the image by radius squared centered on the stellar location.  The angle markings on the right panel are belt centric, so that 0$\degr$ and 180$\degr$ mark the belt's apparent minor axis.  
\label{deprojected}}
\end{figure}

Figure~\ref{deprojected} shows a deprojection of the belt and a radius squared multiplicative scaling of the image centered on the star to normalize for the fall-off in stellar illumination.  The deprojections assumes a circular structure inclined to the line of sight by $66.5\degr$ (Table~\ref{apparent}).  In 2012, Fomalhaut b is 98 AU and 78 AU to the right and below the star, respectively, in this reference frame.  The four-epoch motion is essentially to the right in the +X direction at roughly 0.00260 AU/day.  Assuming Fomalhaut b is coplanar with the belt, it has to travel $\sim$19 AU to reach the inner edge of the belt.  Therefore we might expect to witness the real or projected belt crossing around 2032.  

\begin{figure}[!ht]
\epsscale{1.0}
\plotone{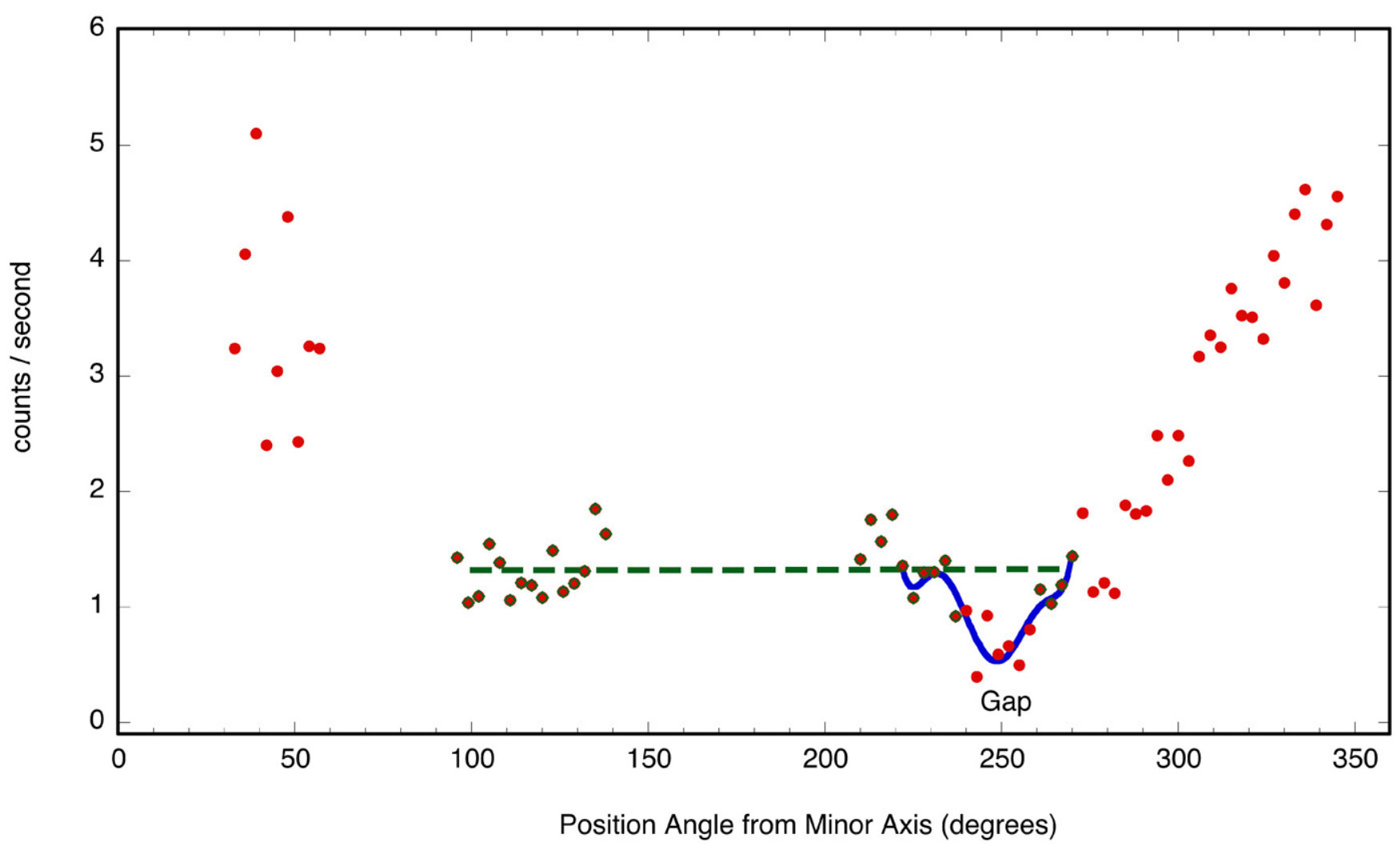}
\caption{\footnotesize
Photometry along the illumination-corrected, deprojected belt (Fig.~\ref{deprojected}) in circular apertures with diameter 0.5''.
The blue line is a 9th order polynomial fit to the gap region.  The
green dashed line is a least square linear fit to the
points excluding the gap, but includes measurements
in the 100$-$140 degree region.  The green dashed line
fit is near horizontal, as expected for the left-right
symmetry argument.  The gap is a significant
depression, but not entirely empty.  The full width
at half-minimum is $\sim$50 AU (note that at 141 AU radius,
1$\degr$ on the plot corresponds to 2.461 AU in
circumference).  
\label{fullpaplot}}
\end{figure}

The  azimuthal brightness asymmetries are due to an asymmetric scattering phase function and the fact that the star is closest to the southern portion of the belt, as discussed in \citet{kalas05a}.  Since the star is 13 AU to the left of the belt center, one effect is that the left hemisphere of the belt receives greater illumination than the right hemisphere, which may account for apparent belt gaps in the right hemisphere.  However, the belt gap is still evident in the illumination-corrected image.  Figure~\ref{fullpaplot} gives photometric measurements along the circumference of the belt in the illumination-corrected image.  The brightness in the gap minimum is approximately 50 per cent of the mirror region in the left hemisphere.  The gap width measured as a full-width at half-minimum is $\sim$50 AU.  We note that the belt minor axis serves as the reference frame for the azimuth (degree) measurements shown in the right panel of Fig.~\ref{deprojected} , which is slightly offset from the reference frame describing the scattering angles relative to the star.  This offset between reference frames is $\sim$5$\degr$ within 30$\degr$ of the minor axis (e.g. between azimuth 150-210$\degr$).   However, at the azimuth of the belt gap ($\sim$250$\degr$) the scattering angle offset is $<$1$\degr$ and therefore the gap cannot be explained by a scattering phase function effect. 

\section{Orbit of Fomalhaut b}
\label{sec:orbit}

\subsection{Astrometry and Uncertainties}
\label{sec:astrometry}

\begin{figure}[!ht]
\epsscale{1.0}
\plotone{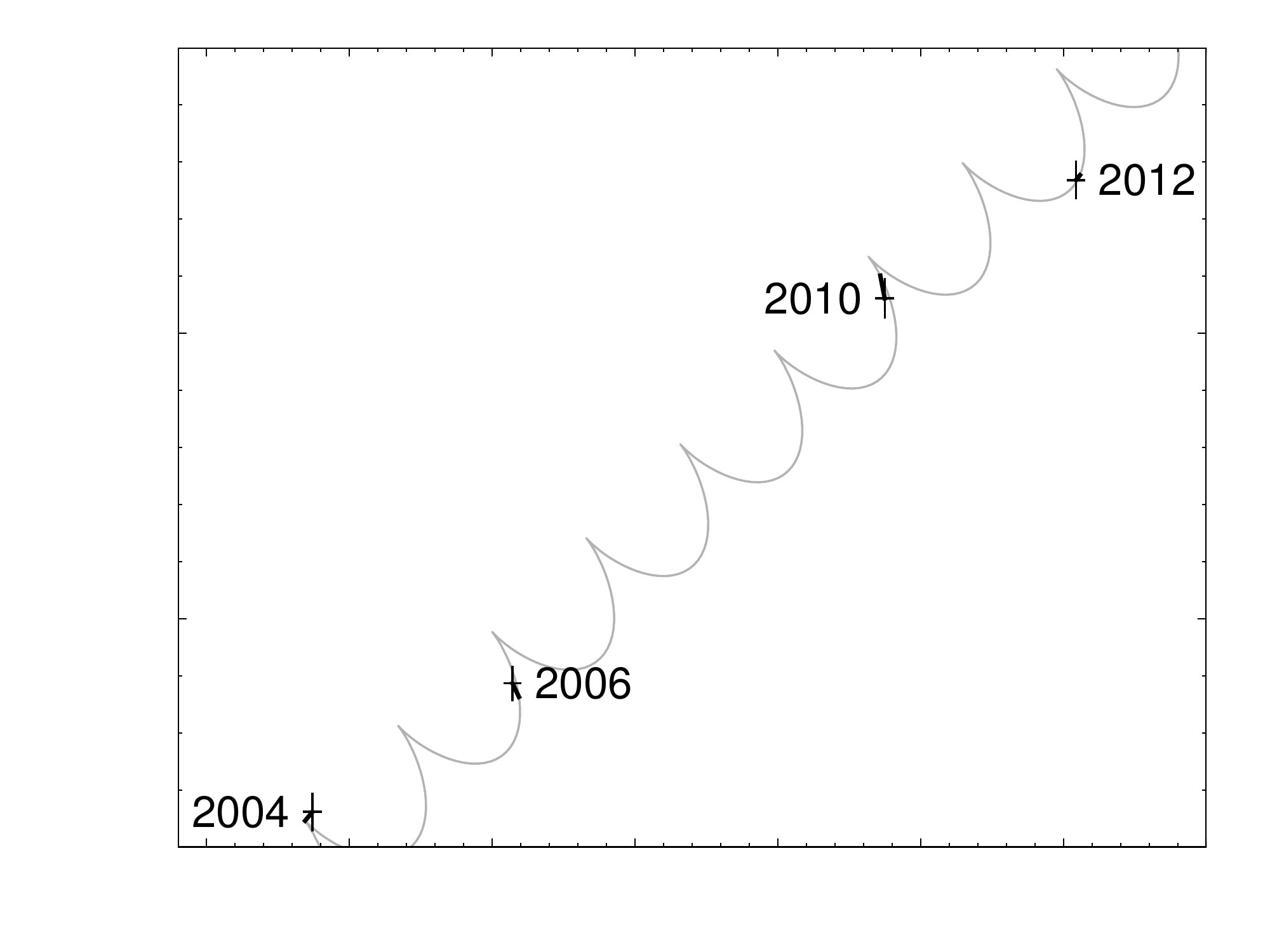}
\caption{\footnotesize
Test of astrometry on a background star in the reference frame of Fomalhaut A.
For the 2010 and 2012 epochs of STIS data, the error bars are
derived assuming 17 mas in residual geometric distortion instead of
the 66 mas from the STIS calibration program (Table~\ref{astrometryerrors}).
\label{backgroundstar}}
\end{figure}

Table~\ref{astrometry} summarizes the four epochs of astrometry with 1$\sigma$ error
bars derived from combining the error terms in Table~\ref{astrometryerrors} in quadrature. 
Determining the position of the star behind occulting spots or wedges, and the residual geometric
distortion in STIS are the two most significant sources of astrometric uncertainty.  To test for possible
systematic errors in any of the epochs, we conducted astrometry on a faint background star
south of Fomalhaut, residing outside of the dust belt boundary.  This is the only background object
detected at all four epochs.  Figure~\ref{backgroundstar} compares our astrometry to the predicted
locations using proper motion and parallax information from the Hipparchos Catalog.  
The residuals between the expected and measured locations 
are $\sim$20 mas, which we take as evidence that the 66 mas value for the residual geometric distortion 
adopted from the STIS calibration program is an overestimate.  
In Figure~\ref{backgroundstar} the error bars plotted are derived from the residual geometric distortion for STIS inferred from the Fomalhaut data (17 mas; Table~\ref{astrometryerrors}).
The residuals are now comparable to the 1-$\sigma$ error bars, justifying adoption of the 17 mas
value for the assumed uncorrected geometric distortion in STIS.  
We note this is a likely upper limit given that the background star is $\sim1\arcsec$ (20 pixels) 
farther from the star than Fomalhaut b in 2012.

\subsection{Kinematics}

\begin{figure}[!ht]
\epsscale{1.}
\plotone{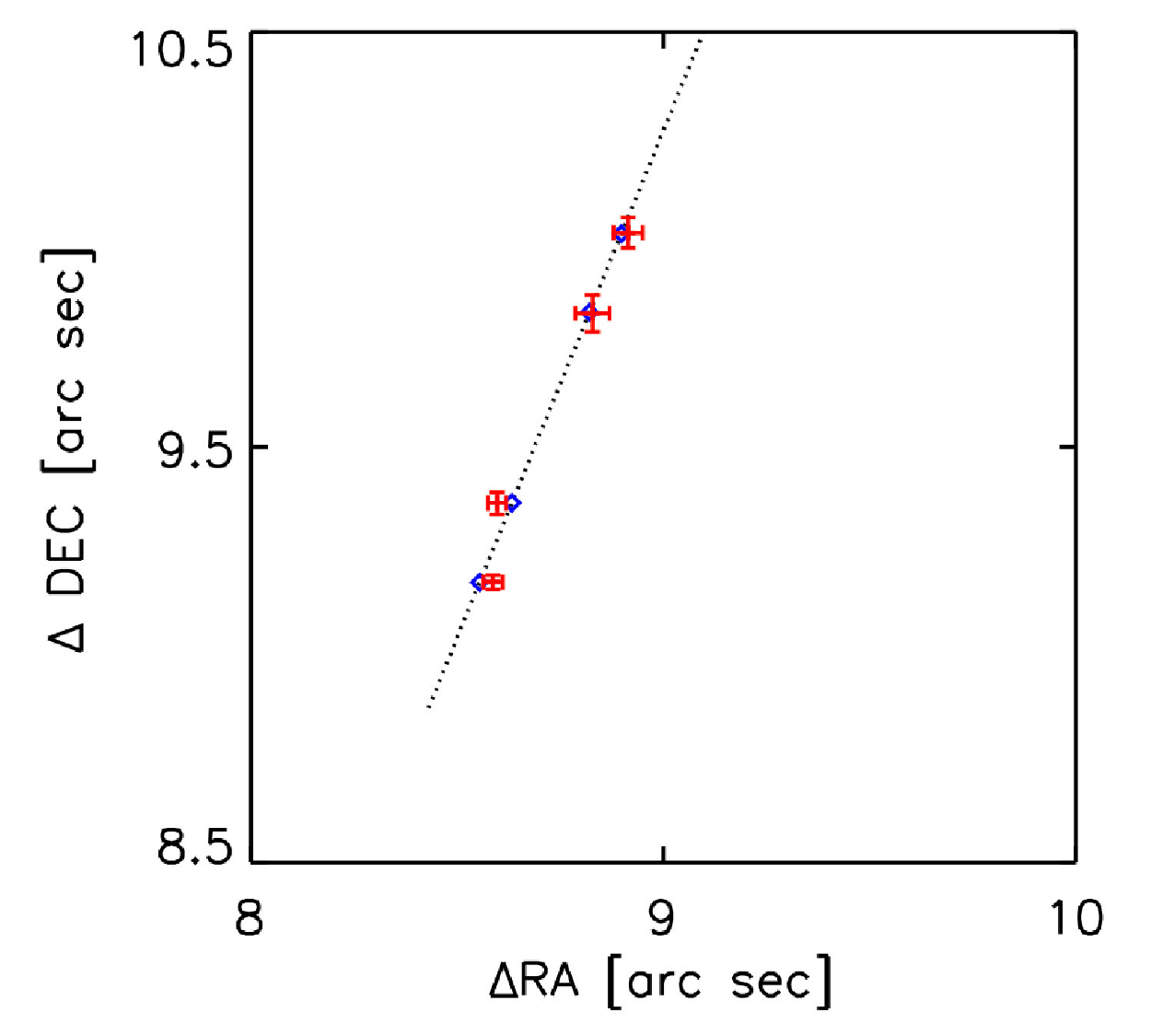}
\caption{\footnotesize
Uniform motion model - the dotted line shows motion at constant velocity $4.36$ km/s; the blue diamonds show the predicted positions for the best-fit model.
\label{uniform_motion}}
\end{figure}

Figure~\ref{uniform_motion} shows the sky plane motion of Fomalhaut b compared with a uniform motion (unaccelerated) model. The corresponding best fit speed is $4.36 \pm 0.17$ km/s with corresponding $\chi^2 = 4.73$ [4 degrees of freedom; the cut-off probability $P(\chi^2 > 4.73) = 0.32$].  The quoted velocity uncertainty is a propagation of the astrometric error (Table~\ref{astrometry}).  We do not know the escape speed if we assume that projection effects are unknown.  However, the escape speed, $v^p_{esc} = \sqrt{2GM_\star/r_p}$ = 5.837 km/s at the mean observed projected separation ($r_p \approx 100$ AU), represents an upper limit. For circular orbits  $v/v^p_{esc} \leq 2^{-1/2} \approx0.707$; hence, the measured value of the ratio  $v/v^p_{esc}=0.747\pm0.03$ implies that the object must be on an elliptical or hyperbolic orbit. 

\begin{figure}[!ht]
\epsscale{0.9}
\plotone{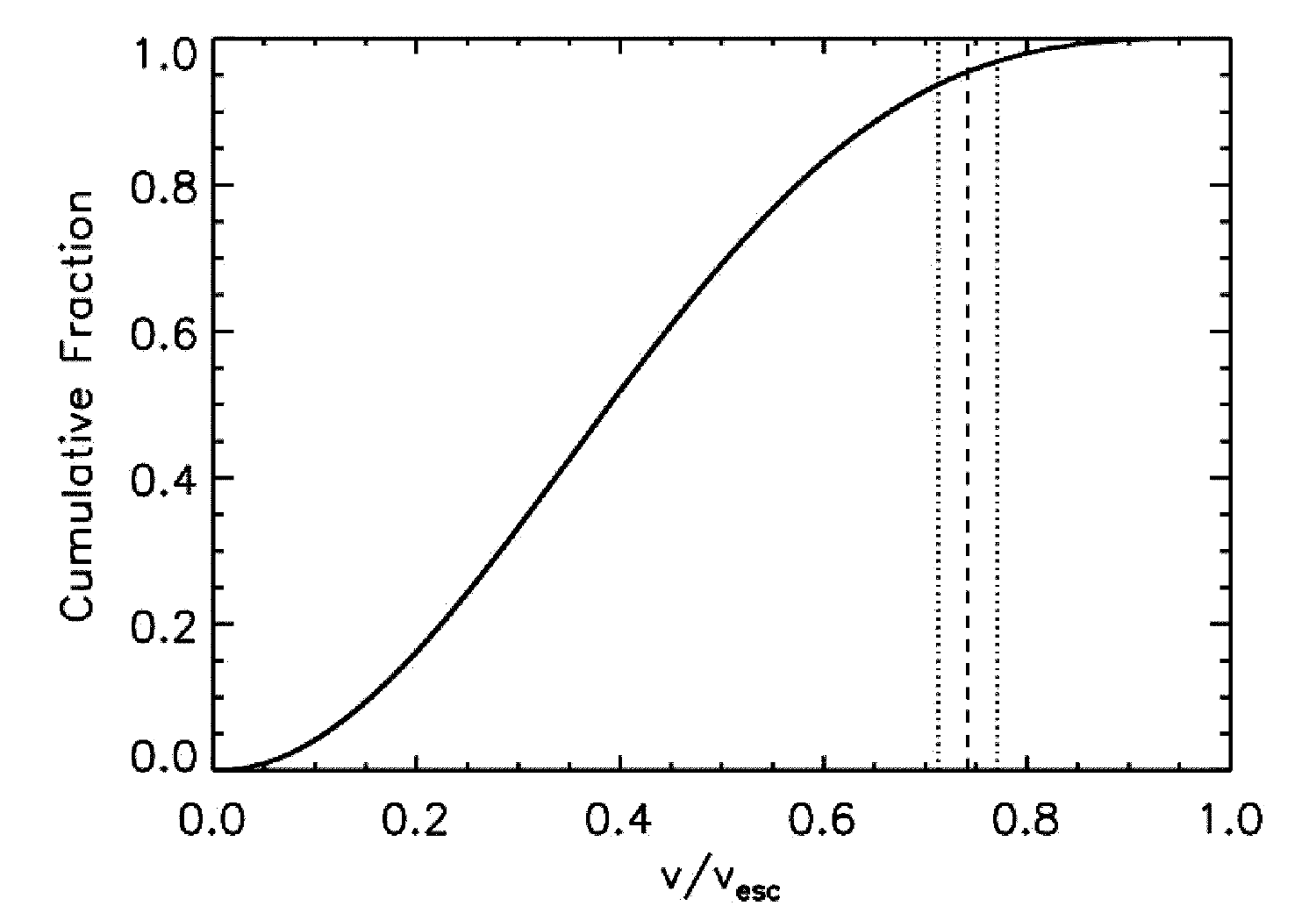}
\caption{\footnotesize
The cumulative distribution of $v/v^p_{esc}$ for randomly oriented orbits with an eccentricity distribution $dp/de=2e$. The vertical dashed line shows the observed value and its $1\sigma$ error bounds. 
\label{random_cum_dist}}
\end{figure}

The large observed value of  $v/v^p_{esc}$ does not mean that the object is unbound. For an ensemble of randomly oriented orbits with $e$ distributed between zero and one,  $v/v^p_{esc} < 1$.  The distribution of $v/v^p_{esc}$ depends on the details of the eccentricity distribution: for a uniform distribution ($0\leq e\leq1$) then  $\langle v/v^p_{esc}\rangle= 0.425\pm0.178$; for a more physically-based ÒthermalÓ distribution ($dp / de = 2e$; \citealt{heggie75a}) the ratio is $0.402\pm0.190$. Figure~\ref{random_cum_dist} shows that cumulative distribution of  $v/v^p_{esc}$ for randomly oriented, elliptical orbits.  Less than 9.2\% (6.2\%) of bound orbits have an observed value of  greater than that allowed by the observations at 99\% (68\%) confidence.  Therefore the allowed phase space is not large for bound orbits in random orientations, but not improbable.


Even though acceleration is not yet detected, the magnitude and direction of motion for a bound object constrain the Keplerian orbital elements.   We assume that mass and heliocentric distance are known and that the errors are sufficiently small that marginalization over the associated uncertainties is unnecessary.  We use standard methods to compute the Cartesian coordinates of an orbiting body by solving KeplerÕs equation for the eccentric anomaly and hence the radius and true anomaly \citep{green85a}.  For hyperbolic orbits ($e>1$) we solved Kepler's equation using the approach of \citet{gooding88a}. 

\begin{figure}[!ht]
\epsscale{0.9}
\plotone{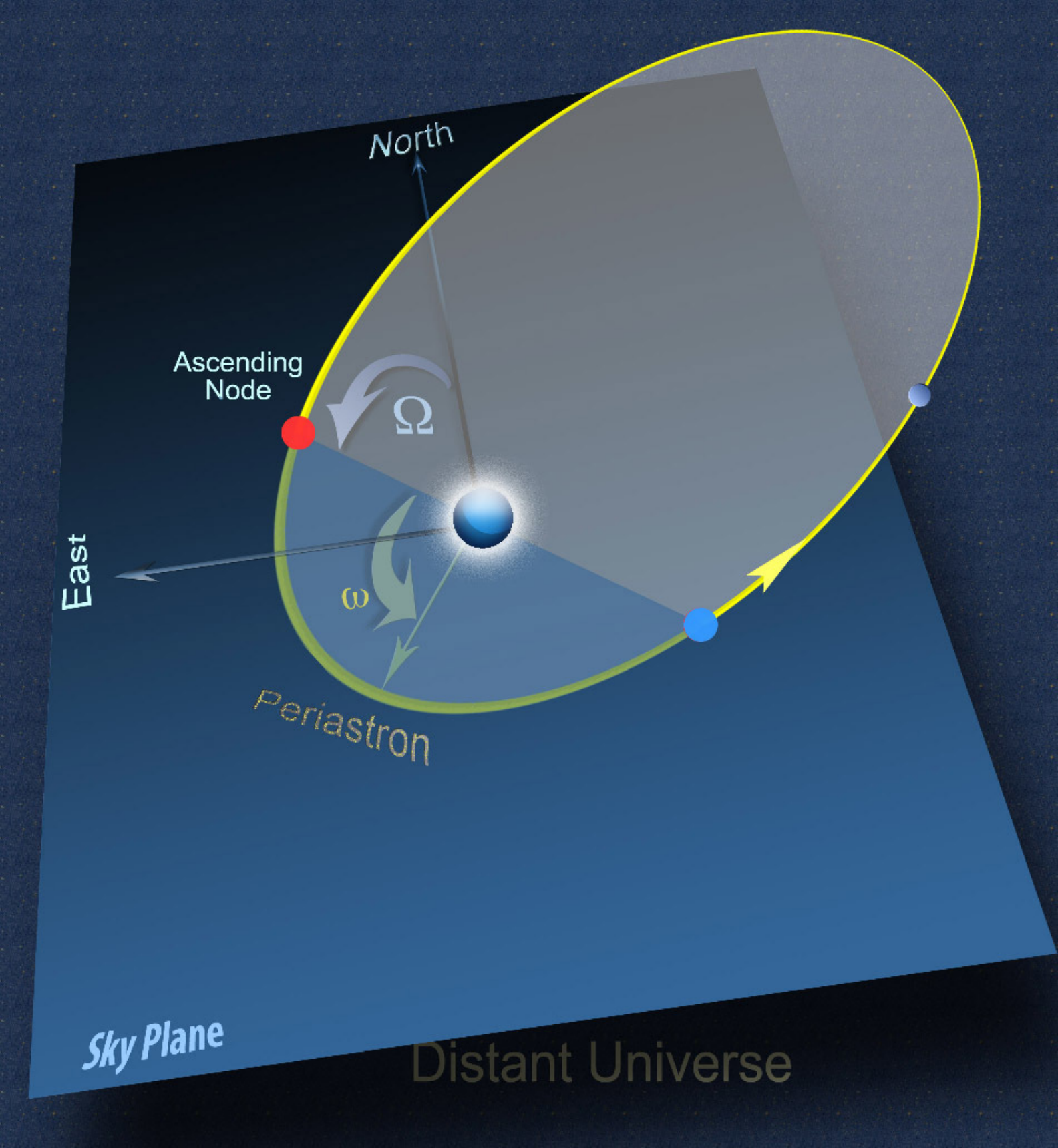}
\caption{\footnotesize
Diagram marking several orbital elements for an orbit inclined relative to the reference plane which is the sky plane.   $\Omega$ and $\omega$ are not coplanar.
We follow the binary star convention where positive Z (not drawn) is into the sky plane (below the sky plane drawn here), such that the ascending node is the point where the orbiting body crosses the reference plane (red circle) toward positive Z \citep{green85a}.
In this particular sketch, the planet lies out of the sky plane (nearest Earth), which means
that the descending node (blue circle) follows periastron passage.  The periastron vector lies in the plane of
the orbit and represents the direction of the true semi-major axis.  This does not necessarily 
correspond to the apparent semi-major axis of an inclined orbit projected onto the sky plane. At the current epoch,
Fomalhaut b has passed through periastron, but it has not yet reached the descending node (i.e. it still
resides behind the sky plane).
\label{orbit_diagram}}
\end{figure}

There are six unknowns in this problem: two describe the shape and size of the orbit (eccentricity, $e$ and semimajor axis, $a$); three angles (argument of perihelion, $\omega$, longitude of the ascending node, $\Omega$, and inclination, $i$, accounts for the orientation in space relative to a reference direction (north and position angle) and reference plane (sky plane); and one describes the orbital phase (epoch of perihelion). Figure~\ref{orbit_diagram} illustrates the astronomical convention where the ascending node is the point where the orbit penetrates into the sky plane away from us; $\Omega$ is an angle in the sky plane measured eastward from north to the ascending node; and $\omega$ is the angle in the orbital plane between the ascending node and periastron, $q$ \citep{green85a}.  We assume that Fomalhaut b is currently observed behind (into) the sky plane such that the inclination of the orbital plane is a negative value.  We note that the main belt is also described by an orbital plane, where the mutual inclination between this and Fomalhaut b's orbital plane is represented by $I$.  

Our data comprise eight measurements: two measurements of position at four distinct epochs (2004, 2006, 2010, and 2012). The problem of finding the orbital elements is therefore over-determined and a statistical approach using, for example, the method of least squares or maximum likelihood is necessary to estimate the orbital elements and their uncertainties.

For initial exploration of the problem we used the Levenberg-Marquardt algorithm to find acceptable sets of parameters \citep{bevington69a}. It is evident from these investigations that the six-dimensional $\chi^2$ surface has many local minima. The Levenberg-Marquardt algorithm finds local minima, not the global minimum; moreover, estimates of the parameter uncertainties, which are derived from a Taylor-series expansion of $\chi^2$ about a local minimum, are untrustworthy.  

We have therefore used a Markov chain Monte Carlo (MCMC) method to sample the posterior probability distributions for the orbital elements. The method employed here computes the likelihood function $-$ assuming that the measurement errors are normally distributed $-$ and the Metropolis-Hastings algorithm to select new members of the chain from a proposal distribution \citep{sivia06a}. The Metropolis-Hastings algorithm guarantees convergence of the Markov chains to the posterior distribution, but convergence is slow when a high rate of rejection ($<<$50\%) of the proposed values occurs; a common circumstance for problems with a large number of parameters. To speed convergence we included an initial phase during which an adaptive proposal distribution is used at each step.  The adaptation is known as simulated tempering \citep{gregory01a}.   We adopt uniform priors for the proposal distributions of the orbital elements and each chain is started with a random value within the prior range.  A burn-in period proceeds and convergence and independence of the Markov chains are establish using the statistical methods of \citet{raftery95a}.

\begin{figure}[!ht]
\epsscale{1.}
\plotone{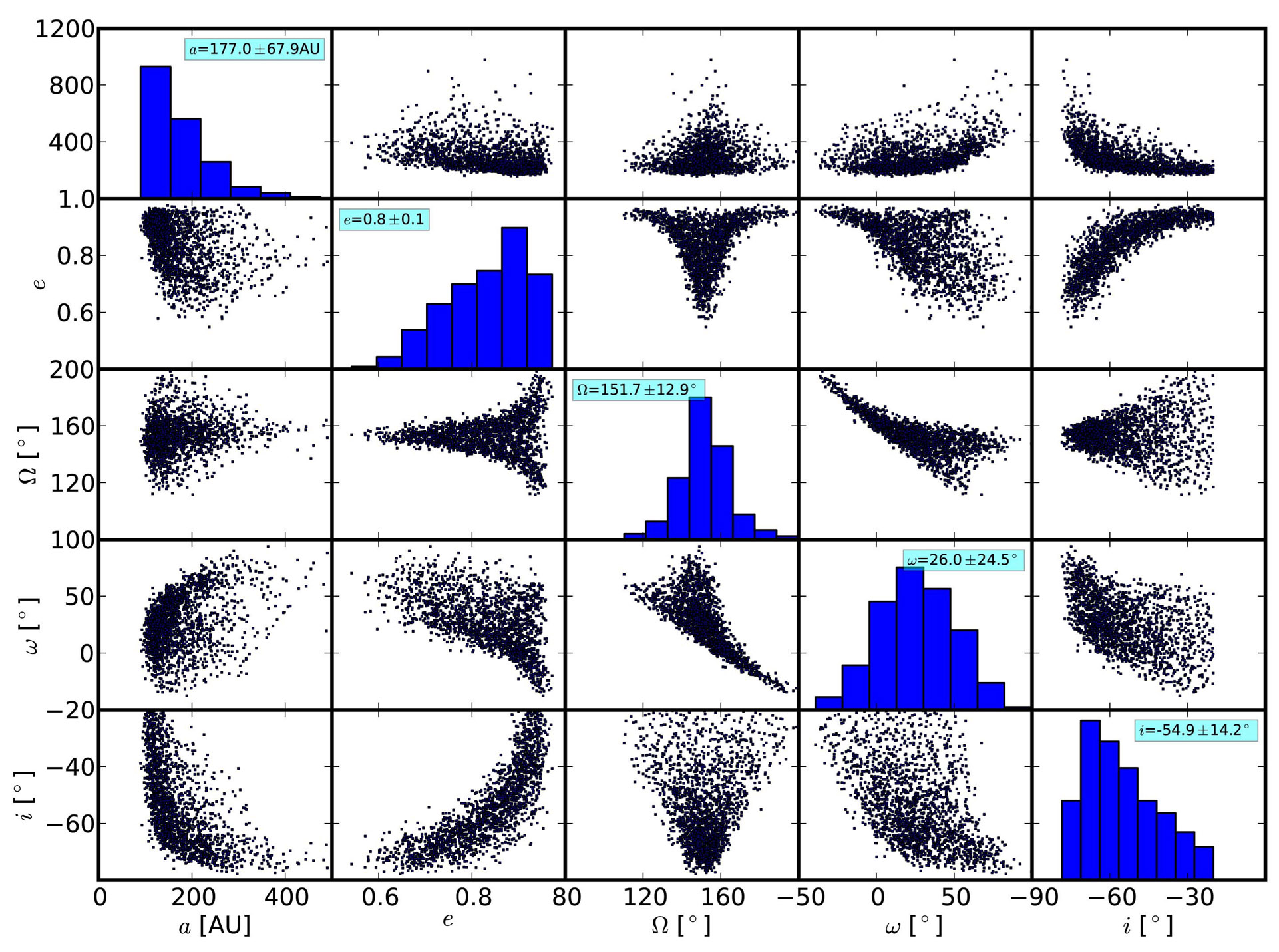}
\caption{\footnotesize
Plots showing the distributions and correlations for orbital elements $a,~e,~\Omega,~\omega$, and $i$ for Fomalhaut b. The histograms along the diagonal show the marginalized probability distribution. The off-diagonal plots show the correlation between the corresponding parameters $-$ each dot represents a Markov chain element. The mean and standard deviation of each marginal distribution are listed in the accompanying legend. 
\label{histogram1}}
\end{figure}

Figure~\ref{histogram1} shows the results of this analysis. The
adopted priors for the free parameters are
$a\in\left[80,800\right],~e\in\left[0.4,1.0\right],~\Omega\in\left[110\degr,200\degr\right]$,
and $i\in\left[0\degr,90\degr \right]$; no priors were imposed on $\omega$ or the epoch of
perihelion. The limits on semimajor axis and longitude of the
ascending node were imposed after extensive exploration of the entire
range of these parameters. No viable solutions were found outside of
these ranges and therefore these priors were adopted as a convenience
to speed the convergence of subsequent Markov chain calculations; the
lower limit of $e = 0.4$ was adopted for the same reason. 
The prior probabilities for semimajor axis and inclination are uniform in $\log a$ and $\cos i$, respectively. 
In each case the posterior distribution is sharply peaked in contrast to the
initial uniform prior and characterized by a standard deviation that
is significantly smaller than the prior range.  For application of the 
Markov chain methods to the determination of the orbital elements
see \citet{ford06a} and \citet{chauvin12a}.

\onecolumn
\twocolumn

\begin{figure}[!ht]
\epsscale{1.0}
\plotone{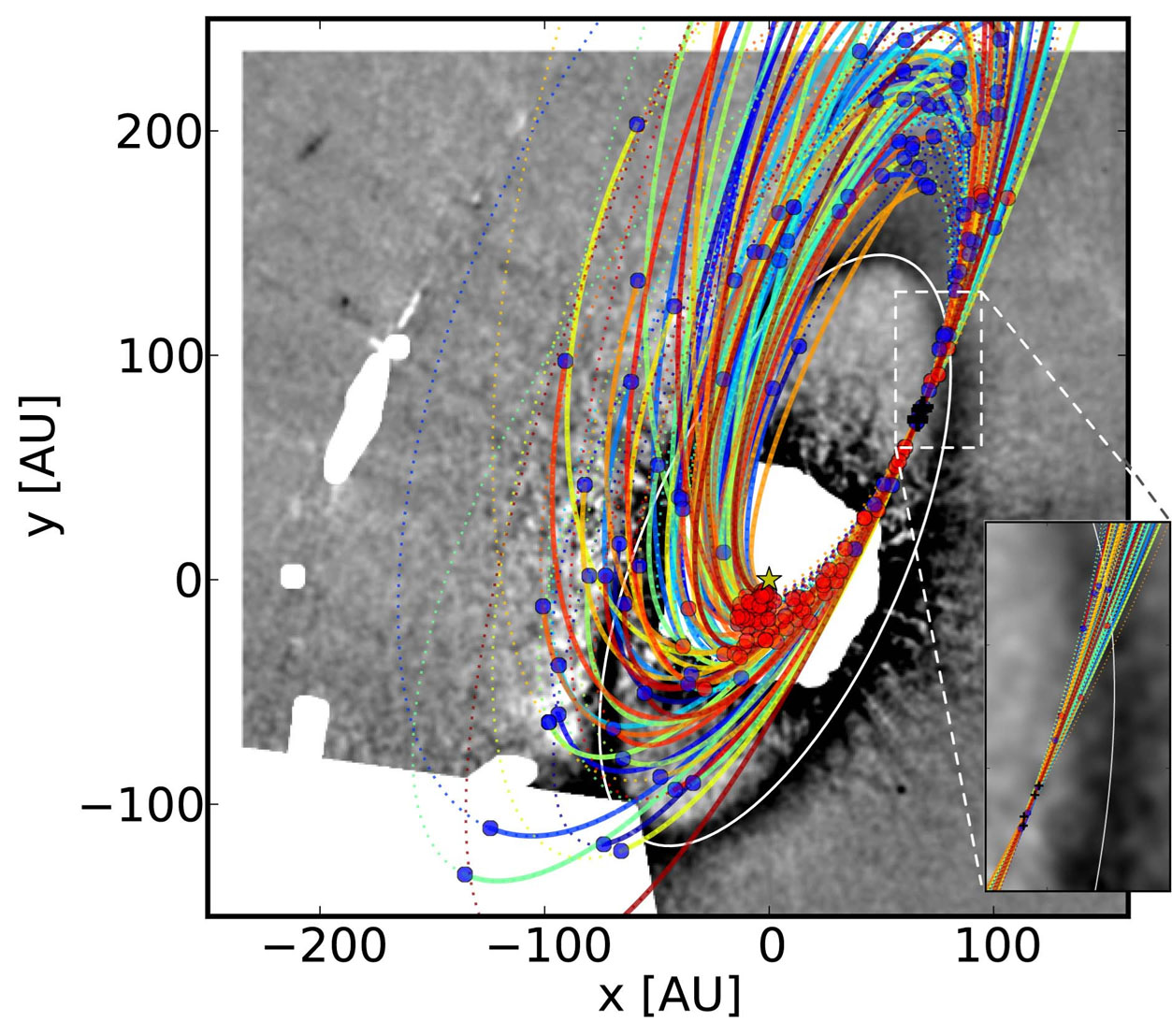}
\caption{\footnotesize
A sample of 100 orbits drawn from the Markov chains, representing orbits that are consistent with the astrometric data ($+$ symbol). The background shows the HST/STIS image; the white line shows the loci of the peak of the main belt.  Orbits are drawn in two segments between the ascending node (red dot) and the descending node (blue dot) {\it with respect to the plane of the main belt}. From the ascending node to the descending node the orbit is drawn as a dashed line (i.e. behind the sky plane); between the descending node and the ascending node the orbit is drawn as a solid line (i.e. in front of the sky plane). The inset shows a zoomed view ($30\times60$ AU) where the astrometric data are plotted with a $+$ symbol. 
\label{2d_orbit}}
\end{figure}

Figure~\ref{histogram1}  shows that the current observations favor an elliptical orbit ($e=0.8\pm0.1$) with large semimajor axis ($a = 177\pm68$ AU); a low eccentricity orbit ($e \approx 0.1$) that is nested within the belt is ruled out.  Figure~\ref{2d_orbit} shows a sample of 100 orbits drawn from the Markov chain, representing orbital elements that are consistent with the astrometric data. This figure demonstrates graphically that the projected motion of Fomalhaut b crosses the main belt. However, because of the mutual inclination of the belt and the orbit Fomalhaut b does not necessarily penetrate the belt.  

\begin{figure}[!ht]
\epsscale{1.0}
\plotone{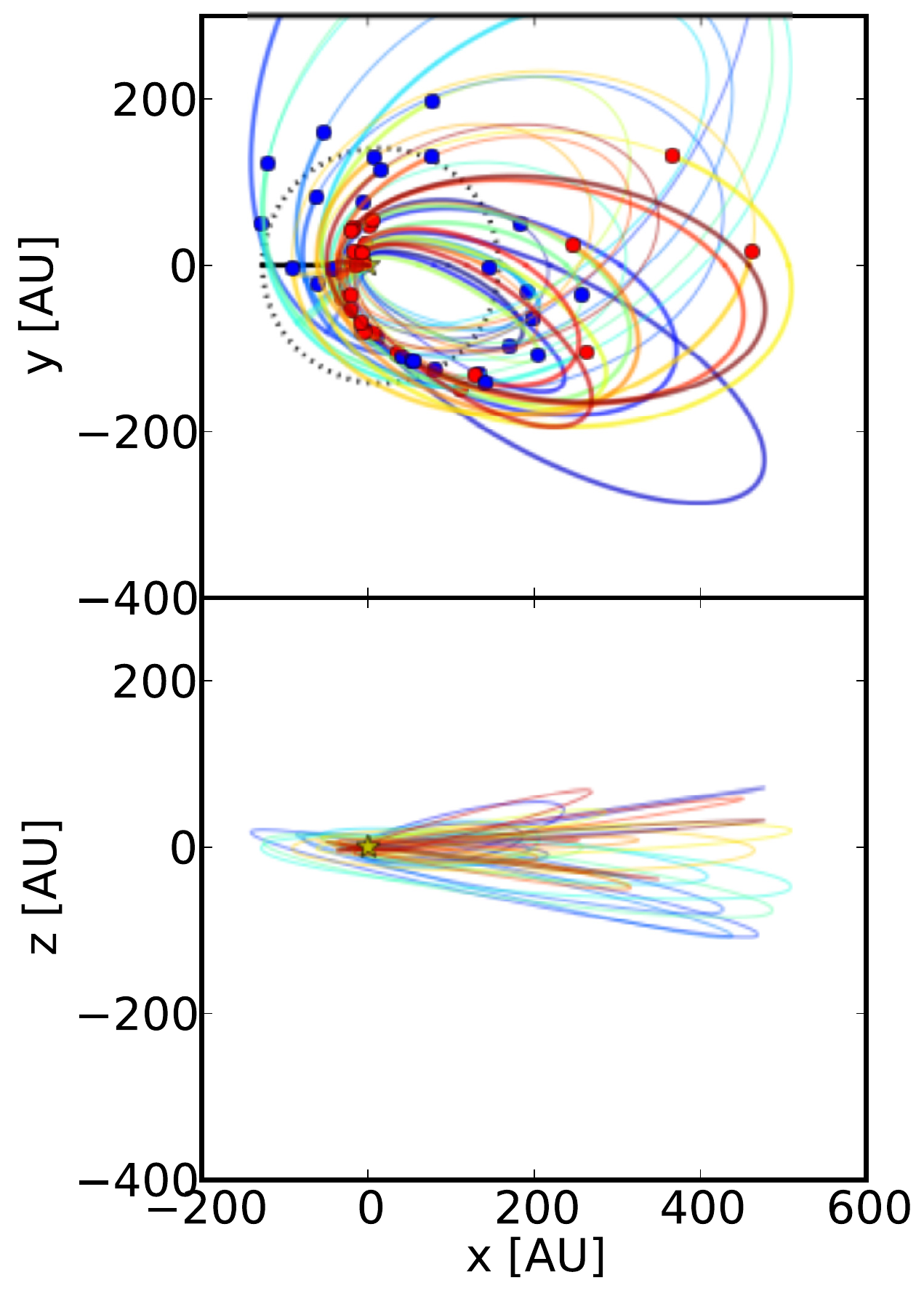}
\caption{\footnotesize
Sample of 30 orbits from Figure~\ref{2d_orbit}, viewed face-on (top) and edge-on (bottom) with the same orientation as the deprojected images shown in Figure~\ref{deprojected}.
The dashed black line represents the main belt, with pericenter to the lower left.  The ascending (red dots) and descending (blue dots) nodes with respect to the belt plane are mostly
concentrated within the perimeter of the belt. 
\label{orbits}}
\end{figure}

\begin{figure}[!ht]
\epsscale{1.0}
\plotone{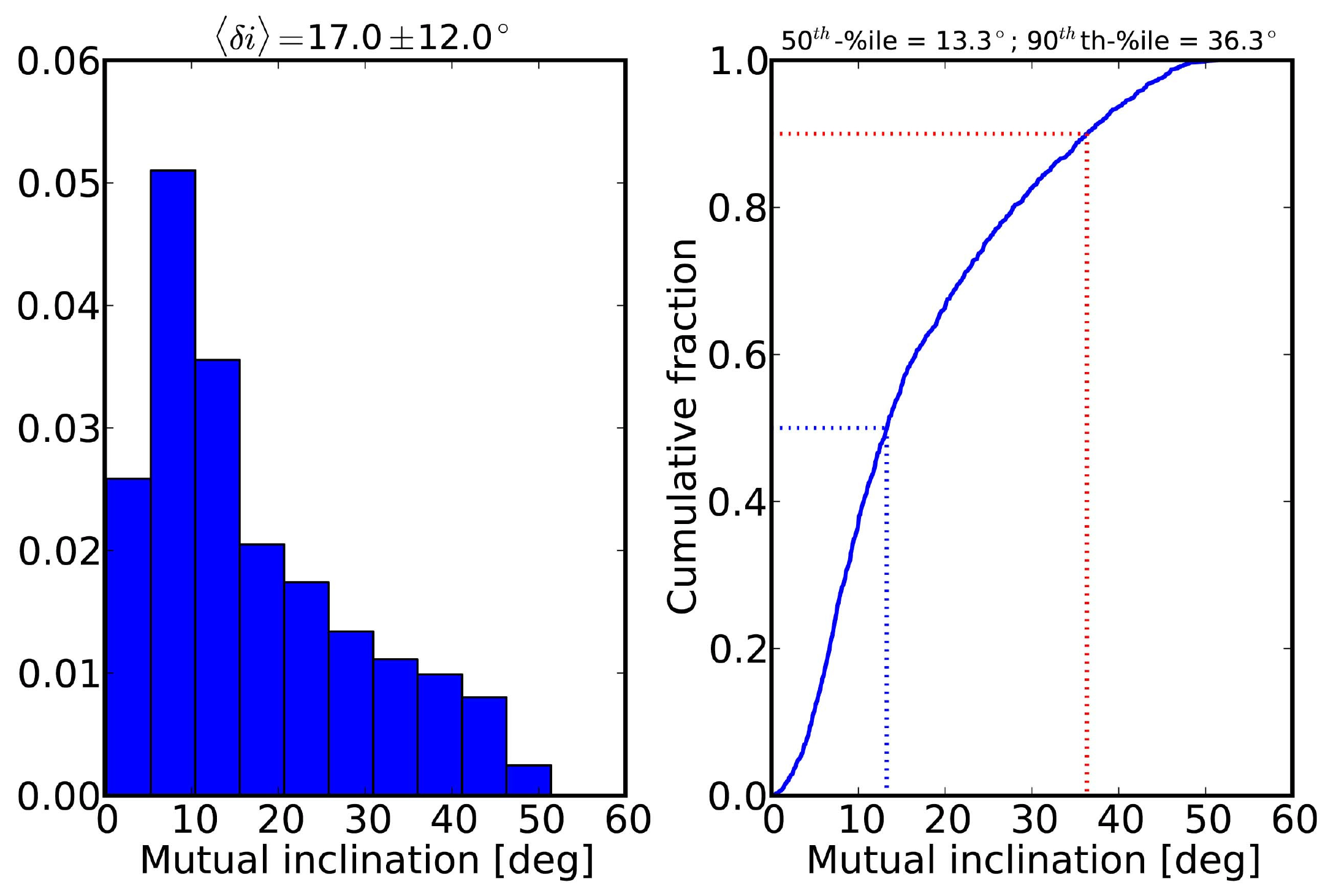}
\caption{\footnotesize
The posterior distribution (left) and cumulative distribution (right) of mutual inclination, $I$, between the orbit of Fomalhaut b and the main belt. The mean difference in inclination is $17.0\degr\pm12.0\degr$. Fifty percent of allowed orbits lie within $13.3\degr$; 90\% of allowed orbits lie within $36.3\degr$. 
\label{mutual}}
\end{figure}

\begin{figure}[!ht]
\epsscale{1.0}
\plotone{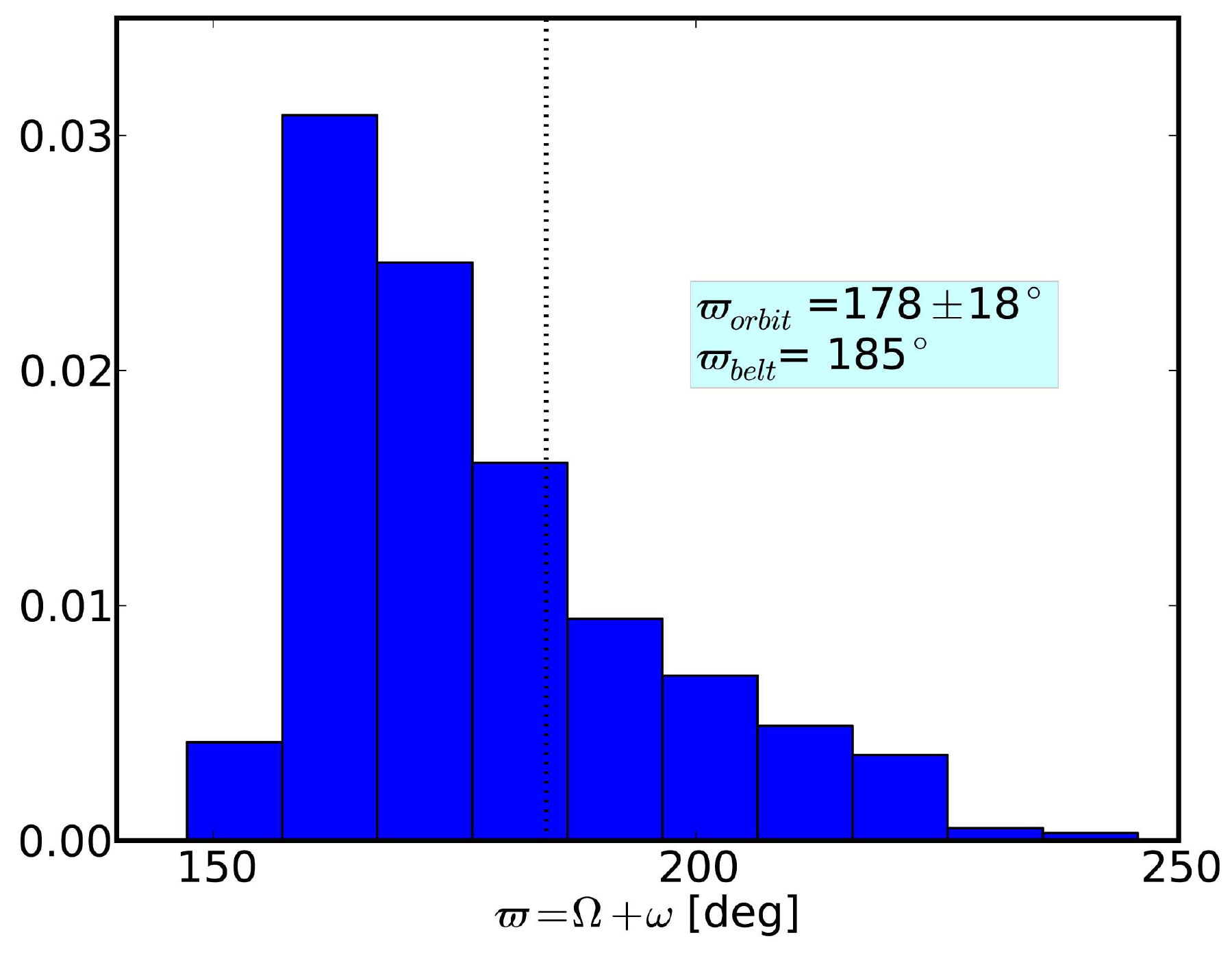}
\caption{\footnotesize
The posterior distribution of the longitude of periapse, $\Omega + \omega$, for Fomalhaut b. The vertical dotted line denotes the longitude of periapse of the main belt. 
\label{vari-pi}}
\end{figure}

Figure~\ref{orbits}  shows the face-on and edge-on views of 30 orbits.  
The majority of ascending nodes relative to the belt are concentrated interior to the belt, near Fomalhaut b's periastron.  
The edge-on view emphasizes that the mutual inclination is most likely $I\lesssim36\degr$ (90\% confidence; Fig. 22).  Fomalhaut b's orbit is unlikely to intersect the main
belt at $a\sim140$ AU because the main belt is relatively flat [the model-dependent opening angle for the belt is 1.5$\degr$ \citep{kalas05a}] and the nodes are distributed at many locations interior and exterior to the belt.  

The probability of Fomalhaut $b$ directly interacting with the main belt depends on how the problem is defined, such as considering the size of Fomalhaut b's Hill sphere at the intersection region (which depends on the planet mass estimate) and the assumed physical boundaries of the belt.  If Fomalhaut b is massive, then it can still gravitationally perturb a portion of the belt without crossing through it.  To quantify the belt crossing probability, we simply calculate the fraction of ascending and descending nodes that occur within various annuli representing the belt, without consideration of Fomalhaut b's mass and Hill radius.  We find that 12\% of nodes occur in the regions $133$ AU $\leq a \leq 158$ AU.  This 25 AU wide annulus was defined in the scattered light observations of \citet{kalas05a} and it is roughly equal to the FWQM of the ALMA radial profile measurements at 870 $\mu$m \citep{boley12a}.  For a wider annulus starting from the belt inner edge at 133 AU, to the newly detected outer edge at 209 AU (Fig.~\ref{combo-collapsed}), the probability is 43\%.  These values suggest that the geometric deformations of the tentative belt detection beyond 209 AU may be dynamically linked to Fomalhaut b, whereas there is a smaller, $\sim$10\% chance that Fomalhaut b interacts with the main concentration of belt mass near $\sim$140 AU.

Inspection of Figure~\ref{2d_orbit} and Table~\ref{derived} suggests that the orientation of the orbit of Fomalhaut b, within the uncertainties, is apsidally aligned with the main belt.  The inclination of the orbit, $i_b = -55\degr\pm14\degr$ is similar to that of the belt ($i_{belt}  =-66\degr$), and the longitude of the ascending node, $\Omega_b = 152\degr\pm13\degr$, is also consistent with that of the belt ($\Omega_{belt} = 156\degr$). The posterior cumulative distribution of mutual inclination between the orbit of Fomalhaut b and the main belt is shown in Figure~\ref{mutual}. The mean difference in inclination is $17\degr \pm12\degr$. Fifty percent of allowed orbits lie within 13$\degr$ and 90\% of allowed orbits lie within $36\degr$; the corresponding solid angles cover 0.8\% and 5\% of the sky respectively, indicating a small chance of this alignment occurring at random. Moreover the longitudes of periapse, $\Omega + \omega$, for Fomalhaut b and the belt are aligned within the errors (Figure~\ref{vari-pi}) .

\begin{figure}[!ht]
\epsscale{1.0}
\plotone{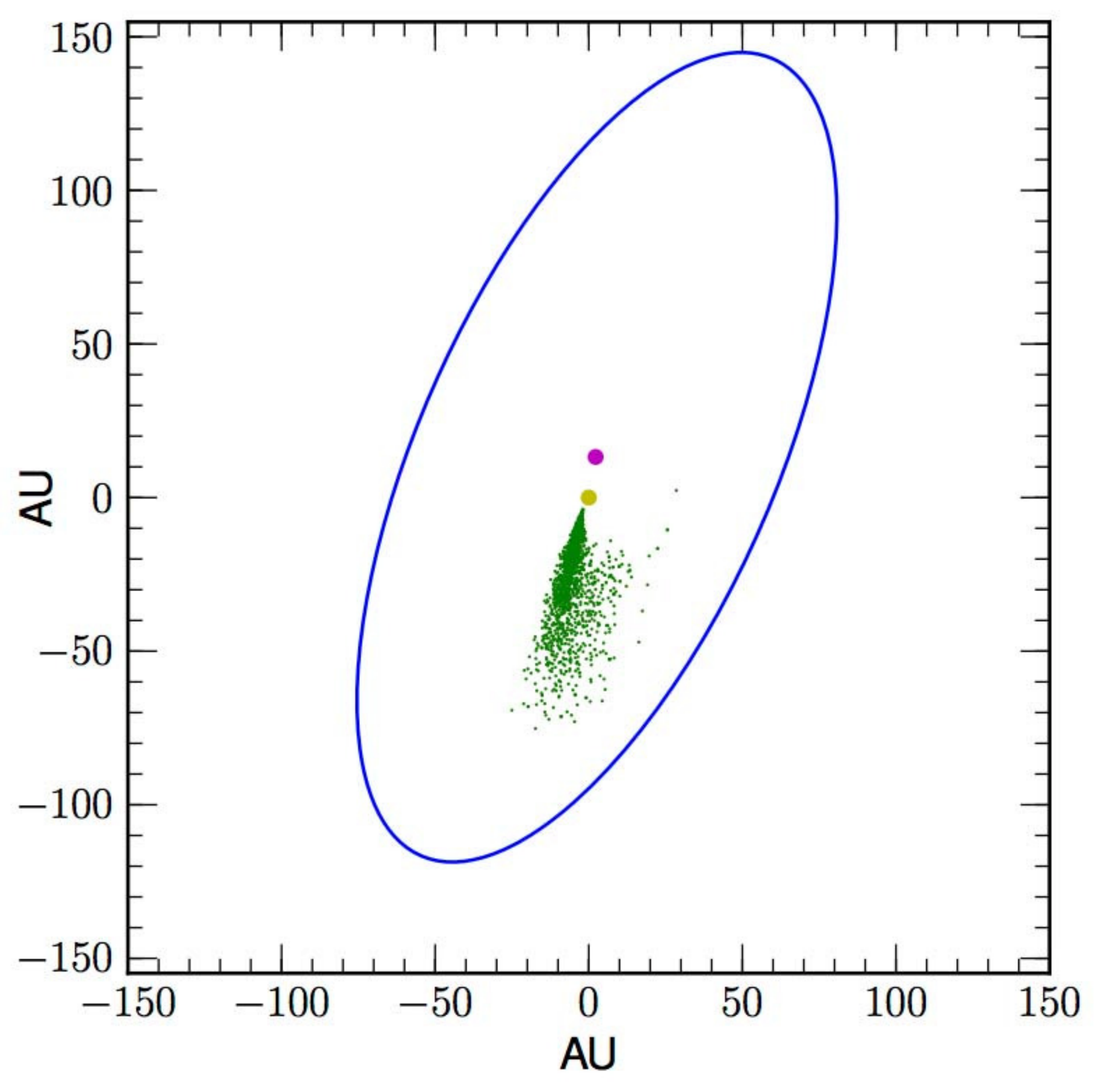}
\caption{\footnotesize
Locations of periastra.  The blue line traces the belt, the red dot is the geometric center of the belt, the yellow dot is the stellar location, and the green points represent the projected pericenters derived from the distribution of orbital elements.
\label{perturber_peri}}
\end{figure}

\begin{figure}[!ht]
\epsscale{1.0}
\plotone{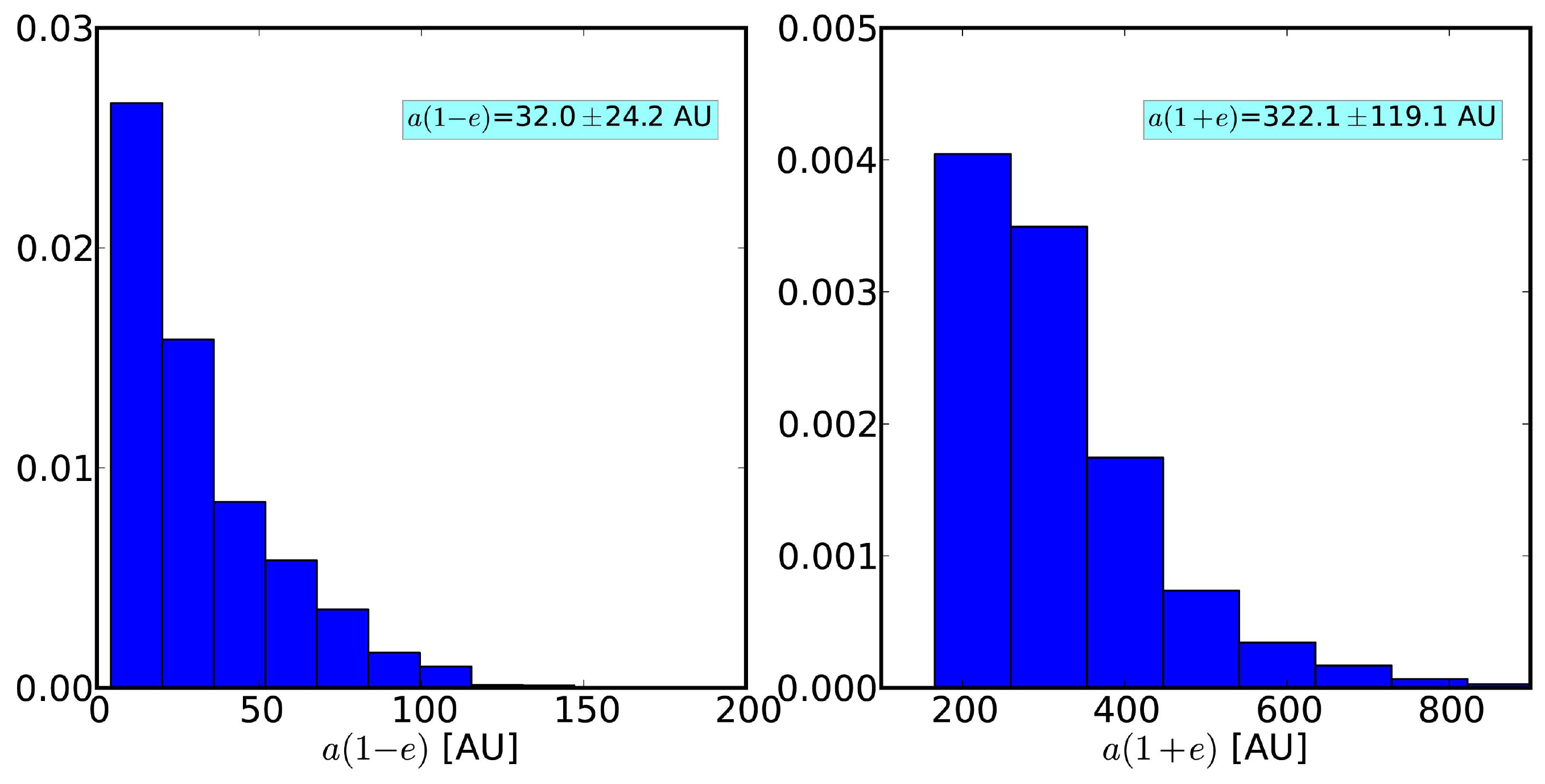}
\caption{\footnotesize
Posterior distributions of the periapse (left) and apoapse (right) distributions.
\label{peri-apo-histo}}
\end{figure}

Figure~\ref{perturber_peri} demonstrates that periapse occurs interior to the main belt, behind the sky plane, and south of the star as projected on the sky plane. The posterior distribution of the periapse distance (Figure~\ref{peri-apo-histo}) has a mean value $32\pm24$ AU. Many of the Fomalhaut b orbits intersect the belt plane near periapse, suggesting that the region near periapse is where Fomalhaut's system may be most dynamically disturbed. 

\section{Discussion}
\label{sec:discussion}

Fomalhaut is emerging as an increasingly complex planetary system. 
It is therefore helpful to review and define its various elements
before assessing the possible nature of Fomalhaut b.

\subsection{Inventory of the Fomalhaut System}  
\label{sec:inventory}

\begin{figure}[!ht]
\epsscale{1.}
\plotone{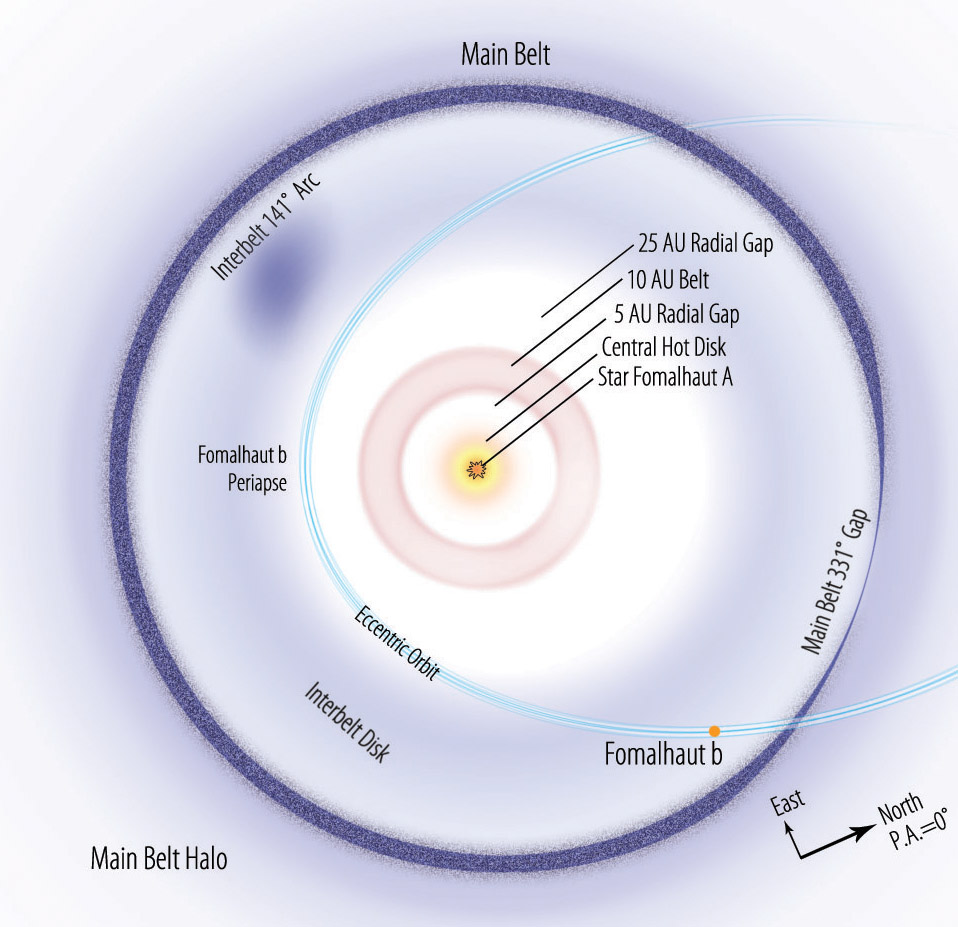}
\caption{\footnotesize
Notional sketch of the Fomalhaut system viewed face-on.  The radial locations
of features are approximate and not to scale.  For example, the 10 AU belt
represents dust near 10 AU radius, but the width of the belt is not precisely known.
\label{sketch}}
\end{figure}

Figure~\ref{sketch} is a notional sketch of the Fomalhaut system, where we
adopt a nomenclature based on the approximate positions of features in
radius and position angle in the sky plane.

{\bf (1) Fomalhaut A} is the central A3V star ($\alpha$ PsA, HD 216956, GJ 881).  \citet{mamajek12a} finds
$T_{eff}$ = 8590$\pm$73 K, L = 16.65$\pm$0.48 L$_\odot$, M = 1.92$\pm$0.02 M$_{\odot}$
and age 440$\pm$40 Myr.   The heliocentric distance is 7.704$\pm$0.028 pc and the angular radius for the stellar
photosphere of 1.01 mas corresponds to 1.84 R$_\odot$ \citep{difolco04a}.  
The position angle of the stellar spin axis is aligned with the minor axis of the main belt \citep{lebouquin09a}.
Given the $\sim16\times$ greater luminosity, the radiation environment at a given radius in the
Fomalhaut system is roughly four
times greater than for the Solar system.
The tidal radius, $a_t$, set by the Galactic tidal field is \citep{tremaine93}::
\begin{equation}
a_t = 1.7\times 10^5 AU \biggl(\frac{M_{\star}}{M_{\odot}}\biggr)^{1/3}
\biggl(\frac{\rho}{0.15~ M_{\odot} ~pc^{-3}}\biggr)^{-1/3}
\end{equation}

Assuming $\rho$ = 0.11 $M_{\odot}~ pc^{-3}$ \citep{holmberg00a}, then
$a_t$ = 234 kAU (1.1 pc).    Therefore Fomalhaut B, discussed below,
is well within the sphere of Fomalhaut A's gravitation influence.
The barycenter is located 15.8 kAU from Fomalhaut A.

\begin{figure}[!ht]
\epsscale{1.0}
\plotone{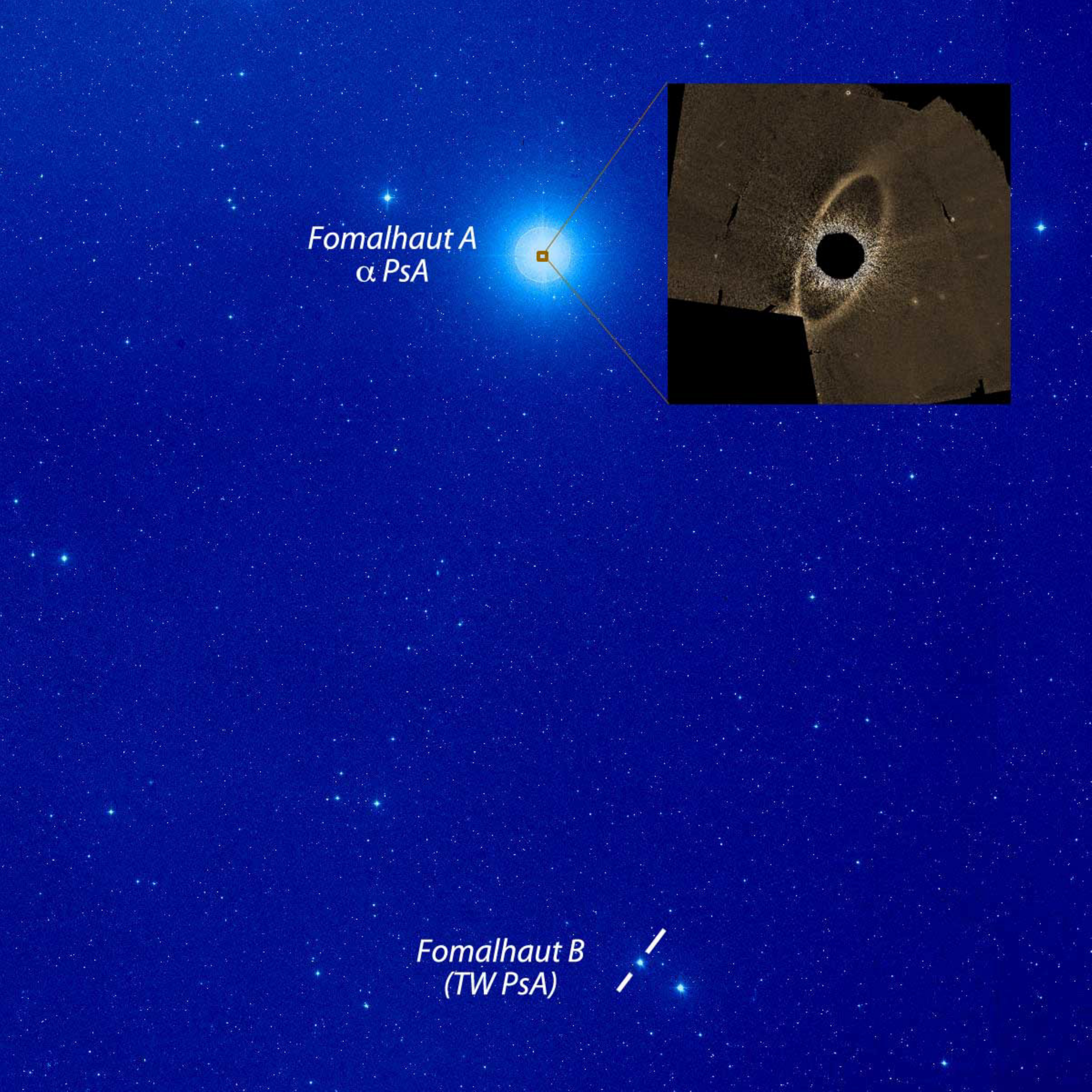}
\caption{\footnotesize
The Fomalhaut system.  North is up, east is left, and the sky-plane separation
between Fomalhaut A and B is 2.0$\degr$.  The background image is a false-color, log-scale,
gnomonic (tan) projection from the optical Digitized Sky Survey 1 (red plates), centered on
$\alpha= 22^h~57^m~36^s,~\delta=-30\degr~25'~08''$ (Credit: National Geographic
Society, Caltech, STScI).  The STIS Fomalhaut image is overlaid.
\label{3degree}}
\end{figure}

{\bf (2) Fomalhaut B} is a common proper motion stellar companion, also known
as TW PsA.  This is a K4Ve star with $T_{eff}$ = 4594$\pm$80 K, L = 0.189$\pm$0.013 L$_\odot$, and
M = 0.73$^{+0.02}_{-0.01}$ M$_{\odot}$.  In the sky plane TW PsA is located 1$\degr$.96 (55 kAU) southwest of Fomalhaut A,
and therefore lacks a projected alignment with the major axis of the belt (Figure~\ref{3degree}).  \citet{mamajek12a}
gives a 3D project separation of 57.4$^{+3.9}_{-2.5}$ kAU.  
The heliocentric distances of Fomalhaut A and B are within 2$\sigma$ of each other.  
No further information is currently available with respect to the possible
orbit of Fomalhaut B.  If 57.4 kAU is adopted as the semi-major axis value, then the orbital 
period is $\sim$8 Myr. We note that given a single astrometric observation of a binary separation,
the most likely value for the orbital semi-major axis is the observed projected separation \citep{savransky11a}.

{\bf (3) Fomalhaut b} could alternately be named Fomalhaut Ab.  The present work revises
the previous notion that Fomalhaut b's orbit is nested within the belt.  Instead, Fomalhaut b's
orbit is highly eccentric.  In the 2012 epoch of observation, Fomalhaut b is 125 AU from the
star assuming an inclination identical to that of the main belt.
Apastron will likely be beyond $300$ AU, where its velocity will be $\sim$1 km/s.  Fomalhaut b's current
blackbody temperature is 50 K, 
whereas a 32 AU periastron and 322 AU apoastron
give temperatures of 99 K and 31 K, respectively.
The mass of Fomalhaut b is $\lesssim1$ M$_J$ due to the non-detection at infrared wavelengths.
Initially reported variability at optical wavelengths is not confirmed.  The possibility that
it is a resolved object also requires future observations for confirmation.  If the optical brightness
of Fomalhaut b is due to circumplanetary dust grain scattering, then compared to the present epoch it was approximately 
16 times brighter at periastron ($\sim$30 AU), and may become undetectable at apastron when it
becomes at least eight times fainter.  

{\bf (4) The Main Belt} is the primary source of far-infrared emission and is also prominent
in optical scattered light \citep{kalas05a}, with a sharp inner edge at a semi-major axis of 133 AU.  The geometric
center of the main belt is offset from the stellar location by 15 AU.  
The eastern hemisphere of the belt is brighter than the western hemisphere because
the former lies out of the sky plane in the forward scattering direction, and the grain surface area
is dominated by $\sim$10 $\mu$m sized grains that are preferentially 
forward scattering.  The main belt has mass $10^{22}-10^{24}$ kg in directly observed grains.  However,
given the age of the system, \citet{wyatt02a} argue that objects as large as a few km participate in the 
collisional cascade, yielding a total main belt mass of 20-30 M$_\earth$.  Including primordial
bodies as large as 1000 km that are not yet collisionally evolved, the belt mass
could be near 1 M$_J$.

{\bf (5) The Main Belt 331$\degr$ Gap} refers to the approximate position angle of the azimuthal
dust depletion detected  in optical scattered light.  In the deprojected reference frame it has FWHM$\approx$50 AU.  
As noted below, the region of the interbelt dust disk contains an arc of 450 $\mu$m emission located 180$\degr$ from
the 331$\degr$ gap.  

{\bf (6) The Main Belt Outer Halo} is a tenuous dust component extending to radii
exceeding 200 AU.  The scattered light color of the belt halo is currently unknown
and the morphology may bend westward at large distances from the main belt.

{\bf (7)  The 10 AU Belt} is inferred from an $unresolved$ component of 24 $\mu$m excess emission detected with
the Spitzer Space Telescope \citep{stapelfeldt04a}.   This mid-infrared bandpass corresponds to blackbody emission with
$T\approx125$ K, which at Fomalhaut is located at $\sim$20 AU radius (and roughly equal to the resolution
limit of the observations).  Re-analysis of these data suggest that the emitting grains can be
constrained to lie between 8 and 12 AU, depending on their size and composition \citep{su13a}.
The significance of this region is that the blackbody grain temperature is 170 K, which is
the canonical ice-line temperature in a circumstellar disk \citep[e.g.,][]{ida05a, kennedy08a}. 
Therefore the 10 AU belt could be called the ``ice-line belt'.

{\bf (8)  The interbelt dust disk}: 
There is evidence for dust located inward from
the main belt.  \citet{kalas05a} referred
to an ``inward intrusion of nebulosity'' from the 
main belt to as close as $\sim$100 AU radius (a
sensitivity limited value).  This inner dust component
is also detected as 24 $\mu$m  \citep{stapelfeldt04a}
and 70 $\mu$m thermal emission \citep{acke12a}.
\citet{acke12a} assumed that the inner edge is $\sim$35 AU,
stating that this is the 100 K water ice line of the system.
Their best-fit model has grain surface density increasing linearly
with radius out to the inner edge of the Main Belt (133 AU).
The total grain mass is $4\times10^{25}$ g, or half
the total grain mass in their model fit to the Main Belt.

{\bf (9) The interbelt 141$\degr$ arc}:  The interbelt dust disk
also contains an arc of 450 $\mu$m emission
consistent with 0.075 lunar mass of dust located at $\sim$100 AU radius from the star
\citep{holland03a}.  In the sky plane, the peak of arc emission
is $\sim4\arcsec$ East and  $\sim5\arcsec$ South of Fomalhaut.
A background galaxy identified in \citet{kalas05a} is $8.3\arcsec$ East 
and  $8.0\arcsec$ South of Fomalhaut in HST observations made in 2004,
approximately 3 years after the SCUBA data were obtained.  The proper
motion of Fomalhaut would place the galaxy even farther away 
from the star in 2001 and therefore the optically detected galaxy is
an unlikely explanation for the 450 $\mu$m arc.
The position angle of peak emission in the arc is $\approx$141$\degr$, close to our
estimate of Fomalhaut b's ascending node (152$\degr \pm 13\degr$), 
$\sim$40$\degr$ smaller than the longitude of periapse ($178\degr \pm 18\degr$),
and $190\degr$ away from the Main Belt 331$\degr$ Gap.  The relative
geometry of these features may help in revealing the active dynamical
mechanisms governing the Fomalhaut system.

{\bf (10)  The 25 AU radial gap} refers to the
radial location where the interbelt dust disk has a minimum
mass, which lies just outside the boundary of the 10 AU belt.  Periastron for Fomalhaut b
it potentially located in this gap region.  It is also notable that
the 15 AU stellocentric offset of the main belt is located near
the inner edge of the gap.

{\bf (11)  The hot disk} is the region within a few AU radius from the
star responsible for excees near-infrared emission \citep{absil09a}.
The 1000-2000 K grain temperature makes it a distinct component
of dust from the 10 AU belt.  Recently reported observations using
the Keck Interferometer Nuller suggest that the hot disk could
be further subdivided into $<0.3~\mu$m sized carbon-rich grains at $\sim$0.1 AU,
and micron sized grains near $\sim$1 AU \citep{mennesson13a}.  

{\bf (12)  The 5 AU radial gap} is the region between the hot disk
and the 10 AU belt.  The habitable zone for Fomalhaut A  lies in the 2$-$5 AU region  \citep{kasting93a}.

Given this inventory information, we evaluate several paradigms based on the
possible orbits of Fomalhaut b.

\subsection{Implications of Fomalhaut b's high $e$, large $a$ orbit}
\label{sec:implications}

The revised, larger values for $e$ and $a$ suggest that Fomalhaut b is not a planet that is solely responsible
for the main belt stellocentric offset and sharp inner edge.
Fomalhaut b's present dynamical state could be a consequence of an interaction
with at least one other massive object that formed in the system.  
However, before exploring the scenarios that predict the existence of other Fomalhaut planets, is there a 
paradigm where Fomalhaut b achieves its dynamically hot state using $only$ the inventory
of observationally confirmed objects and structures presented above?  
\subsubsection{No other undetected massive bodies?}

Fomalhaut B could disturb the Fomalhaut A system either by a close flyby
interaction \citep{larwood01a, kenyon02a, ardila05a, reche09a, malmberg11a} or a secular perturbation \citep{augereau04a, wyatt05b}.  
For example, a flyby interaction studied numerically for the $\beta$ Pic debris disk gives a geometry that 
qualitatively resembles that of Fomalhaut A, Fomalhaut B and the main belt   \citep{larwood01a}.  
An initially symmetric circumprimary disk of material perturbed by a close stellar encounter results in eccentric
belts of material (technically, tightly wound spiral arms) where the apastra of the belts 
point toward the direction of the perturber's periastron.   
This means that the apastron of the perturber (or its post-flyby trajectory in a hyperbolic orbit) and
the apastra of the eccentric belts are pointed in opposite directions.  This simple
geometrical picture is consistent with the present epoch location of Fomalhaut B
south of Fomalhaut A, and the main belt and Fomalhaut b apastra to the north of
Fomalhaut A (Fig.~\ref{3degree}).
However, the critical problem is that if Fomalhaut B is bound to Fomalhaut A,
and given a system age $\approx400$ Myr, there are repeated periastron passages
that would wipe out the belt structure created by the first periastron passage.

Instead of a flyby, Fomalhaut B  may influence the Fomalhaut A system via a secular perturbation.
The Kozai resonance has been invoked as one mechanism to explain highly eccentric exoplanets \citep{wu03a, takeda05a}.  For Fomalhaut B to be responsible for a  Kozai resonance, it must have a mutual inclination of $>39.2\degr$ relative to the orbital planes of either Fomalhaut b or the main belt.  If Fomalhaut b and the main belt are not coplanar, then it is possible that the Kozai mechanism operates only on one component of the Fomalhaut system.  
The approximate period for a Kozai oscillation \citep{ford00a} is:

\begin{equation}
\label{Kozai period}
P_K \simeq P_{b}  ~\frac{m_A + m_b}{{m_B}}~ \biggl(\frac{a_B}{{a_b}}\biggr)^3 ~ (1 - e_B^2)^{3/2}
\end{equation} 

Here the subscripts A, b and B refer to the respective components of the Fomalhaut system.  Using the approximate values of $P_b = 10^3$ yr, $m_A = 2 ~M_\odot, m_B=1  ~M_\odot, m_b = 0, a_B = 6\times10^4$ AU, $a_b=10^2$ AU and e=0.5, the Kozai period is of order $10^{11}$ yr.
The Kozai resonance is therefore relatively ineffective for separations as large as observed between Fomalhaut A and B.  

Instead of Fomalhaut B, the main belt mass may be responsible for a secular perturbation on Fomalhaut b.  
This scenario has been studied by \citet{terquem10a}, but the initial conditions presume that a
planet begins with a mutual inclination $\gtrsim30\degr$ relative to the main belt.  To reach this starting point, the
scenario needs to invoke an additional dynamical interaction with some other body, e.g., a Fomalhaut c, and
therefore the planet-belt Kozai effect does not give a dynamical history consistent with no other massive bodies.

In addition to the problems of explaining Fomalhaut b's dynamically hot orbit, the properties of the
main belt are left without adequate explanation.  The eccentric
orbit of Fomalhaut b tends to exclude the possibility that it dynamically sculpts the inner edge of
the belt since only a small fraction of the belt could be disturbed during each orbital period of Fomalhaut b.
Moreover, the secular perturbation theory invoked to explain the main belt stellocentric
offset is second order with respect to eccentricity, and breaks down at high eccentricity.  
If secular theory is applicable, then Fomalhaut b's high eccentricity would predict that the main belt's eccentricity should be larger
than observed.   On the other hand, apsidal alignment between Fomalhaut b and the main belt
continues to be indicated by the new orbit determination (Table 5).  Future work needs
to determine if the orbital parameter space presented here could be consistent
with secular theory and the observed stellocentric offset.  
In any case, the main belt's sharp inner edge and the azimuthal gap are consistent with the existence of another planet orbiting near the main belt.

To summarize, the observed Fomalhaut inventory does not appear to be sufficient for explaining Fomalhaut b's high eccentricity
and the main belt morphology.
Other perturbing objects must be present in the dynamical history of the Fomalhaut system.  

\subsubsection{Additional Fomalhaut perturbers}
\label{sec:additional}

Permitting the existence of additional perturbers in the past and/or present epochs
allows a variety of plausible dynamical histories that are consistent with the current observables. 
Three classes of dynamical paradigms could focus on endogenic perturbations, exogenic perturbations, or a blend of both.
For example, the dynamical paradigm of Oort cloud comets is a blend that involves
the increasing of minor body semi-major axes and aphelia by close-encounters with gas giant planets, followed by
the raising of perihelia by passing stars and molecular clouds \citep{oort50a, duncan87a}.

In the endogenic class of paradigms, Fomalhaut b's eccentric orbit was produced by an interaction with at least one other planet in the system.
The general idea for planet-planet dynamical interactions is that two or more planets initially form in relative isolation from each other, but subsequent migration mechanisms lead to unstable orbital configurations.  
Two planets may enter within a few times their mutual Hill's sphere \citep{gladman93a, chambers96a, rasio96a, levison98a, marzari02a, adams03a, veras04a, chatterjee08a, juric08a, veras09a}, or their orbits may cross into an unstable resonance due to planetesimal-driven migration \citep{tsiganis05a, thommes08a}.  An instability that modifies the orbital elements of two planets in the system may then lead to unstable orbits for other planets in the system, producing a ``global'' instability.

Because planet-planet scattering evolution is chaotic, the initially closest planet may end up the farthest, and vice versa.  
Overall, the surviving (i.e., not ejected) planet that ends up with the largest semi-major axis will also have higher eccentricity if its mass is less than or equal to the planet it interacted with.  
Unfortunately the $upper$ mass limit of Fomalhaut b ($\leq1 M_J$) is not particularly helpful for constraining the expected mass of another surviving planet.
Moreover, it is also possible that a hypothetical Fomalhaut c was ejected, leaving behind Fomalhaut b as the interior planet.  
For example, numerical tests by \citet{ford08a} involving two planets indicate that the largest eccentricities are obtained for near equal mass planets, where the surviving (bound) planet has $e = 0.624 \pm 0.135$,
but the second planet is lost.  
\citet{juric08a} find that 20\% of simulations that begin with multiple planets, end with only one bound planet.  However, the
majority of systems have at least two surviving planets after chaotic evolution, agreeing with simulations of three-planet
systems conducted by \citet{chatterjee08a}.  Therefore, the detection of Fomalhaut $b$ as a large-$a$, high-$e$ exoplanet makes the existence
of another comparably massive exoplanet in the system more likely than not.
This would also mean that the orbit of Fomalhaut b may undergo further dynamical interactions that will evolve the orbits of both Fomalhaut b and Fomalhaut c.  

The observational avenue for constraining the the problem clearly rests on detecting Fomalhaut c and determining its orbital properties.   
Direct imaging surveys to date have not detected a second companion at infrared wavelengths \citep{kalas08a, marengo09a, kenworthy09a, janson12a, kenworthy13a}.  
Since planet-planet scattering or other instabilities may involve a Fomalhaut c with a Jupiter mass or below, the mass limits explored by these surveys, $>1$ M$_J$, are not adequate to rule out the existence of a Fomalhaut c.

In the exogenic class of perturbations, a third star could have been responsible for pertubing the Fomalhaut system.   \citet{deltorn01a} searched for Fomalhaut ``nemesis'' encounters among 21,497 stars where space motions could be derived from radial velocity and Hipparcos information.  The strongest perturbation was from HD 16895 (SpT=F7V) $474^{+20}_{-19}$ Myr ago.  The age of HD 16895 is $\sim$9 Gyr \citep{ng98a}, and therefore it did not form as part of the Fomalhaut system. The closest approach distance was $1.15^{+0.41}_{-0.34}$ pc, at which time the barycenter was $105^{+21}_{19}$ kAU from Fomalhaut A (approximately twice the {\it current} projected separation between Fomalhaut A and B).  Therefore there is some empirical evidence for a possible exogenic disturbance to the system that could propagate inward, resulting in a global dynamical instability on a secular timescale \citep[e.g.,][]{zakamska04a}.
The availability of expanded position, proper motion and radial velocity catalogs may be used to identify other potential perturbers in future work, and the effect of the galactic tides should also be incorporated in new calculations \citep{kaib13a, veras13a}.

A blend of endogenic and exogenic perturbations requires a comprehensive analytical and numerical analysis.  
To gain a rough picture concerning the dynamical lifetime and outcomes of the current orbital configuration, we used the numerical simulator AstroGrav to evolve the orbits of several test cases for 440 Myr.  The simulations include Fomalhaut A and B, two planets orbiting Fomalhaut A, but have no test particles representing the belt to minimize the simulation times.  One of the test planets represents Fomalhaut b with $a=177$ AU and $e=0.8$, and the mutual inclination with the second planet is either $i = 0\degr$ or $i=20\degr$.  The second planet is either at $a=30$ AU or $a=120$ AU.  We tested a combination of various masses for the two planets representing Jupiter, Saturn and Neptune.  Fomalhaut B has mass $1.45\times10^{30}$ g, $a=57400$ AU, $e=0.0$ and $i=0\degr$. 

The general outcome is that if Fomalhaut b is coplanar with Fomalhaut c, it is ejected from the system on $<10^7$ yr timescales, though there are exceptions where Fomalhaut b survives for the age of the system.  In a small fraction of cases, Fomalhaut b approaches Fomalhaut B as its semi-major axis evolves to large values, but capture is unlikely.  The exogenic influence of Fomalhaut B appears minor given the fixed assumption of $a=57400$ AU, $e=0.0$.  Fomalhaut b remains bound to Fomalhaut A for $>10^7$ yr timescales in the test cases where the mutual inclination is 20$\degr$.  We also found cases where  Fomalhaut c has high eccentricity ($e\sim0.8$) $after$ Fomalhaut b is ejected.  This confirms the previously stated notion that the observed Fomalhaut b (in the simulation it is Fomalhaut c) could have been a planet on a low eccentricity orbit, as originally envisioned to account for the main belt properties, but recently acquired high eccentricity via a planet-planet scattering event.

The overall picture is that given the uncertainties concerning the orbital parameters of Fomalhaut b, Fomalhaut B and the existence of other Fomalhaut A planets, there are configurations where Fomalhaut b obtained its high eccentricity $>10^7$ years ago at early epochs, particularly in the non-coplanar cases.  However, there are circumstances where the configuration is younger than $10^7$ years.  A test of which scenario should be favored could look into how likely the belt is to survive in either case, though in such a scenario the assumed mass of Fomalhaut b is increasingly relevant.  In the next section we consider the belt survival timescales and other physics that would be implied by a co-planar case, given a variety of masses for Fomalhaut b up to one Jupiter mass.

\subsection{Belt Collision Scenario}  

In the coplanar scenario, Fomalhaut b is on a collision course with the main belt.
Fomalhaut b will begin entering the inner edge of the dust belt around 2032 C.E., 
at which point the emergent phenomena would
elucidate the physical nature of Fomalhaut b.  
For example, if Fomalhaut b's optical light is due to a dust cloud, it may
appear to episodically brighten and change color in scattered light as fresh dust rich in smaller grains is produced by collisions with main belt material.  
The direction of the main belt orbital motion may also be ascertained depending on
which direction new features propagate within the belt.
We note that even though the probability of a belt crossing
orbit is of order 10\%, the nodes may precess
or librate, producing intervals where belt crossing occurs.  
Therefore the belt collision scenario is worth studying even if at the
present epoch Fomalhaut b is in a configuration
that does not intersect the belt.

\subsubsection{Planetesimal Dust Cloud Scenario}
\label{sec:planetesimal}

Here we assume that Fomalhaut b is a low-mass planetesimal that
is optically bright because of reflected light from a fresh dust cloud surrounding it.
For example, it could be a planetesimal that was recently disrupted
by forces associated with its recent periaston passage.  Fomalhaut b is unlikely to be
$only$ a dust cloud (i.e. only $\leq$1 mm sized grains) 
because the size of object required to account for the grain scattering
surface area is at least 10 km in size \citep{kalas08a}.  Therefore 
it resides within the gravity regime of planetesimal collision physics,
where ``catastrophic'' collisions are defined as retaining
50\% of the precursor mass in a largest remnant.

An alternative to a collision is tidal, thermal and/or spin breakup of a
weak planetesimal \citep[e.g.,][]{jewitt12a}.  Fomalhaut b's  precursor could be an analog to Shoemaker-Levy 9,
tidally disrupted by passing within the Roche radius of the hypothetical
Fomalhaut c.  Alternately, the analogy may be to a Sun-grazing
comet that breaks up near periastron due to thermal and tidal stresses, or
elsewhere due to spin \citep{marsden05a}.  One empirical test of this idea is to search 
for debris along Fomalhaut b's orbital path (Fig.~\ref{2d_orbit}).  
Unfortunately, the current data are dominated
by speckle noise in most of the region closer to the star than Fomalhaut b's current location.   
One might classify the Fomalhaut b phenomenon as cometary, but the inferred dust mass and 
stellocentric distance places it in the ``giant'' comet category with activity
involving supervolatiles, as inferred
for the activity of comet Halley and other icy objects
at large heliocentric distances \citep{sekanina92a, jewitt09a}.

One major question is whether or not the Fomalhaut b cloud would survive the belt crossing
as it collides with main belt material.  
A key consequence of Fomalhaut b's $e\sim0.8$ orbit is that the relative
velocity of Fomalhaut b with respect to the belt is greater than
previously assumed.
The relative velocity of particles orbiting within the belt is \citep{wyatt02a}:

\begin{displaymath}
v_{rel} = f(e,I) ~v_k = (1.25~e^2 + I^2) ^{1/2}~v_k 
\end{displaymath}
where $v_k$ is the Keplerian orbital velocity.  
The collisional belt model developed analytically by \citet{wyatt02a} assumes that the belt lies between 125 and 175 AU radius, and
the average inclinations and eccentricities of belt particles are $I$ = 5$\degr$ and $e$ = 0.065.
At 150 AU, $v_k=3.4$ km/s, yielding $v_{rel}$ = 0.4 km/s.  

To calculate $v_{rel}$ between Fomalhaut b and the belt as it enters the belt two decades from now, we assume that the incidence
angle is 45$\degr$ in the prograde sense.  The model belt is 50 AU wide and the 
oblique path through the belt has length 71 AU.  The entry point is
150 AU from Fomalhaut A (recall the stellocentric offset),
where the orbital velocity of Fomalhaut b is $\sim$3.7 km/s, and hence
the belt crossing requires $\sim$100 yrs.  
The velocity components of Fomalhaut b
are 2.6 km/s both parallel and orthogonal to the disk velocity 
vector.  Thus, while cutting diagonally through the main belt, Fomalhaut b is rear-ended
by belt material moving faster in the parallel direction at $v_k$ = 3.4 km/s, or
$v_{rel}$=0.8 km/s.  Fomalhaut b will also undergo head-on collisions with an
orthogonal component $v_{rel}$=2.6 km/s.  In the reference frame of
Fomalhaut b, $v_{rel} = \sqrt(0.8^2 + 2.6^2)$ = 2.7 km/s from the lower right ($cf.$ inset of Fig.~\ref{2d_orbit}).

Since the collisional lifetime of particles in the belt is 
$t_{cc} \propto v_{rel}^{-1}$, we can estimate $t_{cc}$ for the Fomalhaut b dust cloud
as it passes through the belt by calculating
$v_{rel}$ between Fomalhaut b and the main belt, and comparing it
to $v_{rel}$ of main belt particles in the \citet{wyatt02a} model.
Given that $v_{rel}$ for Fomalhaut b is six to seven times
greater than  for belt particles, the catastrophic
collision timescale could then be taken as proportionally
shorter for dust grains surrounding the planetesimal 
compared to dust grains colliding with each other in the belt.
The dependence is in fact stronger because smaller and smaller
particles become catastrophic impactors as the relative
velocity increases.  Under the assumption that 
smaller impactors are present, the collision timescale ($s$):

\begin{displaymath}
t_{cc} \propto v_{rel}^{-(1+2\alpha) / 3} \propto v_{rel}^{-8/3}
\end{displaymath}

where $\alpha$ is the exponent for the particle size distribution dependence,
taken here as  $\alpha=7/2$ \citep{wyatt07a, wyatt10a, wyatt11a}.  Therefore the collision timescale for Fomalhaut b
dust cloud particles passing through the belt is (2.7/0.4)$^{8/3} = 163$ times
shorter than the collision timescale of particles within the belt.  This  value has
significant uncertainties that depend on how the grain size distribution
of the planetesimal cloud differs from that of the belt.

Counterbalancing this is the fact that Fomalhaut b spends only 200 years in the belt
per orbital period, which is $\sim$15\% of the orbital period of a belt particle at 150 AU.
Therefore the collision timescale of Fomalhaut b is  one order of magnitude
shorter than a belt particles instead of two orders of magnitude.

\citet{wyatt02a} estimate that the catastrophic collision timescale for 10 $\mu$m
grains within Fomalhaut's belt is  $10^5 \leq t_{cc} \leq 10^6$ yr.
Using the scaling estimated above, the catastrophic collision timescale
for a 10 $\mu$m grain bound to Fomalhaut b is $10^4 \leq t_{cc} \leq 10^5$ yr.
This means that the Fomalhaut b dust cloud would survive for 10 - 100
orbital periods.

The survival of a Fomalhaut b dust cloud after many crossings through the main belt appears counter-intuitive.  
We therefore conduct an order of magnitude check based
on the observables in the optical data, rather than the above extrapolation from the
analytical analysis given by \citet{wyatt02a}.  
We begin by assuming that the lifetime
of Fomalhaut b as a dust cloud is roughly equal to the timescale for intercepting
its own mass in main belt dust grains.  
For this simple scenario we assume that in fact all of the dust cloud interacts
with main belt material, and that both cloud and belt have a uniform
number density of objects for any given grain size.  

Regardless of whether or not
the scattering grains are in a small cloud, large cloud, ring, or any other geometry,
the optical photometry constrains the geometric scattering cross sections
of grains that comprise the structure.
We use the relationship derived
by \citet{kalas08a},

\begin{displaymath}
m_p = -2.5~ \mbox{log}(\sigma_pQ_s) + 70.2 ~\mbox{mag}
\end{displaymath}
where $\sigma_p$ is the projected geometric surface area of scattering grains in m$^2$,
and $Q_s$ is a scattering efficiency factor, such as the geometric albedo.
Observations give $m_p\sim25$ mag in the optical.  
Therefore if the cloud material has a relatively low albedo such that $Q_s$ = 0.1, then the projected geometric
surface area of grains within the Fomalhaut b cloud is  $\sigma_p =1.2\times10^{19}~\mbox{m}^2$. Turning now to the main belt,
\citet{kalas05a} give the model dependent total grain scattering
cross section $\sigma_{mb}$ = 5.5$\times10^{21}~\mbox{m}^2$, which also assumes albedo=0.1.
The planetesimal cloud therefore has a total grain cross section that is $2.2\times10^{-3}$ of the entire optically detected
main belt.

As an aside, these results are testable in the sub-mm, where the flux of the entire main belt is 81 mJy 
\citep{holland98a}, yielding a predicted flux from Fomalhaut b of $\sim$178 $\mu$Jy (if the circumplanetary
grains have similar properties as grains in the main belt).  The current ALMA data have an rms noise
of $\sim$60 $\mu$Jy per beam \citep{boley12a}, which to first order excludes
a circumplanetary dust cloud significantly more massive than in 
our prediction above.
Vetting various dust cloud models will require additional future
observations with ALMA.  

For a single belt crossing, the cloud will encounter only a fraction
of the main belt volume.  
If Fomalhaut b requires 100 years to cross the belt, then the volume of the main belt
that is encountered by the grains within the cloud  is 1.1$\times$10$^{32}$ m$^3$. 
Adopting the assumptions from \citet{wyatt02a} that the belt has an inner and
outer radius of 125 and 175 AU, respectively, and adopting a fixed vertical width of $\sim$10 AU,
the volume of the model belt is 1.6$\times10^{39}$ m$^3$.  
Thus one encounter volume is 6.9$\times10^{-8}$ 
of the total main belt volume.
Multiplying  $\sigma_{mb}$ by this factor, the geometric surface area of main belt grains encountered
by cloud grains in a single belt crossing is $3.8\times10^{14}~\mbox{m}^2$.  For
the dust cloud grains to intercept an equal surface area of belt grains
requires $3\times10^4$ belt crossings. 
Given two belt crossings per 1700 yr orbital period,
the lifetime of the cloud is $\sim10^7$ yr.

To summarise, both the empirically based, order-of-magnitude approach and the analytic approach
of \citet{wyatt02a} confirm that a dust cloud is not destroyed by a single belt crossing,
and in fact survives for a minimum of 10 orbital periods.

Is there a shorter timescale by which the cohesiveness of the dust cloud could be lost?   
If the cloud is not gravitationally bound to itself, then two possibilities are orbital shearing
between the portions of the cloud closest and farthest from the star,
and the velocity dispersion of grains.  For shearing, if the semi-major
axis of one side of the cloud differs from the opposite side by 2 AU (260 mas; 5 STIS pixels)
then the orbital velocities differ by $10^{-2}$ km/s.  In 10 years the separation along
the direction of orbital motion increases by $\sim$0.02 AU (3 mas), which is
not detectable.  However, in one orbital period ($\sim$2000 yrs) the cloud shears by 4.6 AU (600 mas).
A spherically symmetric cloud of grains with no self-gravity therefore spreads into a triaxial
spheroid and eventually into a trail within a few orbital periods.  

\begin{figure}[!ht]
\epsscale{1.0}
\plotone{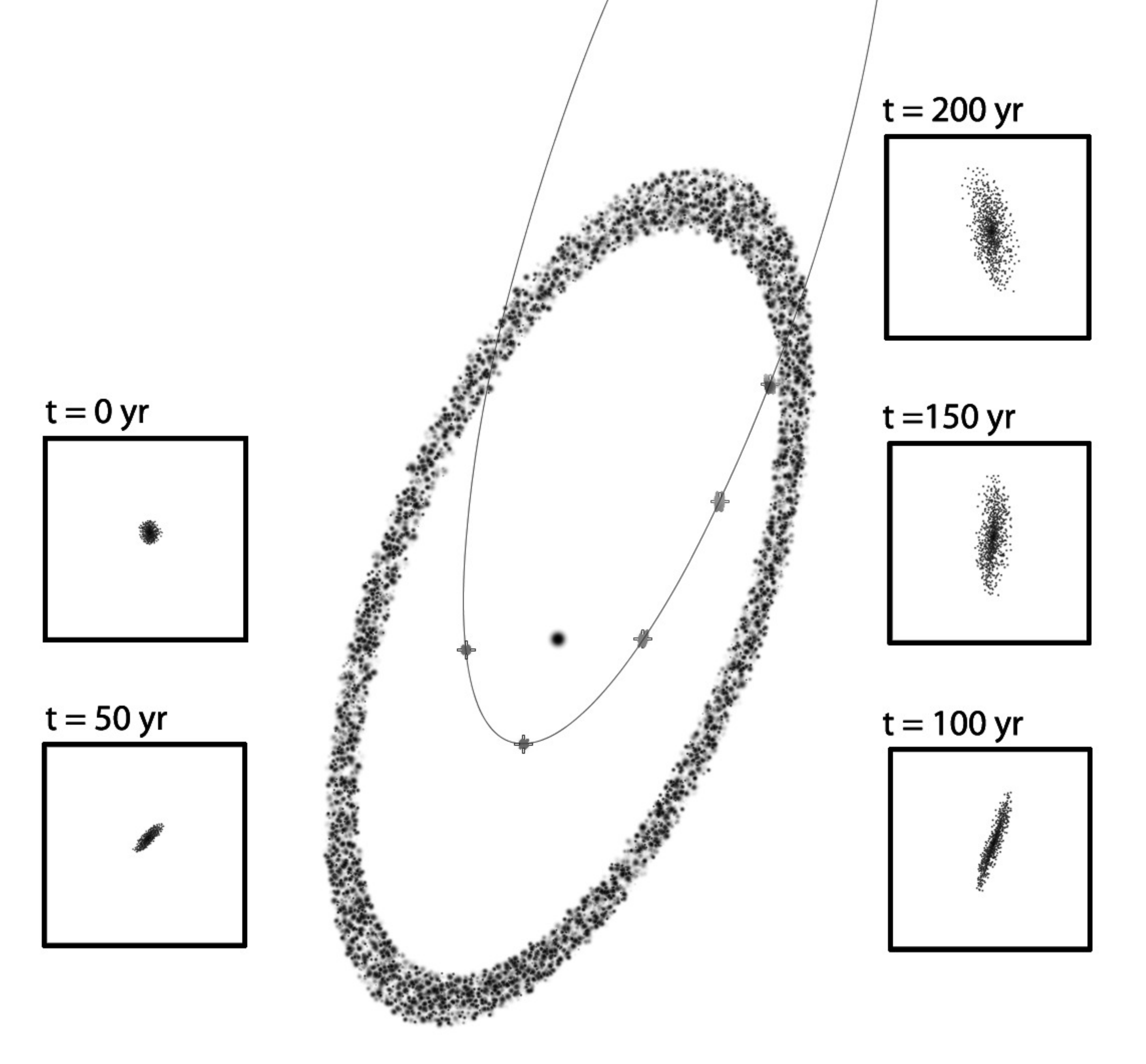}
\caption{\footnotesize
Numerical model of a massless dust cloud which at t=0 has radius 0.05 AU and is located
east of Fomalhaut A.  At t=50 yr it is near periapse and after periapse the structure resembles a triaxial spheroid oriented roughly north-south
in our line of sight.  By t=200 year, the structure is 0.5 AU in diameter (each inset box is 0.82 AU on a side).  
\label{cloud}}
\end{figure}

Figure~\ref{cloud} demonstrates the shearing of a spherical cloud that begins with radius 0.05 AU at a position $-90\degr$ from
periapse.  We use the AstroGrav numerical model to study the evolution of a dust cloud composed of 1000 massless particles, 
with a stellocentric motion that
follows the nominal Keplerian orbit of Fomalhaut b ($a=177$ AU, $e=0.8, i=0\degr$).   Shearing at periapse extends
the structure such that by 200 years when it is in Fomalhaut b's position near belt-crossing, the diameter is 0.5 AU.
This corresponds to 65 mas and would be unresolved by the observations.  However, after the second periapse 
passage the structure is 9 AU in length, resembling a trail of material along the path of the orbit.
Thus, even though Fomalhaut b as a dust cloud could survive many belt-crossings, it shears into a trail of particles
in one orbit if it is not gravitationally bound to a more massive object.  
We therefore consider it unlikely that Fomalhaut b is $only$ a dust cloud, because  it requires a
fortuitous timing in discovering it a few centuries after it was created.

How massive does a central object have to be so that an 0.05 AU radius cloud is not disrupted
by shearing at periapse?  Using the numerical simulation, we found a mass of $1.0\times10^{23}$ kg
is sufficient (1.4 lunar mass).
The cloud can be smaller, since \citet{kalas08a} suggested that a circumplanetary disk with $\sim30~R_J ~(2.1\times10^9$ m or 0.014 AU) is consistent
with the detected optical flux.  For this smaller radius, the cloud is stable against shearing if the central object has mass $5\times10^{21}$ kg (5$\times$ Ceres mass).
Therefore a dwarf planet between Ceres and Pluto in mass may retain a system of satellite dust and moons that is stable against shearing.

The dust cloud model in \citet{kalas08a} specifies that the dust cloud's scattering surface
area corresponds to the disruption of a minimum 10 km radius object.   Since dwarf planets are much larger, 
$\sim$500 km radius objects, the mass in the dust cloud could have been launched
from a single cratering impact. 

Finally, given a scenario that Fomalhaut b is a dwarf planet, is it possible that its mass is still in
the process of increasing significantly due to the accretion of belt material during belt passages?
The mass accretion rate (kg/s) enhanced by gravitational focusing is \citep{kennedy11a}:

\begin{displaymath}
dM/dt = (M_{mb} / V_{mb})~ \pi R_b^2 ~(1+v_{esc}^2 / v_{rel}^2)~ v_{rel}
\end{displaymath}

where the main belt total mass and volume are estimated as $M_{mb}=75 ~M_\oplus$ and $V_{mb} =10^{39}~m^3$, and the dwarf planet
radius and escape velocity are $R_b=10^6$ m and $v_{esc}=1000$ m/s, respectively.  Therefore for a
relative velocity $v_{rel}=2700$ m/s, we find dM/dt = 1.4$\times10^{11}$ kg/yr.  Since each
orbital period consists of only 200 years spent within the belt, the mass accretion rate
is equivalently expressed as 2.8$\times10^{13}$ kg/orbit.  If we assume that 90\% of this
accreted material adds to the mass of the central object and 10\% adds to the
mass of the planetesimal cloud, then we have the following results:  In $\sim$3000 orbits ($5\times10^6$ yr)
Fomalhaut b has accreted $\sim10^{16}$ kg of additional mass into the planetesimal cloud, which is equivalent
to the 10 km sized that was object originally envisioned to explain the grain scattering 
cross section.  However, the central mass has increased by only a factor of $10^{-5}$.  The $5\times10^6$ yr timescale
is of order the lifetime we might expect for the coplanar case where Fomalhaut b crosses
through the planetary region, leading to an eventual strong scattering event with another planet.  
We therefore conclude that even though the coplanar, belt-crossing orbit is most likely
short lived, it is long enough for a dwarf planet to capture a surrounding cloud, but
not long enough for the dwarf planet to increase its mass significantly.

\subsubsection{Planet with satellite system}
\label{sec:satellites}

The circumplanetary dust disk hypothesis presented by  \citet{kalas08a} received a measure of plausibility
with the discovery of Saturn's Phoebe ring at $>$200 R$_{p}$ \citep{verbiscer09a}.
The basic physical mechanism is that the surface of a small (radius$\sim$100 km), distant (a=215 R$_p$)
planetary moon is bombarded by interplanetary meteoroids, launching ejecta that
spirals toward the planet due to Poynting-Robertson drag.  \citet{verbiscer09a}
estimate that the normal optical depth of the Phoebe ring is $\sim2\times10^{-8}$,
which could be attributed to material ejected from a single, 1-km diameter crater.

One consequence of the highly eccentric orbit for Fomalhaut b is that
it is less likely to capture additional outer satellites compared to nested planets such as Saturn because its
velocity relative to nested orbiting objects is high, and stellar tidal forces at periastron are significant.
However, the Phoebe ring scenario only requires the existence of a single distant satellite, and certainly
the prospect that Fomalhaut b previously had a lower eccentricity
orbit is not ruled out.  Moreover, a planet-planet scattering event that could account for
Fomalhaut b's high eccentricity does not necessarily
lead to the loss of the moon orbiting the scattered planet \citep{debes07a,nesvorny07a}.

Instead of a single Phoebe-like satellite, \citet{kennedy11a}
study the capture of many irregular satellites
around Fomalhaut b that collisionally produce a circumplanetary
dust cloud.  This collision concept is different because the Phoebe
ring is produced when a satellite is ``stranded'' far from the next
innermost satellite, and a large fraction of impactors that strike the moon originate from
outside the system.  
If there are numerous irregular satellites like Phoebe, then this ``swarm''
is self-eroding via many mutual collisions. %
The morphology of the dust swarm resembles
an hourglass instead of a ring or torus due to the instability of high inclination moons \citep{hamilton91a,nesvorny03a}.  
\citet{krivov02a} present evidence that Jupiter is surrounded by a cloud
of dust particles between 50 and 300 R$_J$, except that given the
relatively few irregular satellites at this late epoch, the erosion mechanism
is mainly the external meteoroid flux, as with Phoebe, instead of self-erosion.

The $e\sim0.8$,  $a=177$ AU orbit
for Fomalhaut b changes the \citet{kennedy11a} scenario in that the Hill radius
is effectively three times smaller at periastron, and so too is the region of satellite stability.  
The Hill radius depends on the planet and stellar masses, 
the planet semi-major axis, $a_{pl}$, and the planet's orbital eccentricity:

\begin{displaymath}
R_H = a_{pl} (m_{pl} / 3M_\star)^{1/3}
\end{displaymath}

However, in the case of an eccentric orbit, the instantaneous star-planet separation, $\rho$,
should be adopted instead of $a_{pl}$.
At an assumed periastron of $q = \rho=32$ AU, and with $m_{pl}$ = 1 M$_J$ and M$_\star = 2\times10^3$ M$_J$ , the Hill radius 
is 1.76 AU.  For comparison, the Hill radius for our Jupiter and Neptune are 0.33 AU and 0.77 AU, respectively, and a periastron scaling
reduces these values by only a few per cent.  As Fomalhaut b intersects
the belt at $\sim$150 AU in the coplanar case, the instantaneous Hill radius is 8.24 AU. 

Fomalhaut b therefore represents an interesting case study for the
dynamical evolution of moons and the observational consequences
when the host planet has a highly eccentric orbit.
To gain some rough insights, we used the N-body package 
AstroGrav to study the evolution of 500 moons, randomly assigned $0.01 \leq a_{moon} \leq 10$ AU,
$0.0 \leq e_{moon} \leq 0.1$, with a spherical distribution of
orbits around a Jupiter mass planet that has Fomalhaut b's orbital properties.  After $2\times10^5$ years,
approximately 50 moons remain bound to the planet with 
$0.02 \leq a_{moon} \leq 0.91$ AU, with median values $a=0.37$ AU.
The maximum value of 0.91 AU is 52\% of the Hill radius calculated
at periastron.  This result is consistent with previous observational
and theoretical studies concerning the dynamical evolution of distant
satellites orbiting asteroids and planets \citep{hamilton92a, jewitt07a, shen08a}

The majority of moons lost near periastron orbit the star at a reduced semi-major axis and
eccentricity, forming an eccentric disk interior to the main belt, and apsidally aligned with Fomalhaut b.  
It is possible that such lost moons exist if Fomalhaut b's orbit
$before$ the dynamical instability had a larger Hill sphere because the orbit
was initially farther from the star.  An instability that subsequently reduces the star-planet separation
would then result in a smaller Hill sphere and lost moons orbiting the star instead of the planet.
The implication is that the interbelt  dust disk may consist of material with 
different origins:  (a) bodies that formed there, (b) moons lost by planet instabilities,
and (c) material perturbed inward from the outer disk ($cf.$ Sec.~\ref{sec:disruption}).
This scenario also invokes the possibility  that if the perturbed planet is not coplanar with
the main belt, the eccentric disk of lost moons would also orbit and
collisionally evolve in the planet's orbital plane and not the main belt plane.  The system would therefore
appear to have two inclined debris disks.  The $\beta$ Pic system
shows evidence for a secondary, inclined disk that is significantly less
massive and prominent than the primary disk \citep{heap00a, golimowski06a}.
The concept of lost moons after a planetary dynamical
instability could serve as an alternate model to the current paradigm
that inserts an inclined planet into a pre-existing disk, creating
a vertical disk warp that propagates outward \citep{mouillet97a}.  

Since Fomalhaut b and the lost moons continue to have a similar
periapse, a bottleneck of orbits is evident near periapse. Even though
Fomalhaut b's Hill's radius is at a minimum here, the volume number density
of lost moons is greatest near periapse.  We find that moons lost at periastron are recaptured near periastron (ejection from the system or
collision with the planet are also possible).  In fact, the capture epoch begins $after$ periastron when the Hill sphere of
the planet is expanding but the bottleneck is still providing a relatively high volume number density of objects.
Recaptured moons tend to be captured into eccentric ($e\gtrsim$0.3) orbits around the planet, 
which means that they are loosely bound and lost again as the planet approaches the next periastron passage.

Even if the recapture of moons does not occur, the outer moons still bound to Fomalhaut b are dynamically 
heated by the tidal forces at periastron.  The eccentricities found in
the surviving, bound moons are $0.0 \leq e_{moon} \leq 0.9$,
with median $e_{moon} = 0.1$.  Thus there are at least three mechanisms
that could increase collisional dust production surrounding Fomalhaut b at periastron and soon after periastron:
Objects have energetic collisions with Fomalhaut b at periastron, the collisional
grinding of moons bound to Fomalhaut b is enhanced at periastron, and soon after periastron
additional moons may be recaptured on highly eccentric orbits that would collide with
the bound moon system.  Fomalhaut b is currently observed $\sim$120 years after periastron,
and the simulation supports the concept that moon capture may have recently activated the collisional
erosion of a planetary moon system.

Two more epochs of enhanced collisionaly activity may occur at the ascending and descending
nodes relative to the belt plane.  When Fomalhaut b crosses the orbital plane of the belt or the interbelt dust disk,
the external meteoroid flux is enhanced again. 

Observationally, the optical depth of the Fomalhaut b ring/cloud system will increase during enhanced erosion,
the grain size distribution will shift temporarily to small sizes as fresh
dust below the radiation pressure blowout radius is released, 
and as a result we might observe a brighter and
bluer scattered light signature from Fomalhaut b.
On the other hand, a dust cloud could also become optically thick, which would
make light scattered toward the observer sensitive to
viewing geometry.  In other words, shadowing due to too 
rapid dust production could decreases the brightness.

Though this discussion focuses on the erosion of satellites from both circumplanetary and
circumstellar impactors, generating fresh circumplanetary dust that is observable in reflected light,
other processes may be at work that have distinct observable signatures.  
When the bound moons have their eccentricities pumped by 
periastron passage, this could increase the planetary
tidal heating of the hypothetical moons \citep{peale78a, peale79a, cassen79a}.  
Tidal heating and melting has an infrared signature \citep[e.g.][and references therein]{peters12a}.
A clone of Jupiter's moon Io would also generate an optically detectable sodium cloud.  
Jupiter's sodium cloud has been detected in Sun scattered light
to at least 400 R$_J$ \citep[0.2 AU;][]{mendillo90a}.
Thus, in addition to dust scattered light and H$\alpha$ emission as 
possible explanations for Fomalhaut b's anomalously high optical flux \citep{kalas08a}, a sodium cloud
could also contribute at 0.59 $\mu$m.  This lies in the F606W bandpass of the 2004 and 2006 HST/ACS observations.

Finding definitive evidence for such a cloud would have many significant implications, such as showing that Fomalhaut b has a magnetic field similar to Jupiter's.  
Clearly a spectrum of Fomalhaut b is required, but the issue can also be examined via imaging.  For example, Jupiter's sodium cloud is variable in size and brightness.  These variations are correlated to the volcanic activity of Io \citep{mendillo04a}.  Therefore,
the characteristic timescales of variability from imaging Fomalhaut b may be more geophysical than astrophysical $-$
Fomalhaut b may episodically appear brighter and more extended on timescales measured in months.  

We have argued that even though variability and extended morphology exist in the HST optical data (Fig.~\ref{2010falsecolor},~\ref{2012falsecolor}),
they can also be attributed to instrumental noise (Section~\ref{sec:acsphotometry}; Fig.~\ref{2012implants}).  However, \citet{galicher13a} claim that the extended morphology of Fomalhaut b in the 2006 ACS/HRC data is not instrumental in the F814W image.  This is puzzling because the F606W image taken at the same epoch and with greater sensitivity (greater integration time and higher quantum efficiency in the bandpass), does $not$ appear to be extended.  This is difficult to reconcile with a model where optical light arises from grain scattering - the F606W image should also show extended morphology.  However, the more complicated model involving the evolution of atomic and molecular species from moon volcanism to cirumplanetary magnetospheres could yield a solution.  For example, Io is also the source of a circumplanetary potassium cloud that emits at 0.77 $\mu$m \citep{trafton75a}, and this lies within the F814W bandpass.  Jupiter's potassium cloud is significantly weaker and less extended than the sodium cloud, but a different geochemistry for the hypothetical Fomalhaut b moon could conceivably produce a potassium cloud that adds a halo of extended light in the F814W images.

\subsubsection{Disruption of the Main Belt by Planet Crossings}
\label{sec:disruption}

A key question is whether or not a planet mass passing through the belt disrupts the morphology of the belt.
The 25 AU radius of influence (i.e., 3R$_H$ at 150 AU, Section~\ref{sec:satellites}) for a Jupiter mass corresponds to a 
6.5$\arcsec$ diameter.  This size is 3.2 times smaller for a 10 M$_\earth$ planet, but the corresponding 2.0$\arcsec$
angular scale is still resolvable with current instrumentation. 
In principle, the local dynamical stirring should enhance dust production, shifting
the grain size distribution to favor smaller grains with larger surface area, and produce
a transient brightening of the belt in scattered light \citep[e.g.,][]{kenyon01a,dominik03a}.

To explore the cumulative effects of Fomalhaut b's dynamical perturbations on belt parent bodies over many belt crossings and
under a variety of assumptions, we use the 3D, N-body simulator AstroGrav.
Our model of the Fomalhaut system begins with a nested
Jupiter mass planet  (Fomalhaut c) with $a$=120 AU, $e=0.10$, $i=0$ and $\varpi=178\degr$.
We add two populations of main belt objects.  First, an effectively massless population
of 8000 objects are randomly assigned $140\leq a \leq160$ AU, $0.09\leq e \leq 0.11$, and $-1.5\degr \leq i \leq1.5\degr$. 
The orbits are apsidally aligned with the nested planet and randomly distributed in
orbital phase. The second population is 200 objects between Ceres and Pluto in mass
and radius (total mass = 0.057 M$_\earth$), distributed randomly throughout the belt as in the first population.  However, the masses
are not negligible and will gravitationally perturb other objects.  Collisions are treated
as mergers.  

The first part of the simulation does not contain Fomalhaut b (assumed to be on a circular orbit
well within the orbit of Fomalhaut c and dynamically negligible).  The goal is to reach a quasi-steady state with respect to the dynamical sculpting of the
belt's inner edge by Fomalhaut c.  The entire system is coplanar and integrated for $1.5\times10^5$ years
($\sim$1500 orbits).
The model qualitatively reproduces the numerical experiment of \citet{chiang09a}
where Fomalhaut c maintains a sharp inner edge.  A key difference
is that \citet{chiang09a} proceeded to model the dynamics of dust particles
in addition to the parent bodies, where radiation pressure instantaneously increases the
eccentricities of the observed dust population.  Since we
are studying only the parent bodies, the timescales given below may be considered 
upper limits to the eccentricity evolution of observable particles.   Another difference is that particles 
approaching within a few times the Hill radius are not removed from our simulation,
and therefore we observe the capture (and loss) of satellites.  

The second part of the simulation assumes that after $1.5\times10^5$ years, Fomalhaut b is
strongly perturbed from the inner part of the system (by interaction with a third planet) and appears
as a rogue planet with $a=177$ AU, $e=0.8$, $i=0\degr$, and
apsidally aligned with Fomalhaut c.  We assume three cases where the rogue planet has a 
Jupiter, Saturn and Neptune mass. We use the ``roque'' terminology to designate bound
planets with large $a$ and $e$ such that they cross the orbits of other planets
and belts in the system.

The impulse imparted on belt material by a planet crossing through the belt does not 
create a visually noticeable disruption of the overall belt morphology.
Inspection of the belt particle orbital elements shows no statistically
significant difference before and after planet crossing.  For example, after the nested Jupiter
mass Fomalhaut c has been sculpting the belt for the first $1.5\times10^5$ yrs,
the mean eccentricity distribution of belt particles has evolved to $e = 0.1077\pm0.0747$.
Fomalhaut b as a rogue Jupiter mass would have the most significant dynamical effect,
yet after a single crossing the eccentricity distribution is $e = 0.1072\pm0.0746$.
Measured another way, before the belt crossing 8.07\% of belt particles have 
$e>0.20$.  After the single crossing of a Jupiter, 8.17\% have $e>0.20$.

The cumulative effect of many belt crossings is to gradually spread the belt radially. 
Over $10^2-10^4$ belt crossings, the only belt morphology that is noticeably ``disrupted'' is the sharpness of the belt boundaries.
The belt becomes a disk.

\onecolumn
\begin{figure}[!ht]
\epsscale{0.9}
\plotone{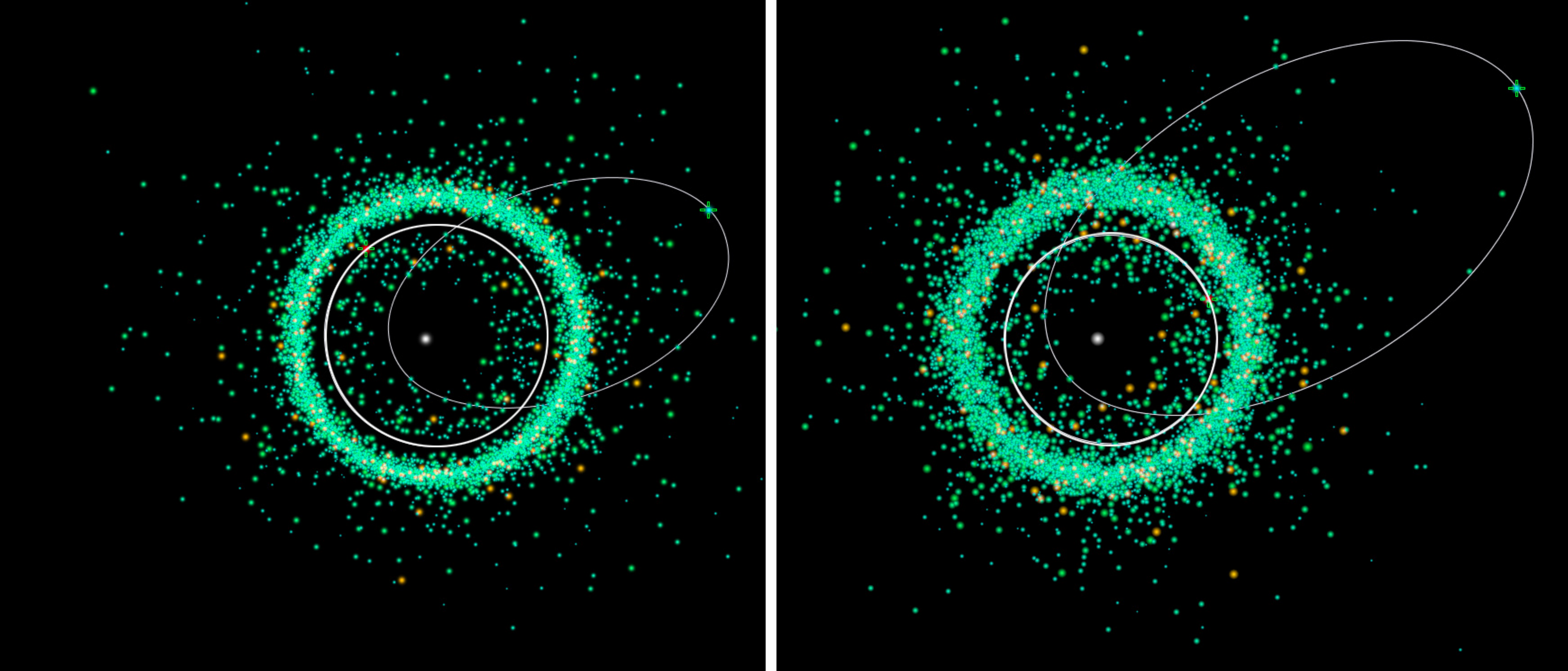}
\caption{\footnotesize
Numerical integration of a Fomalhaut system consisting of a nested Jupiter at 120 AU (white circle), a 20 AU wide main belt,
and a coplanar, belt crossing Fomalhaut b (marked with a cross and white ellipse) that has a Neptune mass (left) and a Saturn mass (right).  
The orbits of the nested and rogue planets are traced by thick and thin white lines, respectively.
The rogue Neptune planet has not eroded the inner belt edge 300 kyr after being introduced in a belt crossing orbit (450 kyr in
the simulation).  Conversely, after only  75 kyr (225 kyr in the simulation), the 
rogue, coplanar Saturn has spread the belt radially, eroding both the inner and outer
edges. The roque Saturn (right) has $a=285$ AU ($e=0.82$) due to a close
encounter with the nested Jupiter (which also has a modified orbit) just before this snapshot.  With the longer orbital period (3500 yrs),
the Saturn simulation will erode the belt on a 75\% longer timescale.  When the rogue Saturn has an initial 20$\degr$
inclination relative to the belt, the belt spreads as in the right panel after $\sim500$ kyr.
\label{coplanar_nep_sat}}
\end{figure}

\begin{figure}[!ht]
\epsscale{0.48}
\plotone{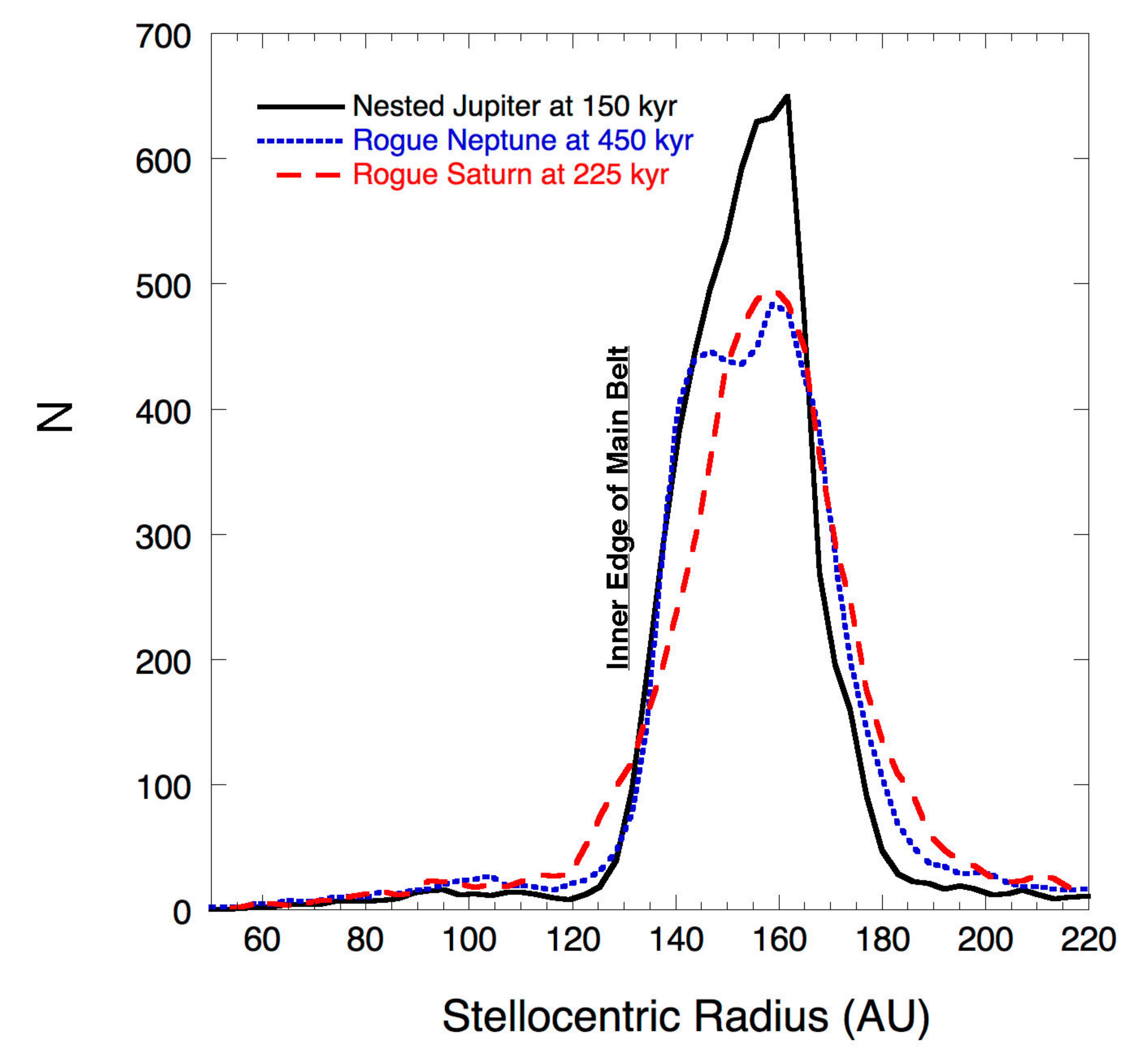}
\caption{\footnotesize
Radial profiles of the model main belt which started as 8000 particles contained in the region 140 - 160 AU.  A Jupiter is placed at $a$=120 AU and after 150 kyr the perturbations widen the belt's width  (black solid line), except that Jupiter maintains a sharp inner edge.  A fraction of particles (6\%) cross inward of Jupiter to produces a the interbelt disk (e.g. shown here between 60 and 120 AU).  We add a rogue, coplanar Neptune ($a=177$ AU, $e=0.8$ and integrate the belt for another 300 kyr years.  By 450 kyr (blue dotted line), perturbations from both Neptune and Jupiter spread the belt more, but Jupiter continues to maintain the sharpness of the inner edge.  Instead of Neptune, we add a rogue Saturn with the same orbit as the Neptune case.  By 225 kyr (75 kyr after it was added into the simulation) the sharpness of the belt inner edge has been significantly eroded.
\label{beltedges}}
\end{figure}

\twocolumn

\citet{kalas06a} noted that debris disks appear either as extended disks or narrow belts.  
The numerical models studied for the Fomalhaut system suggest that when a planet transitions from a nested to a rogue orbit, belts become disks.
The rogue can subsequently evolve away from a belt crossing orbit, and the disk will then resemble a belt again by
interaction with the nested planets, though with reduced mass due to the scattering of objects during the rogue phase.  
The Fomalhaut system could be in transition from belt to disk,
depending on the mass and orbit of Fomalhaut b.

The rogue planet in our model competes with the nested planet in shaping the inner edge (Fig.~\ref{coplanar_nep_sat}).
The qualitative result is that Fomalhaut b as a coplanar Saturn erodes the belt edge significantly after only $10^5$ years (Fig.~\ref{beltedges}).
A coplanar Neptune mass Fomalhaut b, on the other hand, does not erode the belt inner edge on timescales approaching
$10^6$ yr.  One reason we stop the simulations before $\sim10^6$ yr is that the coplanar geometry leads to a significant
planet-planet scattering that alters the orbit of Fomalhaut b and the belt edge erosion timescales have to
be reconsidered.  For example, some encounters reduce Fomalhaut b's
apoastron so that it resides in the belt.  With Fomalhaut b spending a greater
fraction of time in the belt, a Neptune mass becomes a significant
disruptor of pre-existing morphology and the belt becomes a disk in $<10^6$ years.

These timescales are likely lower limits because we have been
assuming that Fomalhaut b is coplanar with the both the main belt and with
Fomalhaut c, maximizing the probability of strong interactions. When
Fomalhaut b's orbital plane is inclined by 20$\degr$ relative
to Fomalhaut c and the main belt, the belt still
spreads radially in the Saturn case, but the
timescale is 500 kyr instead of the 75 kyr observed in the
coplanar case (Fig.~\ref{coplanar_nep_sat}).
Inclining a Neptune mass Fomalhaut b by  20$\degr$ decreases the probability
of a strong interaction with a nested Jupiter.  However,
many weak interactions over  $\sim10^6$ yr timescales
will evolve the Neptune orbit.  For example, by 6 Myr Fomalhaut b's
inclination has increased to $\sim39\degr$ and eccentricity has
decreased to $\sim$0.5.  Kozai-like oscillations
between inclination and eccentricity means that 
the belt tends to be protected from significant
interactions with Fomalhaut b:  high $I$, low $e$ excursions mean
that Fomalhaut b enters the system only near periastron in
the inner regions of the system that miss the belt, wherease low $I$, high $e$
means that Fomalhaut b passes closer to the main belt, but with a shorter
timescale due to the high $e$.

A more thorough exploration of the orbit and mass parameter space
is required and could establish a lower
limit to the age of the current orbital configuration.  Certainly any model must also
test the origin of Fomalhaut b's eccentric orbit, with the possibility that
multiple massive planets are located in the system.
Future observations can also search for evidence that the belt was disturbed by the previous belt crossing
east of the star.  

In summary, a planet mass for Fomalhaut b is not excluded by arguments concerning belt 
disruption because:  (a) Belts are not ``wiped out'' by a single belt crossing of a Jupiter-mass planet, and we 
do not know how many belt crossings have already occurred; (b)  Saturn masses and below can
cross through the belt hundreds of times before the belt edges are eroded; (c) Belt crossings erode the sharpness
of belt edges, but nested planets may compensate by maintaining the edges sharp;
(d) A mutual inclination of Fomalhaut b relative to the main belt increases belt edge erosion timescales significantly.  

\subsubsection{Impacts on the planet}
\label{sec:impacts}

One indirect way to infer the mass of Fomalhaut b is if impacts become evident during a belt crossing.
If Fomalhaut b has the mass of a gas giant, then we could expect phenomena similar to the
Shoemaker-Levy 9 (D/1993F2) impacts with Jupiter \citep[e.g.,][]{graham95a, zahnle95a, anic07a}.
The greater energies involved would manifest as significant optical and infrared variability, 
with a different characteristic spectral energy distribution and time 
dependence than impacts on a dwarf planet.  Careful analysis could
even yield information concerning its atmosphere and composition  \citep[e.g.,][]{bjoraker96a}.

For the SL9 events, \citet{carlson97a} report $\sim0.5\times10^{25}$ erg from the G-impact  fireball in a 60 second interval.
Temperatures are 8000 K at the beginning, cooling to $\sim$1000 K after 80 seconds, giving roughly $8\times10^{22}$ erg/s = $8\times10^{15}$ W.  If this event
were located at the heliocentric distance, $D$, for Fomalhaut, then 

\begin{displaymath}
f_{G} = L / (4~\pi~D^2) = 1.1\times10^{-20}~\mathrm{ W ~m}^{-2}
\end{displaymath}
Relative to the luminosity of Fomalhaut A:
\begin{eqnarray}
\Delta\mathrm{mag} = m_G - m_\star = -2.5~ \mathrm{log} (N\times f_G / f_\star) \nonumber \\
= -2.5~\mbox{log}(N) + 29.8~ \mbox{mag}.
\end{eqnarray}

Here we assume that the received flux from Fomalhaut A is $f_\star = 8.9\times10^{-9}$ W m$^{-2}$, and
$N$ is some tuning factor, such as $N$ fireballs.  For $N>$100, $\Delta$mag $<$ 24.8 mag, 
and Fomalhaut b would appear significantly brighter.  However, the peak of 
emission quickly (i.e., in seconds) shifts from an optical flash to a relatively long-lived
emission at near-infrared wavelengths.  

Are $N>100$ fireballs plausible for Fomalhaut b's encounter with
the belt?  
In Section~\ref{sec:planetesimal} we calculated the accretion rate with gravitational focusing onto a dwarf planet.  If we instead assume a Jupiter mass, with $R_b=6.99\times10^7$ m and $v_{esc}=159$ km/s, then dM/dt = $6.4\times10^{10}$ kg/s, or about one comet Halley per hour.

This rough calculation suggests that optical flashes may be observable as Fomalhaut b
crosses through the belt.  The energies involved will help constrain the mass of Fomalhaut b, but its atmosphere will
also be heated and excavated.
The infrared luminosity would therefore rise and molecular features in a spectrum may become observable
and display variability as conditions change on the planet.

\subsubsection{Recent giant impact as the origin of the main belt}
\label{sec:giantimpact}

Extending the impact theme even further, is it possible that Fomalhaut b collided with a hypothetical second planet, Fomalhaut c,
and the main belt is now the remnant debris of Fomalhaut c?
Such a scenario is attractive because it naturally explains the stellocentric offset of the belt as the elliptical
orbit of the precursor object, Fomalhaut c.  The belt is narrow because it
is recently created and has not had time to collisionally evolve and spread radially.  Fomalhaut b is belt crossing because
the belt would not exist otherwise.  While most of the mass in Fomalhaut c is dispersed
along its orbit, most of Fomalhaut b's mass is retained in a circumplanetary disk in the process of reaccreting onto the planet
or forming moons, but temporarily making it bright in reflected light.  
The fraction of Fomalhaut b's mass that has been lost comprises a more tenuous stream of 
co-orbital material that manifests as the interbelt 141$\degr$ arc of 450 $\mu$m emission. In the next
section we also study whether or not the main belt gap could be explained by this model.

The critical problem given by \citet{boley12a} is that the impact speed
for collisional erosion has to be significantly greater than
the mutual escape speed of the two bodies, but the orbital velocities are 
small at great distances (150 AU) from the star.
Erosive impacts, require $v_i / v_{esc} \geq 1.5$ \citep{asphaug09a, marcus10a, stewart12a, leinhardt12a}, where the impact velocity is $v_i = \sqrt{ v_{esc}^2+v_{rel}^2}$.

For a Moon mass, $v_{esc}$=2.4 km/s and the 45$\degr$
entry of Fomalhaut b into the belt in the prograde
sense gives $v_{rel}=2.7$ km/s.  Therefore, $v_i = 3.6$ km/s, which
is a factor of 1.3 greater than $v_{rel}$.  If the collision
is in the retrograde sense, then $v_{rel}=6.6$ km/s,
and $v_i / v_{esc}$=2.6.

Therefore the collision of two Moon mass objects would be an erosive event 
in the retrograde sense.  For lower mass objects (e.g. Pluto)
the prograde collision becomes erosive ($v_{i} / v_{esc} \approx 3$) but the
objects do not represent enough mass to account for the main belt
mass.  For higher mass objects, $v_{esc}$ becomes too large to
be consistent with erosive impacts.  Therefore the likely mass
ranges of the colliding objects are in the Moon regime, and a retrograde
collision may be necessary.

The timescale for the debris to spread into a circumstellar belt is given by \citet{wyatt02a}:  $\Delta t = 2~\pi / (2 \sqrt{3}~v_\infty ~a) $.
For $a$=150 AU and the range $10$ m/s $< v_\infty < 100$ m/s, we find $10^4 < \Delta\mbox{t} < 10^3$ yr.
Thus the collision may have occurred relatively recently.  Since the collision lifetime
for 10 $\mu$m sized grains is $10^5$ yr \citep{wyatt02a}, the giant impact
scenario allows stripped material from Fomalhaut c to evolve into a belt on shorter
timescales than the collision lifetime of grains. 

\subsubsection{Origin of the Main Belt $331\degr$ Gap}
\label{sec:mainbeltgap}

The current snapshot that Fomalhaut b is about to cross through the belt near the  $331\degr$ gap invokes the idea 
that the gap is caused by material scattered away from the belt when the planet crosses through.  In the previous
sections we argued that this is not true in the case of Fomalhaut, though other astrophysical disks that
are thin, gaseous, self gravitating and/or shadowed by optically thick material closer
to the star may display such morphology (e.g., the azimuthal gap in the circumbinary belt surrounding the  pre-main sequence system GG Tau; \citealp{roddier96a, krist05a}).

One possibility for explaining Fomalhaut's main belt gap is that a planet orbits within the belt, and the gap represents tadpole or horseshoe orbits of
the co-orbital planetesimals and dust.  The analogy is to the dynamics of
the giant planet Trojan populations observed in our solar system \citep{chiang05a, nesvorny09a, lykawka11a}.
The Earth is also known to trap in-spiralling dust grains near a 1:1 resonance, producing a $\sim$1 AU radius
dust ring orbiting the Sun, except near the planet, where grain dynamical lifetimes are short \citep{jackson92a, dermott94a}.

A second possibility is that the gap is related to the giant impact scenario 
(Section~\ref{sec:giantimpact}).
For $\sim$10$^3$ orbital periods all particles created at the collision origin point return back to the collision origin point (a.k.a. ground zero).
A snapshot of debris particle location reveals a gradual pinching of the belt toward and away from ground zero \citep{jackson12a}.
This tapered morphology resembles the main belt gap.  Fomalhaut b would return to the collision origin point if a grazing, hit-and-run collision ejects mantle material, but the planet core stays close to the pre-encounter orbit.  
Therefore, one of Fomalhaut b's two apparent belt crossing per orbital period should occur near this point, which should coincide with the radially tapered section of the belt.  
The recent, giant impact thereby ties together  the apparent proximity of Fomalhaut b to the main belt gap to the north of it.  

Contradicting this scenario is that even though the orbital paths for debris tapers toward and away from the collision origin point, the quantity of dust is not changing.  This scenario therefore does not account for diminishing the scattered light in the belt gap region, though more complex effects mentioned below may come into play.  Another significant problem is that the belt pinch requires a low velocity dispersion of debris (100-150 m/s), which counters the relatively high velocities needed to disrupt the required precursor mass (Alan Jackson, private comm).  

One general solution is to suppose that the giant impact occurred recently ($<10^4$ yr) and the debris is still spreading along the orbit of the precursor object - more time is required to create an azimuthally uniform ring.  We cannot rule this out, and certainly future observations should search for other structure in the belt consistent with a debris field that is dynamically young.  A second solution is that the gap arises from a combination of a relatively young ring ($\sim10^4$ yr) and a period commensurability between the planet and the debris \citep[see Fig. 6 in][]{jackson12a}.

\begin{figure}[!ht]
\epsscale{1.0}
\plotone{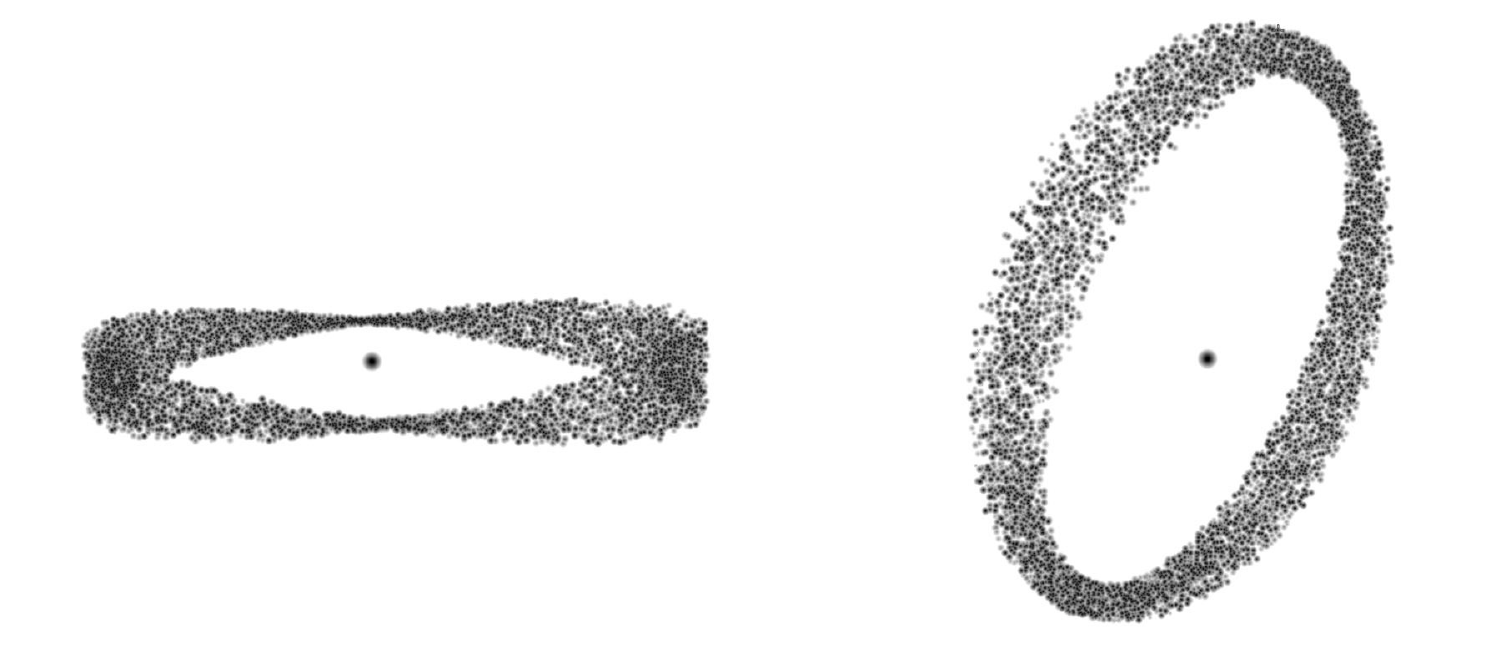}
\caption{\footnotesize
Examination of projected belt geometry when a belt has some vertical thickness and
all of the belt particles are apsidally and nodally aligned.  Here we have taken our N-body model for the main belt and
exagerrated the vertical thickness by a factor of five to emphasize the projected morphologies.  
In the near edge-on view (left panel) the apsidal+nodal alignment produces vertical pinching.  If we were to rotate our viewpoint 90$\degr$ to 
the right or left, the pinch points would not be evident because they are projected within
the ansae of the belt.  The right panel shows the same model belt rotated according to the
geometric elements that we derive for the belt.  The ascending node is to the lower left but
the descending node to the upper right has a tapered morphology as seen in projection.
\label{gapbelt}}
\end{figure}

We propose a third possibility concerning a geometrical effect that results if apsidal alignment is accompanied
by nodal alignment, presumably because of dynamical interactions with a low eccentricity planet.  
If a large ensemble of belt particles have both non-zero orbital inclinations and nodal alignment, the belt pinches vertically
toward the midplane at the ascending and descending nodes.  Figure~\ref{gapbelt} demonstrates this hypothetical configuration.
We note that the 331$\degr$ is roughly 180$\degr$ away from our estimate for the ascending node.
In other words, the belt gap is near the descending node where one of the two pinch areas occurs.
We find that due to projection effects, the pinch area is obscured at the ascending node due to foreground
and background material contained in the line of sight to the ascending node.  
The orbital configuration of particles is similar to the giant impact scenario, 
except that there are two pinch points.

As with the giant impact debris field, the problem with this scenario is that material is confined,
but not necessarily removed from the pinch area.  Therefore the surface brightness should not diminish significantly.
On the other hand, the material at a vertical pinch point has a very flat spatial distribution, so that belt particles
are more likely to be self-shadowing.  Self-shadowing is invoked to explain why Uranus' $\epsilon$ ring 
is fainter at periapse than at apoapse$-$ the ring optical thickness increases at periapse and self-shadows \citep{karkoschka97a}.

Future work is needed to quantitatively study the cumulative effect of these factors on the scattered
light appearance of debris belts near pinchpoints.  The theory of apsidally and nodally locked
planetary rings depends on the interplay between collisions, self-gravity, and the quadropole
field of the planet \citep{chiang03a}.  For a debris belt, collisions are important, but self-gravity
is assumed to be insignificant, and planet-belt dynamics include significant secular
effects.  Observationally, the origin of Fomalhaut's 331$\degr$ main belt gap could be
explored by ALMA observations.  Self-shadowing would be irrelevant at mm wavelengths such that
a belt gap in a mm map would indicate an absence of grains, thereby supporting the horseshoe/tadpole orbit
hypothesis.

\subsection{Comparison to the Solar System}
\label{sec:solarsystem}

\begin{figure}[!ht]
\epsscale{1.0}
\plotone{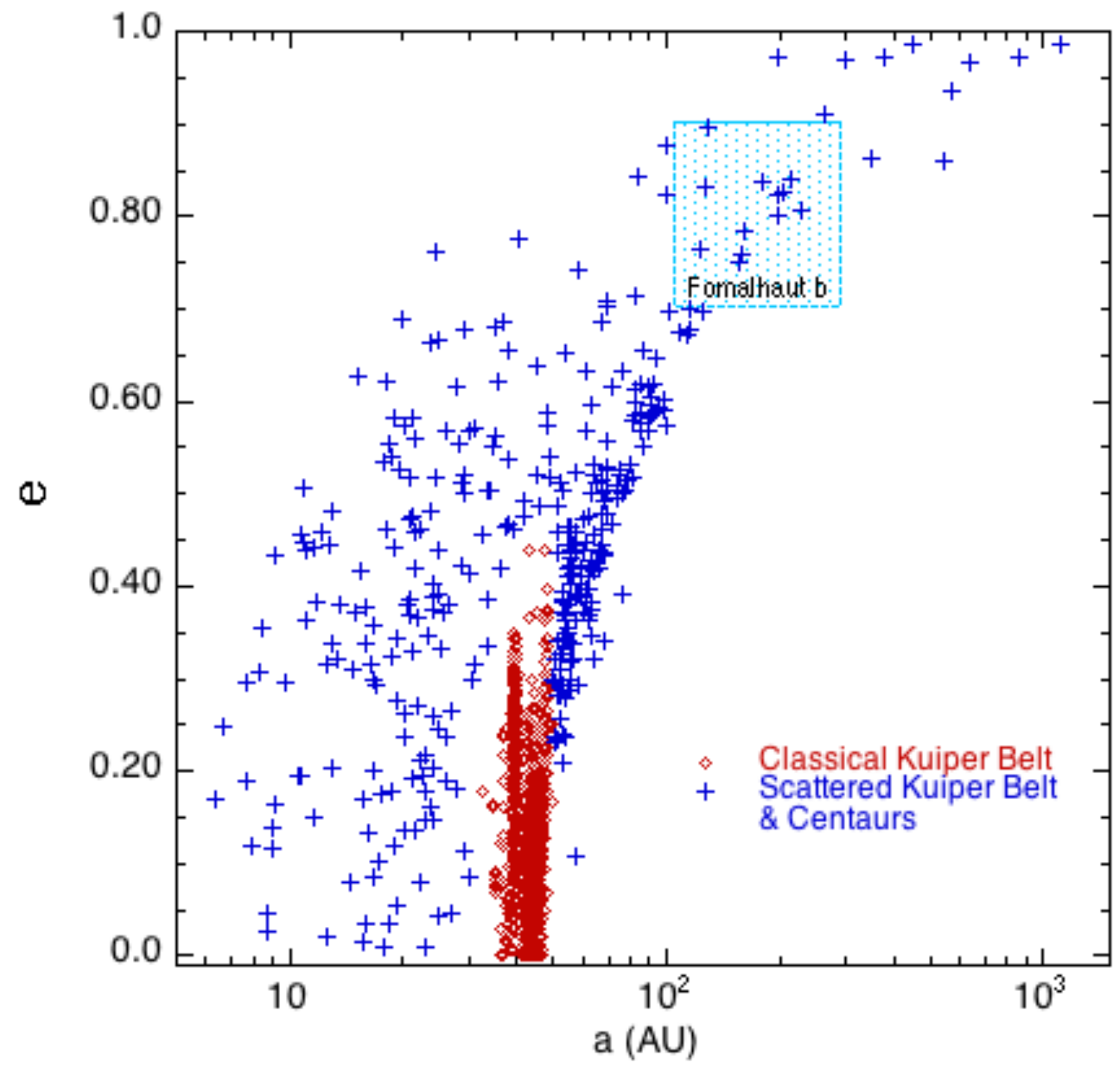}
\caption{\footnotesize
Distribution of $a$ versus $e$ for classical Kuiper Belt objects (red diamonds), scattered Kuiper Belt objects
and Centaurs (blue crosses), and Fomalhaut b (blue shaded region).
\label{skbos}}
\end{figure}

Figure~\ref{skbos} plots the semi-major axes and eccentricities of classical Kuiper Belt objects (KBO's), scattered Kuiper Belt objects (sKBO's), and Centaurs
catalogued by the Minor Planet Center.  Fomalhaut b's orbit lies in a region of $a$-$e$ parameter space occupied by sKBO's \citep{luu97a, levison97a, trujillo00a, brown05a}. 
A key reason for the overlap is that the perihelia of KBO's do not cross inward of Neptune's 30 AU semi-major axis (producing the upward curved boundary on the right side of
the cluster of points), and Fomalhaut b's periastron also
happens to be near 30 AU.  Therefore, the plot merely emphasizes that in terms of eccentricity,
Fomalhaut b is in the domain of the scattered Kuiper Belt instead of the classical Kuiper Belt.
However, because Fomalhaut b's periastron is significantly smaller than particles in
Fomalhaut's main belt, a more apt comparison is to a few known
Centaurs with high eccentricity that cross Neptune's orbit.

One example of such a Centaur is 2001 XA255 with $a=30$ AU, $e=0.7, i=13\degr$ \citep{jewitt02a, marcos12a}.
This object crosses the orbits of three planets (Saturn, Uranus and Neptune) and the dynamical evolution is
chaotic and short \citep[$<10^8$ yr;][]{dones96a, disisto07a, bailey09a}.

This suggests that as an alternative to planet-planet scattering scenarios,  
the dynamical mechanisms that produced the sKBO's and Centaurs may be
active in the Fomalhaut system.  
We calculate the Tisserand parameter of Fomalhaut b relative to a hypothetical, significantly more massive Fomalhaut c 
that serves as the perturber of the main belt inner edge ($a_c=120$ AU), or a hypothetical
Fomalhaut d located near Fomalhaut b's periastron ($a_d=30$ AU).
For example, in the case of an interaction with Fomalhaut c, the Tisserand equation is:

\begin{equation}
\label{tisserand}
T_{bc} = a_{c} / a_{b} + 2~cos(I_{bc})~\sqrt{a_b/a_c (1-e^2_b)}
\end{equation} 

The mutual inclination of the two objects is $I_{bc}$.
If $a_c=a_b$, $e_b$=0, and $I_{bc}=0\degr$, then T = 3.
The Tisserand parameter is approximately conserved for dynamical
interactions in the restricted, three-body problem.  Therefore,
instead of classifying objects in terms of their present-epoch
orbital parameters, the Tisserand parameter is a more useful standard because it tends to be
invariant over the many dynamical interactions with planets
that vary a minor body's orbital parameters over time \citep{levison97a}.

In principle, slow, strong encounters have $T\lesssim3$ and the perturbed
object is dynamically coupled to the massive planet.  An Oort cloud
comet with high mutual inclination has $T<2$.  Objects in the solar system with $2<T<3$ are dynamically 
coupled to the planets.
If $T\gtrsim3$, then the object is not dynamically
coupled to the massive planet.  

In the coplanar case
$T_{bd} = 3.0\pm0.1$ and $T_{bc} = 2.1\pm0.3$  (Fig.~\ref{tisserand}).  Therefore
Fomalhaut b may be dynamically coupled to both planets in the coplanar case.
If we take $I_{bc} = 36\degr$
as the maximum value for mutual inclination ({Fig.~\ref{mutual}), then Fomalhaut d
can have $a_d$ as small as 20 AU for T=3.0.
If the object that scattered Fomalhaut b resides within the main belt (e.g. a planet responsible for
the $331\degr$ gap), then for $a_c=140$ AU, $T_b = 2.1\pm0.3$.

\begin{figure}[!ht]
\epsscale{1.0}
\plotone{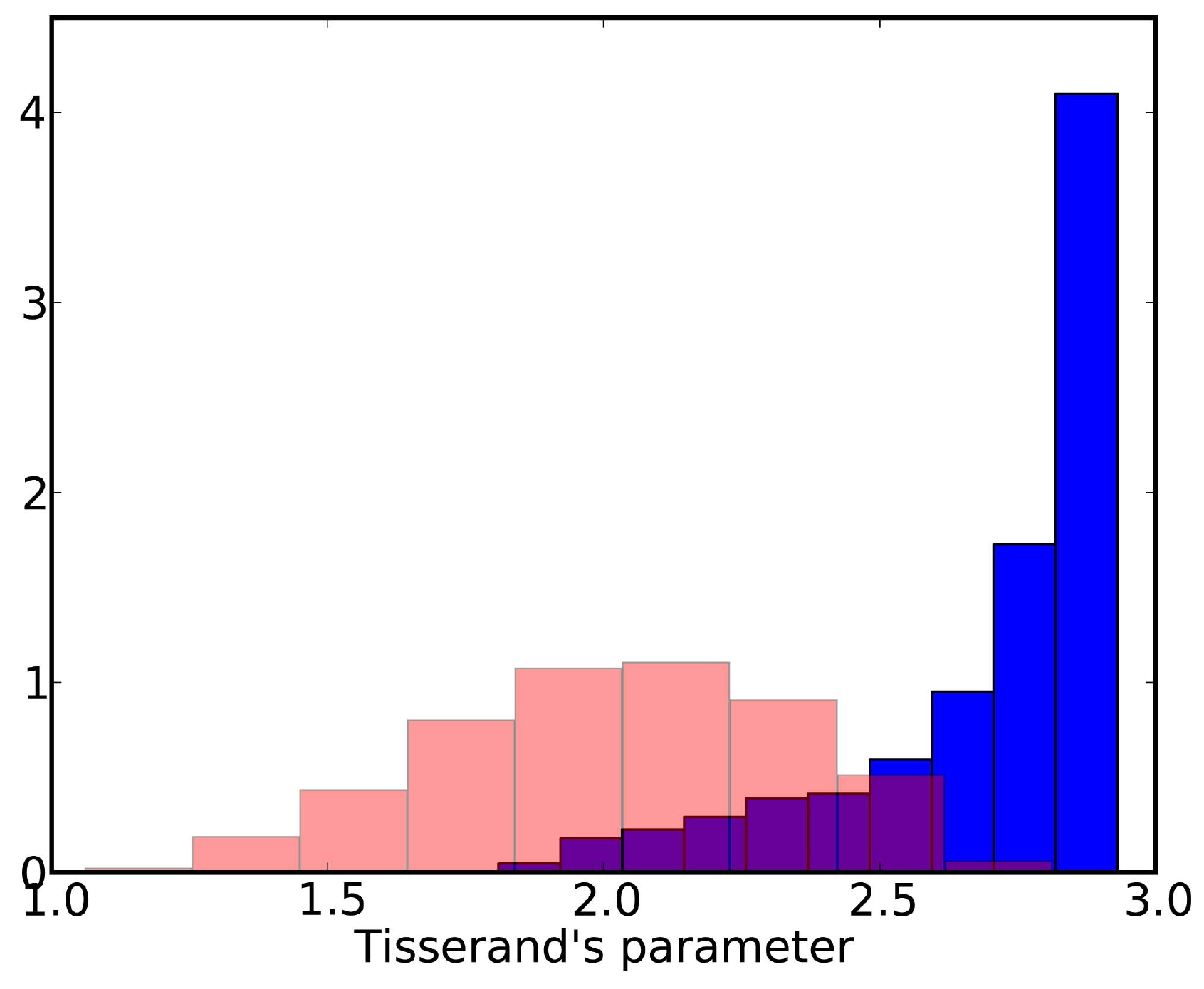}
\caption{\footnotesize
Tisserand parameter for Fomalhaut b assuming that it was scattered by a Fomalhaut c that has $a_d = 30$ AU (blue histogram) or $a_c = 120$ AU (red histogram).   
\label{tisserand}}
\end{figure}

Another potential comparison to the Kuiper Belt concerns mean motion resonances.
If $a_b$ = 177 AU, then $P_b$ = 1700 yrs, whereas the main belt at $a$=133, 143, 153 AU
has $P$ = 1106, 1234 and 1366 yrs, respectively.   Therefore a 3:2 resonance is apparent between Fomalhaut b and the
inner edge of the belt at 133 AU.  This is probably physically irrelevant given that Fomalhaut's main belt has significant width, and Fomalhaut b's 
semi-major axis is sufficiently uncertain, that at least one mean motion resonance can be identified by chance.
Moreover, if Fomalhaut b's orbit is a relatively recent outcome from a planet-planet scattering, then the 
important resonances over the age of the system are those that Fomalhaut b had before it was scattered.

Given the similarity of Fomalhaut b to the dynamics of our solar system's scattered
disk KBO's and Centaurs, is Fomalhaut b better described as an extrasolar dwarf planet rather than an
extrasolar planet?  The main observational constraints aside from the astrometry that gives the orbit, is
the optical brightness that could be reflection from material with a large surface area.  Therefore the analogy to a Kuiper Belt
Object is possible if we invoke Pluto, during the brief ($\sim$100 yr) epoch when its moons were being assembled from a circum-Pluto disk,
or the Haumea family KBO's \citep{brown07a}, at the epoch when a significant collision disrupted the precursor object.  Both scenarios
invoke a ''giant impact'' between dwarf planets, which we considered in Section 9.3.5.  The Charon-forming giant impact accounts for
the small mass ratio between Pluto and Charon \citep{canup05a} and the formation of Nix and Hydra \citep{canup11a}.
The simulations show that low relative velocities ($\sim1$ km/s) can form the Pluto system if the collisions are oblique and the
precursor object is partially differentiated and has an ice shell.  However, other collision scenarios do not necessarily produce 
moons but yield circumplanetary disks of material with mass $\sim10^{20}$ kg.  The Haumea collisionally family currently
consists of one dozen members with moderate eccentricity ($0.1-0.2$), a relatively large inclination ($24\degr - 29\degr$)
and a semi-major axes near 43 AU \citep{lykawka12a}.  The various collision models \citep{volk12a, schlichting09a, leinhardt10a} are broadly similar to the discussion in Section~\ref{sec:giantimpact}.

Given the plausibility of these hypothetical scenarios that address observations of
our own solar system, it is difficult to rule out the hypothesis that Fomalhaut b is
an extrasolar dwarf planet.

Finally, we briefly consider if the solar system could presently contain an object like Fomalhaut b.
The  geometric surface area of grains representing Fomalhaut b, $\sigma_p =1.2\times10^{19}~\mbox{m}^2$, is
$\sim6\times10^3$ greater than the projected surface area of Neptune.  Neptune's apparent visual magnitude is
+7.8 mag, which means that something with a factor of 6000 times greater surface area and the same albedo as
Neptune would be one of the brightest objects in the night sky at -1.6 mag.  Consider now a factor of $\sim$10 increase
in heliocentric distance corresponding to Fomalhaut b's apastron at $\sim$300 AU.  Fomalhaut b is fainter
by a factor of $10^4 $ as viewed from Earth, giving it an apparent magnitude of $+8.4$.  
Clearly no clone of Fomalhaut b exists at the current epoch in our solar system.  

\section{Summary:  Is it a planet?}
\label{sec:summary}

Our finding of a likely periastron passage near 30 AU radius now confers to Fomalhaut b a direct physical connection to the region 
where planetesimals grow to planets because the dynamical timescales are shorter and the primordial disk is denser closer to the star. 
On the other hand, compared to the present-day dynamics of the solar system, the orbit of Fomalhaut b is similar to Centaurs in the solar system.
How do we distinguish between a dwarf planet and a planet in the case of Fomalhaut b?

One $a~priori$ argument $against$ a planet mass for Fomalhaut b is that by crossing through the belt, it would dynamically disrupt the belt.  We give several important reasons why this argument is not definitive:

\begin{description}
\item[1. ] We find that the mutual inclination between Fomalhaut b and the main belt is $17\degr\pm12\degr$, with $\sim$10\% of possible ascending a descending nodes crossing through the main belt.  Therefore it is unlikely, but not ruled out, that Fomalhaut b crosses through the belt at the present epoch, though belt crossing may happen at other epochs due to orbital evolution.  For example, our initial N-body tests indicate that Fomalhaut b may evolve into a significantly different configuration on $10^6-10^8$ yr timescales if a nested, Jupiter-mass planet orbits within the belt perimeter.
\item[2.]  We present N-body simulations that show a planet crossing through a belt does not destroy it, but instead erodes the edges of a belt on timescales that depend on the assumed mass of the planet (e.g., a Saturn mass requires $\sim10^5$ years or $10^2$ crossings).  Since we do not know how long the present orbital geometry has existed in the Fomalhaut system, Fomalhaut b could be a gas giant planet.
\item[3.]  A corollary to [2] is that due to Fomalhaut b's large eccentricity, it passes through the belt quickly ($\sim$100 yrs) in the coplanar case.  Therefore the effectiveness of a belt crossing planet in modifying belt dynamics is diminished.
\item[4.]  We show two new features in the main belt that in fact suggest the dynamics of the system are more complex than previously established.  First we identify a $\sim$50 AU wide gap in the azimuthal structure of the belt north of Fomalhaut b.  Second, the outer edge of the belt is extended to at least 209 AU, and appears warped beyond this radius.
\end{description}

To summarize, the potential belt passage of Fomalhaut b does not exclude any masses up to the 1 $M_J$ mass limit determined by
infrared imaging surveys.  We therefore considered several aspects of a lower mass limit, establishing the following principles:

\begin{description}
\item[5.] Assuming the observed optical light from Fomalhaut b is reflected from dust grains, the mass of the required grains implies a precursor object $>$10 km in size.   The collision physics of objects this large lies in the gravity dominated regime, which means that Fomalhaut b consists of a central planetesimal or family of planetesimals surrounded by a bound dust cloud of greater extent and surface area.
\item[6.]  If the mass of the central planetesimal is too small, a dust cloud surrounding it is sheared away by tidal forces during periastron passages.  We show that for the planetsimal cloud to be stable against shearing, the minimum mass of Fomalhaut b is $5\times10^{21}$ kg, comparable to 5$\times$ Ceres' mass.
\end{description}

Fomalhaut b's mass is therefore in the range between our solar system's dwarf planets and Jupiter.  Unless new dynamical simulations can show otherwise, the main belt's inner edge and stellocentric offset are not definitively linked to Fomalhaut b alone.  However, our current orbit determination shows that Fomalhaut b is apsidally aligned with the main belt.  

To explain all of the various observed features in the system, additional ``nested'' planets may be necessary.  
A comprehensive analytic and numerical study of the possible parameter space is required.  We considered the possibility that Fomalhaut b was scattered from a planet located near its periastron at $\sim$30 AU, that a planet at 120 AU sculpts the inner edge of the main belt, and that a planet orbiting within the main belt could account for the aziumuthal gap.  Current direct imaging studies of the system are ineadequate to exlude any planets less than a Jupiter mass in these regions.

Several additional points are:

\begin{description}
\item[7.] Fomalhaut b's orbital parameters are similar to scattered Kuiper Belt objects and Centaurs, which suggests
the dynamical mechanisms operating in the Fomalhaut system could further our knowledge of the
early solar system's dynamical history.  
\item[8.] For a coplanar orbit, Fomalhaut b will collide with the main belt  two decades from now.  Monitoring for transient phenomena such as
SL-9 impacts on Jupiter would elucidate the mass and composition of Fomalhaut b, and perhaps lead to unique insights of
exoplanet atmospheres as main belt planetesimals excavate and heat the atmosphere.  
\item[9.] We also considered a scenario where the main belt is the partial remnant of a Fomalhaut c that suffered a recent collition with Fomalhaut b.  Due to the low relative velocities at $\sim$140 AU, and the requisite mass of material in the belt, this scenario is more credible for a head on collision (the belt has a retrograde orbit relative to Fomalhaut b).
\item[10.]  If Fomalhaut b has a satellite system that was dynamically disturbed by the recent periastron passage, then the moons may also be tidally heated by the central planet.  By analogy to Io and Jupiter, enhanced volcanic activity could lead to large sodium or potassium clouds around Fomalhaut b that would explain puzzling aspects of the optical data. 
\item[11.] We studied the possibility that Fomalhaut B (TW PsA) may also have a role in the dynamics of the system, but little is known about its orbit except that the period is likely $>$8 Myr.
\end{description}

Future observations and theoretical investigations can therefore address several important open questions:
(1) Does the orbit of Fomalhaut b pass through the belt?
(2) Is the spectrum of Fomalhaut b consistent with reflected light, and are there any features indicating composition?
(3) What is the interconnection between the apsidal alignment of Fomalhaut b with the main belt and the azimuthal belt gap?
(4)  Where is Fomalhaut c?

\acknowledgments
{\bf Acknowledgements}: Based on observations with the NASA/ESA Hubble Space Telescope, obtained at STScI, which is operated by AURA under NASA contract NAS5-26555.
This work received support from the following:  GO-11818 and GO-12576 provided by NASA through a grant from
STScI under NASA contract NAS5-26555; NSF AST-0909188; and the University of California LFRP-118057.  We are grateful to G. Soutchkova, J. Debes, G. Schneider, J. Duval, A. Aloisi and C. Proffit for assistance in the observations.  We thank H. Beust for sharing his insights on the Markov chain orbital element analysis, and A. Jackson, G. Kennedy, H. Levison and M. Wyatt for reviewing the draft manuscript.    We also thank E. Chiang, T. Currie, R. Dawson, E. Ford, B. Hansen, M. Hughes, R. Malhotra, D. Nesvorny, S. Stewart, D. Tamayo for helpful discussions

{\it Facilities:}  \facility{HST (STIS)}.

\clearpage

\clearpage

\begin{deluxetable}{lllll}
\tabletypesize{\scriptsize}
\tablecaption{STIS Observations\label{tbl-1}}
\tablewidth{0pt}
\tablehead{
\colhead{Date} & \colhead{Position} & \colhead{ORIENTAT} & \colhead{Exposures}
}
\startdata
2010-09-13	& WEDGEB2.5		& 193.04$\degr$	& 25 $\times$ 30s\\
			& 				& 201.04$\degr$	& 25 $\times$ 30s\\
			& 				& 209.04$\degr$	& 25 $\times$ 30s\\
			& 				& 217.05$\degr$	& 25 $\times$ 30s\\
2012-05-29	& WEDGEB2.5		&-169.945$\degr$	& 28 $\times$ 30s\\
			& 				&-161.945$\degr$	& 28 $\times$ 30s\\
			& 				&-153.945$\degr$	& 28 $\times$ 30s\\
			& 				&-145.945$\degr$	& 28 $\times$ 30s\\
2012-05-30	& WEDGEB2.5		&-167.945$\degr$	& 28 $\times$ 30s\\
			& 				&-159.945$\degr$	& 28 $\times$ 30s\\
			& 				&-151.945$\degr$	& 28 $\times$ 30s\\
			& 				&-143.945$\degr$	& 28 $\times$ 30s\\
2012-05-31	& WEDGEB2.5		&-165.945$\degr$	& 28 $\times$ 30s\\
			& 				&-157.945$\degr$	& 28 $\times$ 30s\\
			& 				&-149.945$\degr$	& 28 $\times$ 30s\\
			& 				&-141.945$\degr$	& 28 $\times$ 30s\\

\enddata
\tablenotetext{a}{ORIENTAT is the position angle of the image y axis}
\label{log}
\end{deluxetable}
\begin{deluxetable}{lllll}
\tabletypesize{\scriptsize}
\tablecaption{Astrometric Error Terms\label{tbl-1}}
\tablewidth{0pt}
\tablehead{
\colhead{} & \colhead{Value} & \colhead{Source} & \colhead{}
}
\startdata
{\bf Uncorrected Systematic Errors} 	&					&									& \\
Geometric Distortion STIS			& 66 mas	(not used)			& Unpublished Calibration Program	& \\
Geometric Distortion STIS			& 17 mas				& Measured in data	& \\
Stability of optical distortion STIS	& 5 mas			&  STIS astrometric report &\\
Geometric Distortion and Stability ACS/HRC	& 3 mas				&  		Instrument Handbook			& \\
Detector Position Angle			& 0.06$\degr$ (1 mas in X, 1 mas in Y) 			& Measured in data (see text)		&  \\
&&&\\
{\bf Statistical Errors} 							& 					&  					& \\
ACS/HRC 2004 centroiding star						& 16 mas in X, 10 mas in Y		& Measured in data  & \\
ACS/HRC 2004 centroiding on Fom b					& 18 mas in X, 13 mas in Y	& Measured in data	&	 \\
ACS/HRC 2006 centroiding star						& 19 mas in X, 17 mas in Y		& Measured in data  & \\
ACS/HRC 2006 centroiding on Fom b					& 6 mas in X, 21 mas in Y		& Measured in data	&	 \\
STIS centroiding star							& 25 mas in X, 25 mas in Y	& Measured in data	& \\
STIS 2010 centroiding Fom b						& 28 mas in X, 31 mas in Y	&Measured in data	&\\
STIS 2010 recovery of artificial implants				& 8 mas in X, 5 mas in Y	&Measured in data	&\\
STIS 2012 centroiding Fom b						& 16 mas in X, 19 mas in Y	&Measured in data	&\\
STIS 2012 recovery of artificial implants				& 5 mas in X, 10 mas in Y	&Measured in data	&\\
\enddata
\label{astrometryerrors}
\end{deluxetable}
\begin{deluxetable}{lllll}
\tabletypesize{\scriptsize}
\tablecaption{Star--Planet Astrometry\label{tbl-1}}
\tablewidth{0pt}
\tablehead{
\colhead{UT Start/End} & \colhead{Midpoint (JD)} & \colhead{West offset (mas)} & \colhead{North offset (mas)} & \colhead{Detector}
}
\startdata
2004-10-25/26 	& 2453304.2510995 &$8587\pm24$	& $9175\pm17$	& ACS/HRC\\
2006-07-17/20 	& 2453935.3606890 &$8597\pm22$	& $9365\pm19$	& ACS/HRC \\
2010-09-13 	& 2455452.9415740&$8828\pm42$		& $9822\pm44$	& STIS \\
2012-05-29/31 	& 2456078.1699655&$8915\pm 35$	& $10016\pm$37	& STIS \\
\enddata
\label{astrometry}
\end{deluxetable}
\begin{deluxetable}{llllllll}
\tabletypesize{\scriptsize}
\tablecaption{Apparent (Sky-Plane) Belt Geometry\label{tbl-1}}
\tablewidth{0pt}
\tablehead{
\colhead{Data} & \colhead{Semi-Major} &\colhead{PA} & \colhead{Semi-Minor} & \colhead{$\sqrt{1-b^2/a^2}$} & \colhead{Inclination} & \colhead{RA Offset} &  \colhead{Dec Offset}
}
\startdata
\citet{kalas05a}	&140.7$\pm$1.8 AU	&$156.0\degr\pm0.3\degr$	& 57.5$\pm$0.7 AU	& 0.91$\pm$0.01 &  $65.9\degr\pm0.4\degr$ &$-2.2\pm$0.3 AU &$13.2\pm0.9$ AU\\
STIS (bisector)		&141.3$\pm$0.4 AU	&$156.2\degr\pm0.1\degr$	& 57.5$\pm$0.2 AU	& 0.91$\pm$0.01 &  $66.0\degr\pm0.2\degr$ &$-2.66\pm$0.15 AU   &13.06$\pm$0.25 AU \\
STIS (inner edge) 	&135.0$\pm$0.4 AU	&$156.4\degr\pm0.1\degr$	& 52.6$\pm$0.2 AU	& 0.92$\pm$0.01 &  $67.0\degr\pm0.2\degr$ &$-1.13\pm$0.15 AU   &12.75$\pm$0.25 AU \\
\enddata
\label{apparent}
\end{deluxetable}

\begin{deluxetable}{llllllll}
\tabletypesize{\scriptsize}
\tablecaption{Derived Main Belt and Fomalhaut $b$ Keplerian Orbital Elements\label{tbl-1}}
\tablewidth{0pt}
\tablehead{
\colhead{Data} 		& \colhead{a} 			& \colhead{e} 				& \colhead{i} 						& \colhead{$\Omega$} 	& \colhead{$\omega$}
}
\startdata
STIS (bisector)		&$141.77\pm0.28$ AU	& $0.10\pm0.01 $   		&  $-66.1\degr \pm 0.1\degr$  		&  $156.1 \degr \pm 0.1\degr$ 	& $29.6\degr \pm 1.3\degr$ \\
STIS (inner edge) 	&$136.28\pm0.28 $ AU	& $0.12\pm0.01 $		&  $-67.5\degr \pm 0.1\degr$		&  $156.2 \degr \pm 0.1\degr$	& $41.9\degr \pm 1.1 \degr$ \\
Fomalhaut b		&$177\pm68$ AU		& $0.8\pm0.1 $   		&  $-55\degr \pm 14\degr$  		&  $152\degr \pm 13\degr$ 	& $26\degr \pm 25 \degr$ \\
\enddata
\label{derived}
\end{deluxetable}


\begin{thebibliography}{}

\bibitem[Absil et al.(2009)]{absil09a} Absil, O., Mennesson, B., Le Bouquin, J.-B., et al.\ 2009, \apj, 704, 150 

\bibitem[Acke et al.(2012)]{acke12a} Acke, B., Min, M., Dominik, C., et al.\ 2012, \aap, 540, A125 

\bibitem[Adams \& Laughlin(2003)]{adams03a} Adams, F.~C., \& Laughlin, G.\ 2003, \icarus, 163, 290 

\bibitem[Anderson et al.(2010)]{anderson10a} Anderson, D.~R., Hellier, C., Gillon, M., et al.\ 2010, \apj, 709, 159 

\bibitem[Anic et al.(2007)]{anic07a} Anic, A., Alibert, Y., \& Benz, W.\ 2007, \aap, 466, 717 

\bibitem[Ardila et al.(2005)]{ardila05a} Ardila, D.~R., Lubow, S.~H., Golimowski, D.~A., et al.\ 2005, \apj, 627, 986 

%

%

\bibitem[Artymowicz \& Clampin(1997)]{artymowicz97a}  Artymowicz, P., \& Clampin, M.\ 1997, \apj, 490, 863 

\bibitem[Asphaug(2009)]{asphaug09a} Asphaug, E.\ 2009, Annual Review of Earth and Planetary Sciences, 37, 413 

\bibitem[Augereau \& Papaloizou(2004)]{augereau04a} Augereau, J.~C., \& Papaloizou, J.~C.~B.\ 2004, \aap, 414, 1153 

%

\bibitem[Backman \& Gillett(1987)]{backman87a}
{Backman, D.E. \& Gillett, F.C.}, 1987, in {\it Cool Stars, Stellar Systems and
the Sun}, eds. J. L Linsky and R.E. Stencel (Springer-Verlag, Berlin), pp. 340-350.

%

\bibitem[Bailey \& Malhotra(2009)]{bailey09a} Bailey, B.~L., \& Malhotra, R.\ 2009, \icarus, 203, 155 

%

\bibitem[Basri \& Brown(2006)]{basri06a} Basri, G., \& Brown, M.~E.\ 2006, Annual Review of Earth and Planetary Sciences, 34, 193 

\bibitem[Barrado y Navascues et al.(1997)]{byn97a}{Barrado y Navascues, D., Stauffer, J. R., Hartmann, L. \& Balachandran, S.} 1997,\apj, 475, 313

%

%

\bibitem[Bevington(1969)]{bevington69a} Bevington, P.~R.\ 1969, Data reduction and error analysis for the physical sciences, New York: McGraw-Hill, 1969, 

\bibitem[Bjoraker et al.(1996)]{bjoraker96a} Bjoraker, G.~L., Stolovy, S.~R., Herter, T.~L., Gull, G.~E., \& Pirger, B.~E.\ 1996, \icarus, 121, 411 

\bibitem[Boley et al.(2012)]{boley12a} Boley, A.~C., Payne, M.~J., Corder, S., et al.\ 2012, \apjl, 750, L21 

%

\bibitem[Brown et al.(2005)]{brown05a} Brown, M.~E., Trujillo, C.~A., \& Rabinowitz, D.~L.\ 2005, \apjl, 635, L97 

\bibitem[Brown et al.(2007)]{brown07a} Brown, M.~E., Barkume, K.~M., Ragozzine, D., \& Schaller, E.~L.\ 2007, \nat, 446, 294 

%

%

%

%

\bibitem[Canup(2005)]{canup05a} Canup, R.~M.\ 2005, Science, 307, 546 

\bibitem[Canup(2011)]{canup11a} Canup, R.~M.\ 2011, \aj, 141, 35 

\bibitem[Carlson et al.(1997)]{carlson97a} Carlson, R.~W., Drossart, P., Encrenaz, T., et al.\ 1997, \icarus, 128, 251 

\bibitem[Cassen et al.(1979)]{cassen79a} Cassen, P., Reynolds, R.~T., \& Peale, S.~J.\ 1979, \grl, 6, 731 

\bibitem[Chambers et al.(1996)]{chambers96a} Chambers, J.~E., Wetherill, G.~W., \& Boss, A.~P.\ 1996, \icarus, 119, 261 


\bibitem[Chatterjee et al.(2008)]{chatterjee08a} Chatterjee, S., Ford, E.~B., Matsumura, S., \& Rasio, F.~A.\ 2008, \apj, 686, 580 

\bibitem[Chauvin et al.(2012)]{chauvin12a} Chauvin, G., Lagrange, A.-M., Beust, H., et al.\ 2012, \aap, 542, A41 


\bibitem[Chiang \& Culter(2003)]{chiang03a} Chiang, E.~I., \& Culter, C.~J.\ 2003, \apj, 599, 675 

\bibitem[Chiang \& Lithwick(2005)]{chiang05a} Chiang, E.~I., \& Lithwick, Y.\ 2005, \apj, 628, 520 

\bibitem[Chiang et al.(2009)]{chiang09a}{Chiang, E., Kite, E., Kalas, P., Graham, J.R. \& Clampin, M.} 2009, \apj, 693, 734

\bibitem[Currie et al.(2012)]{currie12a}{Currie, T., Debes, J., Rodigas, T. et al.} 2012, \apj, in press.

\bibitem[Debes \& Sigurdsson(2007)]{debes07a} Debes, J.~H., \& Sigurdsson, S.\ 2007, \apjl, 668, L167 


%

\bibitem[Deltorn \& Kalas(2001)]{deltorn01a} Deltorn, J.-M., \& Kalas, P.\ 2001, Young Stars Near Earth: Progress and Prospects, 244, 227 

\bibitem[Dent et al.(2000)]{dent00a} Dent, W.~R.~F., Walker, H.~J., Holland, W.~S., \& Greaves, J.~S.\ 2000, \mnras, 314, 702 

\bibitem[Dermott et al.(1994)]{dermott94a} Dermott, S.~F., Jayaraman, S., Xu, Y.~L., Gustafson, B.~{\AA}.~S., \& Liou, J.~C.\ 1994, \nat, 369, 719 

\bibitem[Di Folco et al.(2004)]{difolco04a} Di Folco, E., Th{\'e}venin, F., Kervella, P., et al.\ 2004, \aap, 426, 601 

\bibitem[di Sisto  \& Brunini(2007)]{disisto07a} di Sisto, R.~P., \& Brunini, A.\ 2007, \icarus, 190, 224 

\bibitem[Digby et al.(2006)]{digby06a}{Digby, A. P., Hinkley, S., Oppenheimer, B. R., et al.} 2006, \apj, 650, 484

\bibitem[Dominik \& Decin(2003)]{dominik03a} Dominik, C., \& Decin, G.\ 2003, \apj, 598, 626 

\bibitem[Dones et al.(1996)]{dones96a} Dones, L., Levison, H.~F., \& Duncan, M.\ 1996, Completing the Inventory of the Solar System, 107, 233 

%

\bibitem[Duncan et al.(1987)]{duncan87a} Duncan, M., Quinn, T., \& Tremaine, S.\ 1987, \aj, 94, 1330 

\bibitem[Duncan \& Levison(1997)]{duncan97a} Duncan, M.~J., \& Levison, H.~F.\ 1997, Science, 276, 1670 

%

\bibitem[Ford et al.(2000)]{ford00a} Ford, E.~B., Kozinsky, B.,  \& Rasio, F.~A.\ 2000, \apj, 535, 385 

\bibitem[Ford(2006)]{ford06a} Ford, E.~B.\ 2006, \apj, 642, 505 

\bibitem[Ford \& Rasio(2008)]{ford08a} Ford, E.~B., \& Rasio, F.~A.\ 2008, \apj, 686, 621 

%

\bibitem[Galicher et al.(2013)]{galicher13a} Galicher, R. et al. 2013, \apj, in press.

\bibitem[Gladman(1993)]{gladman93a} Gladman, B.\ 1993, \icarus, 106, 247 

%

\bibitem[Gomes et al.(2006)]{gomes06a} Gomes, R.~S., Matese, J.~J., \& Lissauer, J.~J.\ 2006, \icarus, 184, 589

\bibitem[Gooding \& Odell(1988)]{gooding88a} Gooding, R.~H., \& Odell, A.~W.\ 1988, Celestial Mechanics, 44, 267 

%

\bibitem[Golimowski et al.(2006)]{golimowski06a} Golimowski, D.~A., Ardila, D.~R., Krist, J.~E., et al.\ 2006, \aj, 131, 3109 

\bibitem[Graham et al.(1995)]{graham95a} Graham, J.~R., de Pater, I., Garrett Jernigan, J., Liu, M.~C., \& Brown, M.~E.\ 1995, Science, 267, 1320 

\bibitem[Green(1985)]{green85a} Green, R.~M.\ 1985, ``Spherical Astronomy'', Cambridge and New York, Cambridge University Press

\bibitem[Gregory(2001)]{gregory01a} Gregory, P. 2001, ``Bayesian Logical Data Analysis for the Physical Sciences'', Cambridge and New York, Cambridge University Press

%

\bibitem[Hamilton \& Burns(1991)]{hamilton91a} Hamilton, D.~P., \& Burns, J.~A.\ 1991, \icarus, 92, 118 

\bibitem[Hamilton \& Burns(1992)]{hamilton92a} Hamilton, D.~P., \& Burns, J.~A.\ 1992, \icarus, 96, 43 

\bibitem[Hartmann \& Davis(1975)]{hartmann75a} Hartmann, W.~K., \& Davis, D.~R.\ 1975, \icarus, 24, 504 

\bibitem[Heap et al.(2000)]{heap00a} Heap, S.~R., Lindler, D.~J., Lanz, T.~M., et al.\ 2000, \apj, 539, 435 

\bibitem[H{\'e}brard et al.(2008)]{hebrard08a} H{\'e}brard, G., Bouchy, F., Pont, F., et al.\ 2008, \aap, 488, 763 

\bibitem[Heggie(1975)]{heggie75a} Heggie, D.~C.\ 1975, \mnras, 173, 729 

\bibitem[Holland et al.(1998)]{holland98a} Holland, W.~S., Greaves, J.~S., Zuckerman, B., et al.\ 1998, \nat, 392, 788 

\bibitem[Holland et al.(2003)]{holland03a} Holland, W.~S., Greaves, J.~S., Dent, W.~R.~F., et al.\ 2003, \apj, 582, 1141 

\bibitem[Holmberg \& Flynn(2000)]{holmberg00a} Holmberg, J., \& Flynn, C.\ 2000, \mnras, 313, 209 

\bibitem[Ida \& Lin(2005)]{ida05a} Ida, S., \& Lin, D.~N.~C.\ 2005, \apj, 626, 1045 

\bibitem[Jackson \& Zook(1992)]{jackson92a} Jackson, A.~A., \& Zook, H.~A.\ 1992, Lunar and Planetary Institute Science Conference Abstracts, 23, 595

\bibitem[Jackson \& Wyatt(2012)]{jackson12a} Jackson, A.~P., \& Wyatt, M.~C.\ 2012, \mnras, 425, 657 

\bibitem[Janson et al.(2012)]{janson12a} Janson, M., Carson, J.~C., Lafreni{\`e}re, D., et al.\ 2012, \apj, 747, 116 

\bibitem[Jewitt et al.(2002)]{jewitt02a} Jewitt, D.~C., Sheppard,  S.~S., Kleyna, J., 
\& Marsden, B.~G.\ 2002, Minor Planet Electronic Circulars, 85 

\bibitem[Jewitt \& Haghighipour(2007)]{jewitt07a} Jewitt, D., \& Haghighipour, N.\ 2007, \araa, 45, 261 

\bibitem[Jewitt(2009)]{jewitt09a} Jewitt, D.\ 2009, \aj, 137, 4296 

\bibitem[Jewitt(2012)]{jewitt12a} Jewitt, D.\ 2012, \aj, 143, 66 

\bibitem[Johnson et al.(2011)]{johnson11a} Johnson, J.~A., Winn, J.~N., Bakos, G.~{\'A}., et al.\ 2011, \apj, 735, 24 

\bibitem[Juri{\'c} \& Tremaine(2008)]{juric08a} Juri{\'c}, M., \& Tremaine, S.\ 2008, \apj, 686, 603 

\bibitem[Kaib et al.(2013)]{kaib13a} Kaib, N.~A., Raymond, S.~N., \& Duncan, M.\ 2013, \nat, 493, 381 

\bibitem[Kalas et al.(2005)]{kalas05a}{Kalas, P., Graham, J.R. \& Clampin} 2005,\textit{Nature}, 435, 1067

\bibitem[Kalas(2005)]{kalas05b} Kalas, P.\ 2005, \apjl, 635, L169 

\bibitem[Kalas et al.(2006)]{kalas06a} Kalas, P., Graham, J.~R., Clampin, M.~C., \& Fitzgerald, M.~P.\ 2006, \apjl, 637, L57 

\bibitem[Kalas et al.(2008)]{kalas08a}{Kalas, P., Graham, J.R., Chiang, E., et al.} 2008, \textit{Science}, 322, 1345

\bibitem[Kalas et al.(2010)]{kalas10a}{Kalas, P., Graham, J.R., Fitzgerald, M., \& Clampin, M.} 2010, in \textit{In the Spirit of Lyot 2010}, ed. A. Boccaletti.

\bibitem[Kalas (2011)]{kalas11a}{Kalas, P.} 2011, in \textit{The Astrophysics of Planetary Systems}, IAU Symposium, Vol. 276, p. 279.

\bibitem[Karkoschka(1997)]{karkoschka97a} Karkoschka, E.\ 1997, \icarus, 125, 348 

\bibitem[Kasting et al.(1993)]{kasting93a} Kasting, J.~F., Whitmire, D.~P., \& Reynolds, R.~T.\ 1993, \icarus, 101, 108 

\bibitem[Kennedy \& Kenyon(2008)]{kennedy08a} Kennedy, G.~M., \& Kenyon, S.~J.\ 2008, \apj, 673, 502 

\bibitem[Kennedy \& Wyatt(2011)]{kennedy11a}{Kennedy, G. M. \& Wyatt, M. C.} 2011, \mnras, 412, 2137

\bibitem[Kennedy et al.(2012)]{kennedy12a} Kennedy, G.~M., Wyatt, M.~C., Sibthorpe, B., et al.\ 2012, \mnras, 421, 2264 

\bibitem[Kenyon \& Bromley(2001)]{kenyon01a} Kenyon, S.~J., \& Bromley, B.~C.\ 2001, \aj, 121, 538 

\bibitem[Kenyon \& Bromley(2002)]{kenyon02a} Kenyon, S.~J., \& Bromley, B.~C.\ 2002, \aj, 123, 1757 

\bibitem[Kenworthy et al.(2009)]{kenworthy09a} Kenworthy, M.~A., Mamajek, E.~E., Hinz, P.~M., et al.\ 2009, \apj, 697, 1928 

\bibitem[Kenworthy et al.(2013)]{kenworthy13a} Kenworthy, M.~A., Meshkat, T., Quanz, S.~P., et al.\ 2013, \apj, 764, 7 

\bibitem[Krist et al.(2005)]{krist05a} Krist, J.~E., Stapelfeldt, K.~R., Golimowski, D.~A., et al.\ 2005, \aj, 130, 2778 

\bibitem[Krist et al.(2011)]{krist11a} Krist, J.~E., Hook, R.~N., \& Stoehr, F.\ 2011, \procspie, 8127

\bibitem[Krivov et al.(2002)]{krivov02a} Krivov, A.~V., Wardinski, I., Spahn, F., Kr{\"u}ger, H., \& Gr{\"u}n, E.\ 2002, \icarus, 157, 436 

\bibitem[Larwood \& Kalas(2001)]{larwood01a} Larwood, J.~D., \& Kalas, P.~G.\ 2001, \mnras, 323, 402 

\bibitem[Le Bouquin et al.(2009)]{lebouquin09a}{Le Bouquin, J.-B., Absil, O., Benisty, M., Massi, F., Merand, A. \& Stefl, S.} 2009, \aap, 498, L41

\bibitem[Leinhardt et al.(2010)]{leinhardt10a} Leinhardt, Z.~M., Marcus, R.~A., \& Stewart, S.~T.\ 2010, \apj, 714, 1789 

\bibitem[Leinhardt \& Stewart(2012)]{leinhardt12a} Leinhardt, Z.~M., \& Stewart, S.~T.\ 2012, \apj, 745, 79 

\bibitem[Levison \& Duncan(1997)]{levison97a} Levison, H.~F., \& Duncan, M.~J.\ 1997, \icarus, 127, 13 

\bibitem[Levison et al.(1998)]{levison98a} Levison, H.~F., Lissauer, J.~J., \& Duncan, M.~J.\ 1998, \aj, 116, 1998 

%

%

%

\bibitem[Lykawka et al.(2011)]{lykawka11a} Lykawka, P.~S., Horner, J., Jones, B.~W., \& Mukai, T.\ 2011, \mnras, 412, 537 

\bibitem[Lykawka et al.(2012)]{lykawka12a} Lykawka, P.~S., Horner, J., Mukai, T., \& Nakamura, A.~M.\ 2012, \mnras, 421, 1331 

%

\bibitem[Luu et al.(1997)]{luu97a} Luu, J., Marsden, B.~G., Jewitt, D., et al.\ 1997, \nat, 387, 573 

\bibitem[Marcus et al.(2010)]{marcus10a} Marcus, R.~A., Sasselov, D., Stewart, S.~T., \& Hernquist, L.\ 2010, \apjl, 719, L45 

%

\bibitem[Malmberg et al.(2011)]{malmberg11a} Malmberg, D., Davies, M.~B., \& Heggie, D.~C.\ 2011, \mnras, 411, 859 

\bibitem[Mamajek(2012)]{mamajek12a} Mamajek, E.~E.\ 2012, \apjl, 754, L20 

\bibitem[Maness et al.(2009)]{maness09a} Maness, H.~L., Kalas, P., Peek, K.~M.~G., et al.\ 2009, \apj, 707, 1098 

\bibitem[de la Fuente Marcos \& de la Fuente Marcos(2012)]{marcos12a} de la Fuente Marcos, C., \& de la Fuente Marcos, R.\ 2012, \aap, 547, L2 

\bibitem[Marsh et al.(2005)]{marsh05a} Marsh, K.~A., Velusamy, T., Dowell, C.~D., Grogan, K., \& Beichman, C.~A.\ 2005, \apjl, 620, L47 

\bibitem[Marengo et al.(2009)]{marengo09a} Marengo, M., Stapelfeldt, K., Werner, M.~W., et al.\ 2009, \apj, 700, 1647 

\bibitem[Marsden(2005)]{marsden05a} Marsden, B.~G.\ 2005, \araa, 43, 75 

\bibitem[Marzari \& Weidenschilling(2002)]{marzari02a} Marzari, F., \& Weidenschilling, S.~J.\ 2002, \icarus, 156, 570 

%

\bibitem[Mendillo et al.(1990)]{mendillo90a} Mendillo, M., Baumgardner, J., Flynn, B., \& Hughes, W.~J.\ 1990, \nat, 348, 312 

\bibitem[Mendillo et al.(2004)]{mendillo04a} Mendillo, M., Wilson, J., Spencer, J., \& Stansberry, J.\ 2004, \icarus, 170, 430 

\bibitem[Mennesson et al.(2013)]{mennesson13a} Mennesson, B., Absil, O., Lebreton, J., et al.\ 2013, \apj, 763, 119 

\bibitem[Min et al.(2010)]{min10a}{Min. M., Kama, M., Dominik, C. \& Waters, L.B.F.M.} 2010, \aap, 509, L6

%

%

\bibitem[Moro-Martin \& Malhotra(2002)]{moromartin02}Moro-Martin, A. \& Malhotra, R. 2002, \aj, 124, 2305

\bibitem[Mouillet et al.(1997)]{mouillet97a} Mouillet, D., Larwood, J.~D., Papaloizou, J.~C.~B., \& Lagrange, A.~M.\ 1997, \mnras, 292, 896 

%

\bibitem[Nesvorn{\'y} et al.(2003)]{nesvorny03a} Nesvorn{\'y}, D., Alvarellos, J.~L.~A., Dones, L., \& Levison, H.~F.\ 2003, \aj, 126, 398 

\bibitem[Nesvorn{\'y} et al.(2007)]{nesvorny07a} Nesvorn{\'y}, D., Vokrouhlick{\'y}, D., \& Morbidelli, A.\ 2007, \aj, 133, 1962 

\bibitem[Nesvorn{\'y} \& Vokrouhlick{\'y}(2009)]{nesvorny09a} Nesvorn{\'y}, D., \& Vokrouhlick{\'y}, D.\ 2009, \aj, 137, 5003 

\bibitem[Ng \& Bertelli(1998)]{ng98a} Ng, Y.~K., \& Bertelli, G.\ 1998, \aap, 329, 943 

\bibitem[Oort(1950)]{oort50a} Oort, J.~H.\ 1950, \bain, 11, 91 

\bibitem[Ozernoy et al.(2000)]{ozernoy00a} Ozernoy, L.~M., Gorkavyi, N.~N., Mather, J.~C., \& Taidakova, T.~A.\ 2000, \apjl, 537, L147 

\bibitem[Peale \& Cassen(1978)]{peale78a} Peale, S.~J., \& Cassen, P.\ 1978, \icarus, 36, 245 

\bibitem[Peale et al.(1979)]{peale79a} Peale, S.~J., Cassen, P., \& Reynolds, R.~T.\ 1979, Science, 203, 892 

\bibitem[Peters \& Turner(2012)]{peters12a} Peters, M.~A., \& Turner, E.~L.\ 2012, arXiv:1209.4418 

\bibitem[Quillen(2006)]{quillen06a}{Quillen, A.} 2006, \mnras, 372, L14

%

\bibitem[Raftery \& Lewis(1995)]{raftery95a} Raftery \& Lewis, in ÒPractical Markov Chain Monte CarloÓ, D.J. Spiegelhalter W.R. Gilks, \& S. Richardson, eds., Chapman and Hall, London.

\bibitem[Rasio \& Ford(1996)]{rasio96a} Rasio, F.~A., \& Ford, E.~B.\ 1996, Science, 274, 954 

\bibitem[Raymond et al.(2012)]{raymond12a} Raymond, S.~N., Armitage, P.~J., Moro-Mart{\'{\i}}n, A., et al.\ 2012, \aap, 541, A11 

\bibitem[Reche et al.(2009)]{reche09a} Reche, R., Beust, H., \& Augereau, J.-C.\ 2009, \aap, 493, 661 

\bibitem[Ricci et al.(2012)]{ricci12a} Ricci, L., Testi, L., Maddison, S.~T., \& Wilner, D.~J.\ 2012, \aap, 539, L6 

\bibitem[Roddier et al.(1996)]{roddier96a} Roddier, C., Roddier, F., Northcott, M.~J., Graves, J.~E., \& Jim, K.\ 1996, \apj, 463, 326 

\bibitem[Ryan \& Melosh(1998)]{ryan98a} Ryan, E.~V., \& Melosh, H.~J.\ 1998, \icarus, 133, 1 

%

\bibitem[Savransky (2011)]{savransky11a} Savransky, D., 2011, "Estimation Theory Applications for Planet Finding", Ph.D. Thesis, Princeton University

\bibitem[Schlichting \& Sari(2009)]{schlichting09a} Schlichting, H.~E., \& Sari, R.\ 2009, \apj, 700, 1242 

\bibitem[Schneider et al.(2011)]{schneider11a} Schneider, J., Dedieu, C., Le Sidaner, P., Savalle, R., \& Zolotukhin, I.\ 2011, \aap, 532, A79 

\bibitem[Sekanina et 
al.(1992)]{sekanina92a} Sekanina, Z., Larson, S.~M., Hainaut, O., Smette, A., \& West, R.~M.\ 1992, \aap, 263, 367 

\bibitem[Shen 
\& Tremaine(2008)]{shen08a} Shen, Y., \& Tremaine, S.\ 2008, \aj, 136, 2453 


\bibitem[Sivia \& Skilling(2006)]{sivia06a} Sivia, S. \& Skilling, J. \ 2006, ``Data Analysis: A Bayesian Tutorial'', Oxford University Press.

%

\bibitem[Stapelfeldt et al.(2004)]{stapelfeldt04a} Stapelfeldt, K.~R., Holmes, E.~K., Chen, C., et al.\ 2004, \apjs, 154, 458 

\bibitem[Stewart \& Leinhardt(2012)]{stewart12a} Stewart, S.~T., \& Leinhardt, Z.~M.\ 2012, \apj, 751, 32 

\bibitem[Soter(2006)]{soter06a} Soter, S.\ 2006, \aj, 132, 2513 

%

\bibitem[Su et al.(2013)]{su13a} Su, K.~Y.~L., Rieke, G.~H., Malhotra, R., et al.\ 2013, \apj, 763, 118 

\bibitem[Takeda \& Rasio(2005)]{takeda05a} Takeda, G., \& Rasio, F.~A.\ 2005, \apj, 627, 1001 

\bibitem[Terquem \& Ajmia(2010)]{terquem10a} Terquem, C., \& Ajmia, A.\ 2010, \mnras, 404, 409 

%

\bibitem[Thommes et al.(2008)]{thommes08a} Thommes, E.~W., Bryden, G., Wu, Y., \& Rasio, F.~A.\ 2008, \apj, 675, 1538 

\bibitem[Trafton(1975)]{trafton75a} Trafton, L.\ 1975, \nat, 258, 690 

\bibitem[Tremaine(1993)]{tremaine93}Tremaine, S.  1993, in {\it Planets Around Pulsars}, ASP Conf. Series, Vol. 35, p. 335

\bibitem[Trujillo et al.(2000)]{trujillo00a} Trujillo, C.~A., Jewitt, D.~C., \& Luu, J.~X.\ 2000, \apjl, 529, L103 

\bibitem[Tsiganis et al.(2005)]{tsiganis05a} Tsiganis, K., Gomes, R., Morbidelli, A., \& Levison, H.~F.\ 2005, \nat, 435, 459 

\bibitem[Veras \& Armitage(2004)]{veras04a} Veras, D., \& Armitage, P.~J.\ 2004, \icarus, 172, 349 

\bibitem[Veras et al.(2009)]{veras09a} Veras, D., Crepp, J.~R., \& Ford, E.~B.\ 2009, \apj, 696, 1600 

\bibitem[Veras \& Evans(2013)]{veras13a} Veras, D., \& Evans, N.~W.\ 2013, \mnras, 430, 403 

\bibitem[Verbiscer et al.(2009)]{verbiscer09a}{Verbiscer, A.J., Skrutskie, M.F., \& Hamilton, D.P.} 2009, \textit{Nature}, 461, 1098

\bibitem[Volk \& Malhotra(2012)]{volk12a} Volk, K., \& Malhotra, R.\ 2012, \icarus, 221, 106 

\bibitem[Weidenschilling(1975)]{weidenschilling75a} Weidenschilling, S.~J.\ 1975, \aj, 80, 145 

\bibitem[Woodgate et al.(1998)]{woodgate98a} Woodgate, B.~E., Kimble, R.~A., Bowers, C.~W., et al.\ 1998, \pasp, 110, 1183 

\bibitem[Wu \& Murray(2003)]{wu03a} Wu, Y., \& Murray, N.\ 2003, \apj, 589, 605 

\bibitem[Wyatt et al.(1999)]{wyatt99a}{Wyatt, M.C., et al.}, 1999, \apj, 527, 918

\bibitem[Wyatt \& Dent(2002)]{wyatt02a} Wyatt, M.~C., \& Dent, W.~R.~F.\ 2002, \mnras, 334, 589 

\bibitem[Wyatt et al.(2005)]{wyatt05a} Wyatt, M. C., Greaves, J.S., Dent, W.R.F. \& Coulson, I. M. 2005, \apj, 620, 492

\bibitem[Wyatt(2005)]{wyatt05b} Wyatt, M.~C.\ 2005, \aap, 440, 937 

\bibitem[Wyatt et al.(2007)]{wyatt07a} Wyatt, M. C., Smith, R., Su, K.Y.L., et al.  2007, \apj, 663, 365

\bibitem[Wyatt et al.(2010)]{wyatt10a} Wyatt, M.~C., Booth, M., Payne, M.~J., \& Churcher, L.~J.\ 2010, \mnras, 402, 657 

\bibitem[Wyatt et al.(2011)]{wyatt11a} Wyatt, M. C., Clark, C.J., \& Booth, M. 2011, Celest. Mech. Dyn. Astr, 111, 1

\bibitem[Zahnle \& Mac Low(1995)]{zahnle95a} Zahnle, K., \& Mac Low, M.-M.\ 1995, \jgr, 100, 16885 

\bibitem[Zakamska \& Tremaine(2004)]{zakamska04a} Zakamska, N.~L., \& Tremaine, S.\ 2004, \aj, 128, 869 

\bibitem[Zuckerman \& Becklin(1993)]{zuckerman93a} Zuckerman, B. \& Becklin, E.E., 1993, \apj, 414, 793


\end{thebibliography}
\end{document}